\documentclass[pre,amsmath,amssymb,floatfix,superscriptaddress]{revtex4}
\usepackage{graphicx}
\usepackage{epsfig}
\usepackage{dcolumn}
\usepackage{bm}
\usepackage{subfigure}
\usepackage{bbm}

\let\a=\alpha \let\b=\beta \let\g=\gamma \let\d=\delta
  \let\h=\eta 
\let\l=\lambda \let\m=\mu   
 \let\t=\tau  
   \let\G=\Gamma
 \let\L=\Lambda  
\let\Si=\Sigma   \let\Y=\Upsilon
 \let\r=\rho  \let\io=\infty

\def\ie{{i.e. }}\def\eg{{e.g. }}

\def\PP{{\cal P}}

\def\ZZ{{\cal Z}}

\def\to{\rightarrow} \def\la{\left\langle} \def\ra{\right\rangle}

 \newcommand{\wt}{\widetilde}
\newcommand{\Tr}{\text{Tr}}

\def\(({\left(}
\def\)){\right)}
\def\[[{\left[}
\def\]]{\right]}

\def\un{{\underline{n}}}

\def\u0{{\underline{0}}}

\def\bn{{\boldsymbol{n}}}
\def\bh{{\boldsymbol{h}}}
\def\bq{{\boldsymbol{q}}}

\def\ubn{{\underline{\boldsymbol{n}}}}
\def\uby{{\underline{\boldsymbol{y}}}}
\def\ubnu{{\underline{\boldsymbol{\nu}}}}

\def\by{{\boldsymbol{y}}}

\def\bY{{\boldsymbol{\Y}}}
\def\ubY{{\underline{\boldsymbol{\Y}}}}
\def\bnu{{\boldsymbol{\nu}}}
\def\bs{{\boldsymbol{s}}}
\def\uby{{\underline{\boldsymbol{y}}}}

\def\Ns{{N_{\rm s}}}
\def\Nt{{{\cal N}_{\rm traj}}}

\def\hW{\widehat{W}}

\def\Tr{{\rm Tr}}

\def\bh{{\boldsymbol{h}}}

\def\I{{\rm 1\hspace{-0.90ex}1}}

\def\Z{{\cal Z}}
\def\di{{\partial i}}
\def\dimj{{\partial i \setminus j}}
\def\ij{{\langle i,j\rangle }}
\def\ik{{\langle i,k\rangle }}
\newcommand{\beq}{\begin{equation}} 
\newcommand{\eeq}{\end{equation}}
\newcommand{\bea}{\begin{eqnarray}} 
\newcommand{\eea}{\end{eqnarray}}

\begin{document}

\title{The quantum Biroli-M\'ezard model: \\ glass transition and superfluidity
in a quantum lattice glass model}

\author{Laura Foini}
\affiliation{Laboratoire de Physique Th\'eorique, Ecole Normale Sup\'erieure,  CNRS UMR 8549,
 24 Rue Lhomond, 75005 Paris, France}
\affiliation{
SISSA and INFN, Sezione di Trieste, via Bonomea 265, I-34136 Trieste, Italy
}

\author{Guilhem Semerjian}
\affiliation{Laboratoire de Physique Th\'eorique, Ecole Normale Sup\'erieure,  CNRS UMR 8549,
 24 Rue Lhomond, 75005 Paris, France}

\author{Francesco Zamponi}
\affiliation{Laboratoire de Physique Th\'eorique, Ecole Normale Sup\'erieure,  CNRS UMR 8549,
 24 Rue Lhomond, 75005 Paris, France}

\begin{abstract}
We study the quantum version of a lattice model whose classical counterpart 
captures the physics of structural glasses. 
We discuss the role of quantum fluctuations in such systems and in 
particular their interplay with the amorphous order developed 
in the glass phase. We show that quantum fluctuations might facilitate
the formation of the glass at low enough temperature. We also show
that the glass transition becomes a first-order transition
between a superfluid and an insulating glass at very
low temperature, and is therefore accompanied by phase coexistence between
superfluid and glassy regions.
\end{abstract}

\maketitle

\section{Introduction}

It is a well established fact that certain
liquids form a glass if cooled sufficiently 
fast to avoid crystallization. The simplest examples of such glass-formers
are models of particles interacting via Lennard-Jones--like~\cite{KA95a,KA95b} 
and hard spheres~\cite{PV86} potentials, but a similar phenomenology 
is observed in many real systems like Silica, molecular systems such as 
Glycerol and OTP, and many polymers~\cite{De96,BK05}. The precise meaning of 
``sufficiently fast to avoid crystallization'' is strongly system-dependent,
with time scales ranging from nanoseconds for three-dimensional monoatomic
systems, to hours or centuries for more complex systems like binary mixtures
or molecular and polymeric systems. If crystallization is avoided, below
the melting temperature $T_m$ the system
enters a ``supercooled liquid'' state~\cite{De96}, which
can be treated as a metastable equilibrium state and hence described using tools
from equilibrium statistical mechanics. The transition to the glass happens at
a lower temperature $T_g$, at which the relaxation time of the metastable
liquid exceeds the experimentally accessible time scales and the system behaves
as a solid for any practical purpose~\cite{De96,BK05}. 
The glass transition can be driven by lowering the temperature $T$
or increasing the density $\r$, and one can define a line
of transitions $T_g(\r)$ which delimits the ``glassy'' region of the phase
diagram.
It is interesting to note that although the glass transition
happens out of equilibrium, its phenomenology is very close to that of a thermodynamical
second order phase transition, signaled by
a jump in specific heat in absence of any latent heat of solidification. It has been
conjectured by many authors that a true thermodynamic phase transition happens at a
lower temperature $T_K(\r)$ and is responsible for the non-equilibrium glass transition
at $T_g$.

It is reasonable to expect that at low enough temperatures and high enough density
(the precise scales depending on the details of the interaction potential), 
quantum fluctuations 
(quantified for instance by the ratio $\G=\L/a$, where $\L$ is the De Broglie
thermal wavelength and $a$ is the interparticle distance)
becomes relevant. A typical example is that of He$^4$,
that forms a solid at temperatures so low that quantum effects are very 
important~\cite{Wi67}.
Another interesting example is that of hydrogen at very high pressure and low 
temperature~\cite{BSOG04}. Yet, the possible 
presence of glassy phases in these quantum
systems has been proposed only very recently~\cite{BPS06,BCZ08}, and its
experimental observation is still debated~\cite{HPGYBD09,BC08}.
These preliminary observations raise the natural question
whether the classical picture of the glass transition is strongly
modified by quantum fluctuations. In particular, one would like to know
what is the nature of the transition (does it remain second order?), what
is the shape of the glass transition line in the $(T, \r, \G)$ space, what happens
to the quantum dynamics in the vicinity of the glass transition (is tunnelling
between distinct glassy metastable states possible?), and so on. Due to the very complex nature
of the glass transition, that is characterized by the breaking of translational
invariance towards an amorphous ordering and by very slow dynamics, it is very
hard to address these questions using numerical methods such as Quantum Monte
Carlo~\cite{BPS06} or exact diagonalization.

As already mentioned, for classical systems a widespread approach to the glass
transition consists in treating it as a true phase transition, using tools
of equilibrium statistical mechanics. After several decades of efforts, a coherent
theory of the glass transition based on this approach has emerged and now goes
under the name of Random First Order Transition (RFOT) theory~\cite{LW07}.
According to this theory, the glass transition belongs to a new classification: it
is a second order transition from a thermodynamic point of view, but its order parameter
(that we will define later on) jumps at the transition and many physical properties around
the transition are related to some nucleation-like physics (hence the transition looks more
like a first order transition in this respect, which is at the origin of the name RFOT).
The main inspiration for this theory came from density functional theory~\cite{SSW85,KW87},
from Mode-Coupling Theory (a dynamic theory of supercooled liquids that emerged
in the context of critical phenomena)~\cite{KW87,MCT}
and from the exact solution of a class
of mean-field spin glass models (the so-called $p$-spin glasses)~\cite{KT87,KTW89,CK93} 
that display a phenomenology extremely
close to that of glass-forming liquids. 
A complete review of RFOT theory goes well beyond
the scope of this paper, and we refer the reader 
to the excellent reviews~\cite{LW07,Ca09,BB09} for more details.
What is relevant for the present discussion is the existence of mean field 
lattice models
of glasses~\cite{BM01,RBMM04,PTCC03} that are extremely simple (they are formulated
in terms of hard particles on a lattice with simple local many body repulsion),
exactly solvable (thanks to the mean field nature of the lattice one consider) and
at the same time display the full RFOT phenomenology: in particular for these models
one can compute the glass transition line $T_K(\r)$, and one can also study the 
dynamics on approaching the glass transition from the liquid. These models are a very
natural starting point for the investigation of the glass transition in the quantum case.
Moreover, the RFOT construction can be implemented under controlled approximations
to compute {\it quantitatively} the glass transition in realistic systems, such as
Lennard-Jones~\cite{MP99,MP09} or hard spheres~\cite{PZ10}.

The extension of RFOT theory to quantum systems has been initiated long time ago
by studying the quantum version of the $p$-spin glasses~\cite{Go90,NR98,BC01,CGS01,JKKM08},
stripe glasses~\cite{WSW03} and of some
optimization problems~\cite{JKSZ10} that share the same qualitative phenomenology. 
The main result of these studies is that the high temperature (RFOT-like) glass
transition is turned into a first order transition between the liquid and the glass
at very low temperature.
However, the picture provided by these studies 
is not complete, since all the models that have been studied up to now are in the
glass phase at low enough $T$ in absence of quantum fluctuations; in other words,
none of these models displays, in the classical limit $\G=0$, 
a glass transition at low temperature as a function
of another control parameter (e.g. the density).
The possibility of studying the glass transition at low $T$ as a function of $\r$
adds an important ingredient to the discussion, which is entropy. In fact, the 
models studied in \cite{Go90,NR98,BC01,CGS01,JKKM08,WSW03,JKSZ10} have a non-extensively
degenerate ground state, hence their entropy at $T=0$ vanishes. Conversely,
lattice glass models might have a finite entropy even at $T=0$ (think for instance
to hard spheres~\cite{PZ10}) and in this case the glass transition is completely driven by
entropy. A preliminary attempt to take into account the role of entropy in this class of problems
has been reported for a toy model in~\cite{FSZ10}, where it has been shown that due to the 
interplay of classical entropy and quantum fluctuations, the latter may induce the formation of a 
glass: from a liquid phase at $\G=0$ the system can fall in a glassy situation as soon as
$\G>0$. A similar phenomenon has been found in~\cite{MMBMRR10} for a system of quantum hard spheres.

Finally, a very interesting point that has been raised in the study of quantum glasses
is the possibility that a glass display superfluidity~\cite{BPS06,BCZ08}. 
This would lead to a particular supersolid phase (a ``superglass''). However,
the existence of such a phase has by now only been shown numerically~\cite{BPS06}
or analytically in a quite special model~\cite{BCZ08}.
For systems displaying a standard second order spin glass transition~\cite{MPV87} (which is quite
different from a RFOT), a superglass phase has been shown to exist 
generically~\cite{CTZ09,TGIGM10}. However, the physics of RFOT is quite different
from that of a second order spin glass transition, therefore the extension of these
results to the RFOT scenario is not straightforward.

In this work we shall study a Bosonic quantum extension of the Biroli-M\'ezard 
lattice glass model~\cite{BM01}. 
Our aim is to study the effect of quantum fluctuations 
in this system and in particular their interplay with the amorphous order 
developed in the glass phase, and with the large degeneracy of states leading to a 
finite entropy at $T=0$. 
Specifically, we want to compute the full phase diagram
of the system to find the liquid-glass transition line, and look for a superglass phase.
We will do this by means of the recently developed quantum cavity
method~\cite{LSS08,KRSZ08,STZ09}.
We find a complex phase diagram at low temperature,
characterized by two unexpected phenomena: {\it i)} A re-entrance of the glass transition
line towards lower density on increasing quantum fluctuations, driven by entropy;
{\it ii)} A first order phase transition between superfluid and glass at zero temperature.
Our results suggest that a true superglass phase is not present in the model, but we can establish
the presence of phase coexistence between superfluid and glass, which might have interesting
phenomenological consequences.

The rest of the paper is divided as follows. In section \ref{classic} we introduce the classical model 
clarifying its importance in the description of classical glasses.  In section \ref{classic_rs}
we derive the so-called ``cavity-equations'' describing the system in the replica symmetric 
ansatz, 
which provides the exact solution of the liquid phase. Then, in section \ref{classic_rsb} 
and \ref{sec:methods_classic}, we 
describe how to access its glass phase within the replica symmetry breaking scheme.
A summary of the results for the classical model is given in section \ref{results_classic}.
In section \ref{quantum_model} we generalize the model as a quantum many-body system, made 
of hard-core bosons, with the same interaction potential as
the classical one but where quantum effects are induced by an hopping term. 
The definition of the model is given in section \ref{quantum_model_definition}. The rest of 
Sec. \ref{quantum_model} is more technical and can be skipped by readers not interested in the 
formalism of the quantum cavity method used to solve the model; in section \ref{qcavity_method}
we derive the ``cavity equations'' which describe the physics of the
system and in sections \ref{qcavity_equations_rrg} and \ref{resolution_qcavity_equations} we sketch their numerical resolution.
In section \ref{results_quantum} we present the main results for the quantum system.
In section \ref{sec:order_parameters} we define the order parameters of the different phases;
in section~\ref{sec:details_cavity} we discuss the details of the cavity computations;
in section \ref{quantum_Tfinite} we report the main results on the phase diagram of the model.
In section \ref{sec:quantum_dynamics} we investigate the quantum dynamics of the model: we show
that at the glass transition there is ergodicity breaking and we discuss its relation with the spectrum
of the quantum Hamiltonian. In section \ref{sec:ED} we discuss some exact diagonalization results.
Finally, in section~\ref{sec:conclusions} we summarize and discuss our results.

\section{The classical model}
\label{classic}

The classical model considered in the following is a generalization of a lattice glass model 
proposed in \cite{BM01,RBMM04} and widely used as a prototype for the description of
particle systems undergoing a glass transition. Its degrees of freedom are occupation numbers
$n_i=\{0,1\}$ of the sites $i=1,\dots,N$ of a graph, a particle being present on site $i$ if and
only if $n_i=1$. If a site $i$ is occupied it feels a repulsive interaction with the particles
on the neighboring sites of the graph, the set of which shall be denoted $\di$. More precisely,
each particle can have at most $\ell$ neighbors without any energy cost and it is 
subject to an interaction $V>0$ for every neighboring particle in excess.
The Hamiltonian describing this interaction is the following:
\beq
H = V \sum_i n_i \, q_i \, \theta(q_i)- \mu \sum_i n_i \ ,
\label{eq_Hclass}
\eeq
where $q_i=\sum_{j\in\di} n_j-\ell$ is the number of neighbors ``in excess''.
Here and in the following $\theta(x)=0$ if $x<0$ and $\theta(x)=1$ if $x\geq0$. We included
in $H$ the chemical potential $\mu$ conjugated to the total particle number.
The original model of \cite{BM01,RBMM04} corresponds to the case $V\to\io$; configurations
where one particle has more than $\ell$ neighboring sites occupied are then strictly forbidden.
This finite $V$ version will be more convenient for the quantum generalization
described later on.

The model above can be defined on any finite graph. Its finite dimensional version has been
studied numerically by Monte Carlo simulations in~\cite{BM01,DLGMZT02,DRB10}, demonstrating
that it exhibits the phenomenology of glassy systems. Note that there is no quenched disorder 
in the Hamiltonian; as in real liquids the disorder and frustration that yield an amorphous glassy 
order are in fact self-generated by the different possible arrangements of particles.
In order to obtain analytical results we will study the model on the Bethe lattice,
as in \cite{BM01,RBMM04}. This underlying 
geometry is known to give a mean field description of the system, with the advantage of 
preserving the notion of distance and that of finite connectivity~\cite{cavity,STZ09}. 
More precisely, Bethe lattices are intended here
as random $c$-regular graphs, i.e. graphs chosen uniformly among the set of 
graphs with $N$ vertices where all sites have precisely $c$ neighbors. These graphs converge
locally to trees in the thermodynamic (large $N$) limit.
There are several motivations to study the problem on this geometry~\cite{cavity}.
Due to the local tree-like structure, in fact, on this kind of graphs one is able to take 
explicitly the thermodynamic limit and derive the exact solution of the model.  
On the other hand, having no boundary and all sites playing statistically the same role, 
random regular graphs allow to get rid of boundary problems which affect trees. Still, even if locally
tree-like, they do have loops, which are an essential ingredient for the frustration induced in 
the system. Moreover, their randomness forbids crystalline ordering, making easier the study of glassy phases.
All these properties make the Bethe lattice a convenient playground for the study 
of frustrated systems.
More generally, however, the Bethe lattice can also give insight into
Euclidean lattices. In fact, the solution that one gets on random regular graphs
can be thought of as a ``refined'' (with respect to the standard fully-connected model) 
mean field approximation for a finite dimensional lattice with the same connectivity.
Finally the free energy of the Bethe lattice has the important property of being self-averaging, 
in the thermodynamic limit, with respect to the uniform measure which defines the ensemble.
This ensures that for large enough size already one exemplar is a good representative 
of the entire ensemble.

For completeness let us give here the definition and the relations between thermodynamic
observables, i.e. the partition sum at temperature
$T=1/\b$, $Z(\beta,\mu) = \sum_{\{{n\}}} e^{- \beta H(\{{n\}})} = e^{- \beta N f(\mu,\beta)}$,
the Gibbs-Boltzmann distribution $\mu_B(\{{n\}})=e^{- \beta H(\{{n\}})} / Z(\beta,\mu)$,
the free energy per particle
$f(\mu,\beta) = e -  \mu \rho - T s$, where
$e = V \partial_V f$, $s = - \partial_T f$, $\rho = - \partial_\mu f$
are respectively the average energy and entropy per particle, and the density.

\subsection{The liquid phase: RS solution}
\label{classic_rs}

\begin{figure}
\includegraphics[width=8cm]{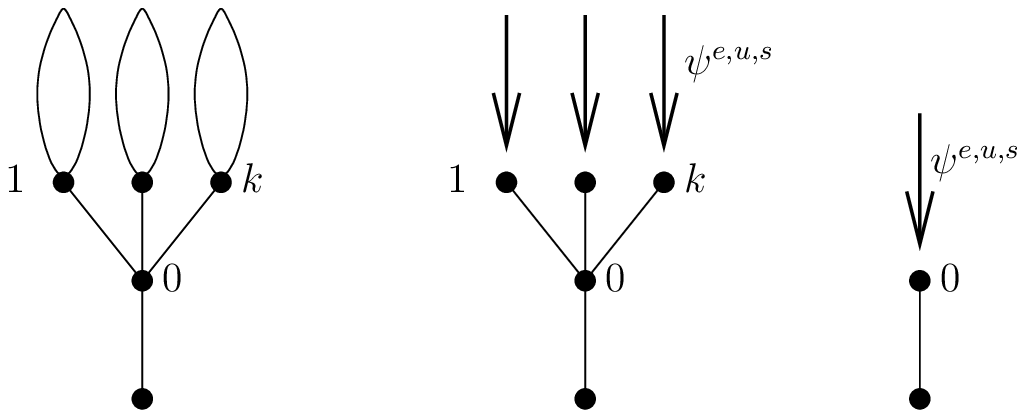}
\caption{
Illustration of a cavity iteration for the classical system.
The sites $1,\cdots,k$ are described, in absence of site $0$, 
by cavity fields $\psi^{e,u,s}_l$, $l=1,\cdots,k$.
Integrating out these sites leads to the expression of 
$\psi^{e,u,s}_0$ in Eq.~(\ref{psi_iter}).
}
\label{fig:iter}
\end{figure}

We now proceed to discuss the cavity method at the 
replica symmetric (RS) level, extending the derivation of~\cite{RBMM04} to the
soft (finite $V$) model. This calculation provides the exact solution of the problem in the liquid,
homogeneous, phase. As we mentioned above we consider the thermodynamic limit of random regular graphs 
with connectivity $c=k+1$, which locally converge to regular trees. The derivation of the RS cavity equations
presented in this section starts by considering the solution of the model on a finite tree (a slightly different
method will be used in section~\ref{qcavity_method}). As represented pictorially in Fig.~\ref{fig:iter},
trees admit naturally a recursive description. Consider the sites $1,\dots,k$ in absence of their common
neighbor $0$ (they become so-called ``cavity sites''). Because of the tree structure the Gibbs-Boltzmann 
probability distribution factorizes over the $k$-subtrees rooted at $1,\dots,k$. Each ``cavity site'' $i$
can be found in three different states: \textit{empty} if $n_i=0$,
\textit{saturated} if $n_i=1$ and if the number of occupied neighbors \emph{above} it (which is $k$ at most) 
is $\geq \ell$ and \textit{unsaturated} if $n_i=1$ and if the number of neighboring particles above is $< \ell$.
We shall denote $\{\psi_i^e, \psi_i^u, \psi_i^s\}$ the respective probabilities of these three states with
respect to the Gibbs-Boltzmann measure. Thanks to the factorization properties of trees one can obtain
recursion relations between these cavity probabilities~\cite{cavity,RBMM04}:
\beq\label{psi_iter}
\begin{split}
\psi^e_{0} & = \frac 1 {Z_{\rm iter}} \\
\psi^u_{0} & = \frac{e^{\beta\mu}}{Z_{\rm iter}} \left(\prod_{i=1}^k \psi^e_{i}\right) 
\sum_{j=0}^{\ell-1} \sum\limits_{\substack{1\leq i_1<\dots<i_j\leq k}}~
\prod_{p=1}^j  \frac{\psi_{i_p}^u+\psi_{i_p}^s ~e^{- \beta V} }{\psi_{i_p}^e } \\
\psi^s_{0} & = \frac{e^{\beta\mu}}{Z_{\rm iter}} \left( \prod_{i=1}^k \psi^e_{i}\right)\sum_{j=\ell}^{k}
 e^{(\ell-j)\beta V} 
\sum\limits_{\substack{1\leq i_1<\dots<i_j\leq k}}~
\prod_{p=1}^j  \frac{\psi_{i_p}^u+\psi_{i_p}^s ~e^{- \beta V} }{\psi_{i_p}^e }
\end{split}
\eeq
where  $Z_{\rm iter}$ is a normalization constant ensuring $\psi^e_{0} + \psi^u_{0} +\psi^s_{0} = 1$.
Note that the $j=0$ term in the sum in the second line of the above equation (which is the only one
for $\ell=1$) should be interpreted formally as giving a contribution equal to 1. 
Deep inside a very large regular tree, for an homogeneous liquid phase, the cavity probabilities converge
to the fixed-point solution of Eq.~(\ref{psi_iter}),
\beq\label{psi_rs}
\begin{split}
\psi^e & = \frac 1 {Z_{\rm iter}} \\
\psi^u & = \frac{e^{\beta\mu}}{Z_{\rm iter}} \sum_{i=0}^{\ell-1}  {k \choose i} (\psi^e)^{k-i} 
(\psi^u+\psi^s e^{- \beta V})^i   \\
\psi^s & = \frac{e^{\beta\mu}}{Z_{\rm iter}} \sum_{i=\ell}^{k}  {k \choose i} (\psi^e)^{k-i} 
(\psi^u+\psi^s e^{- \beta V})^i  e^{(\ell-i)\beta V}
\ .
\end{split}
\eeq
The solution of Eq.~(\ref{psi_rs}) can easily be found numerically for any choice of the parameters 
$\beta, \mu, V$. 

The recursive equations written above were strictly valid for trees. The local tree-like character of
random graphs allows however, under the correlation decay hypothesis of the RS cavity method, to apply 
these results to the thermodynamic limit of Bethe lattices. One finds in particular for
the free energy density~\cite{cavity,RBMM04}:
\beq\label{f_rs}
f=  \Delta f_{\rm site} - \frac{c}{2} \Delta f_{\rm link}  =  -\frac{1}{\beta}\log{Z_{\rm site}} + \frac {c}{2\beta} \log{Z_{\rm link}}
\eeq
with 
\beq
\begin{split}\label{rs_z_sl}
Z_{\rm link} & = (\psi^e)^2  + 2\psi^e \psi^u + (\psi^u)^2 + 2 \psi^e \psi^s + 2 \psi^s \psi^u e^{-\beta V} + (\psi^s)^2 e^{-2\beta V} \ , \\
Z_{\rm site} & = 1+ e^{\beta\mu} \left[ \sum_{i=0}^{\ell-1} {k+1 \choose i} (\psi^e)^{k+1-i} (\psi^u+\psi^s e^{-\beta V})^i 
 +\sum_{i=\ell}^{k+1} {k+1 \choose i} (\psi^e)^{k+1-i} (\psi^u+\psi^s e^{-\beta V})^i  e^{(\ell-i)\beta V} \right] \ .
\end{split}
\eeq
All thermodynamic observables can then be obtained from Eq.~(\ref{f_rs}) by derivation with respect to the
parameters $\beta,\mu,V$. Actually only their explicit dependence has to be derived, the order parameter 
equations (\ref{psi_rs}) being stationary point conditions of $f$ with respect to $\psi^{e,u,s}$.

The replica symmetric solution provides the natural description for the 
homogeneous, liquid (or paramagnetic) phase where all correlations are short range.
However, the correctness of the RS ansatz breaks down when one approaches the glass phase, 
where the system develops amorphous density profiles and long range correlations. 
In this model this happens at low enough temperature, when the competition between the 
chemical potential $\mu$ and the interaction energy $V$ is the strongest. One evidence
for the inconsistency of the RS assumption is the fact that the associated entropy becomes
negative in this region of parameters, which is impossible for a discrete model.
Another criterion for the incorrectness of the RS ansatz is the divergence of the so-called ``spin-glass'' 
susceptibility, defined as
$\chi_{\rm SG} (\beta,\mu) = N^{-1} \sum_{i,j}\langle n_in_j\rangle_c^2$. For all parameters we
investigated we found no divergence of the susceptibility. 
This means that the resolution of the \textit{entropy crisis} requires a phase transition which has to
be discontinuous since it is not associated to a diverging susceptibility. 
This scenario is typical for models displaying \textit{1 step replica symmetry breaking} (1RSB) \cite{RBMM04}.

\subsection{The glass phase: 1RSB solution}
\label{classic_rsb}

The replica-symmetric approach stated above was based on an hypothesis of correlation decay, which allowed
to neglect the presence of long loops in random graphs and to use the results derived on a tree. This 
assumption fails at low temperature, and must then be replaced by the first step of replica symmetry breaking
(1RSB). In the latter description one partitions the configuration space of the model in pure states,
identified by an index $\alpha$, each of them having free-energy $f_\alpha$. One assumes then that the 
Gibbs-Boltzmann measure restricted to the pure state $\alpha$ enjoys the correlation decay property of the RS 
ansatz, in other words that there is a (non-homogeneous) solution of the equations (\ref{psi_iter}) for
each pure state $\alpha$, and that its free-energy $f_\alpha$ is given by the RS expression~(\ref{f_rs}).
The cavity method at the 1RSB level corresponds then to a statistical treatment of these pure states.
One introduces the ``complexity'' or ``configurational entropy'' $\Sigma(f)$, such that the number of pure
states $\alpha$ with free-energy $f_\alpha \approx f$ is $\sim e^{N\Sigma(f)}$. The computation of this function
is done through the partition function~\cite{Mo95,cavity,RBMM04}
\beq\label{z}
Z(m;\beta,\mu) = \sum_{\alpha} e^{-\beta N m f_\alpha} \simeq 
\int_{f_{\rm min}}^{f_{\rm max}} df e^{N( \Sigma(f) - \beta m f)} \simeq 
e^{ - N \beta \min_{f \in [f_{\rm min},f_{\rm max}]}  (m f - \frac{1}{\beta} 
\Sigma(f))} = e^{- N \beta m \phi(m;\beta,\mu)}
\ ,
\eeq
where the sum is over all the pure states, and the interval $[f_{\rm min},f_{\rm max}]$ corresponds to
the domain where $\Sigma(f)\ge0$. The parameter $m$, which is 
known in the literature as the Parisi parameter, plays the role of a Lagrange parameter conjugated to $f$
(an effective temperature). Indeed the ``replicated free energy''$ \phi(m)$ and $\Sigma$ are Legendre 
transforms one of the other~\cite{Mo95},
\beq\label{sigma}
m \, \phi(m) = \min_{f \in [f_{\rm min},f_{\rm max}]} \{ mf - T \Sigma(f) \} \ ,
\eeq
and from the properties of the Legendre transform one has
\beq\label{properties}
\partial_f \Sigma(f) = m \beta \ ,\hspace{1cm} \Sigma = \beta \, m^2 \partial_m \phi(m) \ , \hspace{1cm} f = 
\partial_m [m \, \phi(m)] 
\ .
\eeq

The usual partition function is recovered from Eq. (\ref{z}) by setting $m=1$. However it can happen that
the saddle-point in (\ref{z}) occurs for $m=1$ outside the authorized interval $[f_{\rm min},f_{\rm max}]$.
In that case the dominant contribution to the partition function corresponds to the pure states with 
free-energy $f=f_{\rm min}$, which are in sub-exponential number as $\Sigma(f_{\rm min})=0$.

The computation of $\phi(m)$ is performed through a statistical study of the solutions of (\ref{psi_iter})
among the various pure states.
We call $\h=\{\psi^e,\psi^u,\psi^s\}$ the set of the (normalized) 
fields that appear in Eq.~(\ref{psi_iter}) and $\mathcal{P}(\h)$ their joint 
probability distribution. 
Eq.~(\ref{psi_iter}) defines a map $\h_0 = \hat{\h}(\h_1,\cdots,\h_k)$ that allows
to construct a new field from a set of $k$ fields.
Using this map, the 1RSB equations read:
\beq\label{rsb_eq}
\mathcal{P}(\h) = \frac{1}{\mathcal{Z}} \int \prod\limits_{i=1}^k~d\h_i~\mathcal{P}(\h_i) ~ \delta(\h-\hat{\h}(\h_1,...,\h_k)) ~
Z_{\rm iter}(\h_1,...,\h_k)^m
\ ,
\eeq
where $\mathcal{Z}$ is the normalization constant and $Z_{\rm iter} = \exp(-\b \Delta f_{\rm iter})$,
which is defined by the denominator of Eq.~(\ref{psi_iter}),
takes into account the shift in 
the free energy after a cavity iteration~\cite{cavity}.
Once that  $\mathcal{P}(\h)$ is known it is possible to compute all thermodynamic potentials, 
in particular the ``replicated free energy'', $\phi(m;\beta,\mu)$.
In fact, one has:
\beq
\phi = \Delta\phi_{\rm site} - \frac{c}{2} \Delta\phi_{\rm link}
\eeq
with
\bea
\Delta\phi_{\rm site} & = & - \frac{1}{\b m} \ln \int \left[ \prod\limits_{i=1}^{k+1}~d\h_i\mathcal{P}(\h_i) \right] ~ Z_{\rm site}(\h_1,...,\h_{k+1})^m \\
\Delta\phi_{\rm link}  & = &  - \frac{1}{\b m} \ln \int~d\h_1d\h_2~\mathcal{P}(\h_1)\mathcal{P}(\h_2) ~ Z_{\rm link}(\h_1,\h_{2})^m
\eea
and
\beq\label{fsite_flink}
\begin{split}
&Z_{\rm site}(\h_1\cdots \h_{k+1}) = 1 + e^{\beta\mu}\left( \prod_{i=1}^{k+1} \psi^e_{i}\right) \(( 
\sum_{j=0}^{\ell-1} \ \sum\limits_{\substack{1\leq i_1<\dots<i_j\leq k+1}}~
\prod_{p=1}^j  \frac{\psi_{i_p}^u+\psi_{i_p}^s ~e^{- \beta V} }{\psi_{i_p}^e } \right. \\
& \hspace{6cm} + \left.
\sum_{j=\ell}^{k+1} e^{(\ell-j)\beta V} 
\sum\limits_{\substack{1\leq i_1<\dots<i_j\leq k+1}}~
\prod_{p=1}^j  \frac{\psi_{i_p}^u+\psi_{i_p}^s ~e^{- \beta V} }{\psi_{i_p}^e }
\)) \ ,
\\
&Z_{\rm link}(\h_1,\h_2)= \psi^{e}_1  \psi^{e}_2 +  \psi^{e}_1  \psi^{u}_2 +  \psi^{u}_1  \psi^{e}_2 +   \psi^{u}_1  \psi^{u}_2 +  \psi^{e}_1  \psi^{s}_2 +  \psi^{s}_1  \psi^{e}_2 + \psi^{u}_1  \psi^{s}_2 e^{-\beta V}  + \psi^{s}_1  \psi^{u}_2 e^{-\beta V} + \psi^{s}_1  \psi^{s}_2 e^{-2\beta V} \ .
\end{split}
\eeq
Then, from $\phi(m;\beta,\mu)$, using (\ref{sigma}) and (\ref{properties}) with $m$ as a parameter, 
it is possible to compute the curve $\Sigma(f)$ for every value of $(\beta,\mu)$~\cite{Mo95}.

Depending on the value of the external parameters $\b, \mu, V$ and $m$ three situations can occur:
\begin{itemize}
\item Only the trivial replica symmetric solution exists at $m=1$, which is recovered when  
$\mathcal{P}(\h) = \delta(\h-\h_{rs})$. In this regime the RS approach is correct and the system is in its liquid phase.
\item  If at $m=1$ there is a non trivial solution of the 1RSB equations with positive complexity 
the Gibbs measure is dominated by 
an exponential number of states and the system is in its \textit{clustered} phase.
\item  If a non trivial solution exists at $m=1$, but the associated complexity is negative, this means that those 
states are absent in the thermodynamic limit. Instead the total free energy is dominated by the states of free energy
$f_{\rm max}$, that
satisfy $\Sigma(f_{\rm max})=0$. This condition is equivalent from Eq. (\ref{properties}) to the extremization condition 
$\partial_m \phi(m) = 0$. We call $m^\ast$ the value of $m$ at which the extremum is attained (note that we are only
interested in the region $m^\ast < 1$).
In this phase the system is said to be in its \textit{condensed} phase, as its equilibrium measure is supported
by a sub-exponential number of pure states.
\end{itemize}
The passage between these three regimes defines two transitions which are well-known in the context of 
glasses (see e.g.~\cite{CC05} for a review). 
The first one represents a dynamical transition at which the phase space of the system gets 
fragmented into an exponential number of states. At this point ergodicity breaking takes place, but the 
thermodynamics of the system is completely blind to the transition~\cite{CC05}. 
On the other hand, the second transition towards the condensed phase 
corresponds to the truly thermodynamic phase transition. 
This phase transition, which is usually called Kauzmann transition, is second order according to the 
Ehrenfest classification, but still it has peculiar features with respect to usual second order phase 
transitions, since the order parameter jumps at the transition and the physics around it is dominated
by nucleation in finite dimension. For these reasons it is also known under the name of 
Random First Order Transition (RFOT)~\cite{LW07}.

\subsection{Methods of solution of the cavity equations}
\label{sec:methods_classic}

The RSB equations (\ref{rsb_eq}) can be solved through the population dynamics method~\cite{cavity},
based on the representation of the probability distribution $\mathcal{P}(\h)$ as a weighted sample.
Starting with an initial representation  (\textit{population}) of $\mathcal{P}(\h)$ in  the form
\beq\label{class_pop_dyn}
\mathcal{P}(\h) = \sum_{i=1}^{{\cal N}} w_i \, \d(\h-\h_i)
\text{\hspace{0.5cm}such that \hspace{0.5cm}}
 \sum_{i=1}^{{\cal N}} w_i = 1
 \ ,
\eeq
the fixed point of Eq.(\ref{rsb_eq}) is found by iterating three main steps:
\begin{enumerate}
\item sample $k$ elements $\{\h_{i_1},\dots,\h_{i_k}\}$ from $\mathcal{P}(\h)$, independently with
the probabilities $w_{i}$;
\item compute a new field $\h=\hat{\h}(\h_{i_1},\dots,\h_{i_k})$ according to Eq.(\ref{psi_iter});
\item add to the new population the element $\h$ with weight $\propto Z_{\rm iter}(\h_{i_1},...,\h_{i_k})^m$.
\end{enumerate}
The update of the new population can be either parallel or serial. In the first case the same population
is used for ${\cal N}$ iterations, after which a new representation is available and it is substituted to 
the older one. Conversely, in the serial update the new generated element is immediately substituted 
in a random position in the pre-existing population. In some cases the parallel update can fail to find
a fixed point solution of the self-consistency equation (\ref{rsb_eq}) and enters instead a periodic orbit 
(modulation phenomenon); this spurious effect can be avoided using the serial update version of the 
population dynamics method.

Once convergence is reached all thermodynamic observables can be computed sampling from the
distribution. All the results of this section 
have been derived within this scheme where we fixed the size of the population 
equal to $2^{17}$ and we checked that finite population size effects were negligible.
Both in the classical and in the quantum case we used the serial update.
The population dynamics technique sketched here is in fact a particular case of the quantum population 
dynamics algorithm, and more details are presented in Sec.\ref{qcavity_equations_rrg}.

The numerical accuracy of the population dynamics obviously depends on the number $\cal N$ of representatives
of the distribution ${\cal P}(\h)$. However, a large value of $\cal N$ is not sufficient to ensure a good
precision of this discretized representation: if only a few weights $w_i$ dominate the sum in 
(\ref{class_pop_dyn}) one should consider that the effective size of the population is the number of
such dominant elements. To be more quantitative one can define an inverse participation ratio
\beq
\text{IPR} = \frac{1}{{\cal N}} \left(\sum\limits_{i=1}^{{\cal N}} w_i^2\right)^{-1}
\ ,
\eeq
where $w_i$ are the normalized weights of the sampled population defined in (\ref{class_pop_dyn}).
Its value is easily evaluated in the two limits of a perfectly balanced population and of a single
dominant weight:
\beq
\text{IPR} = 
\left\{ 
  \begin{array}{l l}
    {\cal O}(1) & \quad \text{if $w_i\sim{\cal O}(1/{\cal N})$}\\
    {\cal O}(1/{\cal N}) & \quad \text{if $w_i\sim \d_{i,i^{\ast}}$}\\
  \end{array} \right. \ ,
\eeq 
and one can thus define an effective population size as ${\cal N}_{\rm eff} = {\cal N} \times \text{IPR}$.
The inverse population ratio should thus be maintained as close as possible to one to achieve a good
numerical precision. We observed that the implementation of the population dynamics method with
the definition of the fields given in Eqs.(\ref{psi_iter}) leads to very small values of the IPR (this
can be traced back to the presence of the factor $e^{\b\m}$ which takes very high values for the relevant
values of the parameter). It turns out that a redefinition of the fields allows to bypass this problem.
We found much larger values of the IPR using
\beq
\phi^e_i = \frac{1}{Z'_{\rm iter}} \psi^e_i \qquad \phi^u_i = \frac{ e^{-\b\m r_u} 
}{Z'_{\rm iter}} \psi^u_i \qquad \phi^s_i = \frac{e^{-\b\m r_s}}{Z'_{\rm iter}}  \psi^s_i 
\ ,
\eeq
where $r_u$ and $r_s$ are two a priori arbitrary parameters (a good choice for the case $c=3$, $\ell=1$ 
was $r_u=1/5$ and $r_s=2/5$).
In terms of the transformed fields, the local self consistent equations (given here only for $\ell=1$
for simplicity) become:
\beq\label{psi_iter_transformed}
\begin{split}
\phi^e_{0} & = \frac 1 {Z'_{\rm iter}}  \prod_{i=1}^k \((\phi_{i}^e +
\phi_{i}^u e^{\beta\mu r_u}+\phi_{i}^s ~e^{\b\m r_s} \)) \\
\phi^u_{0} & = \frac{e^{\beta\mu(1-r_u)}}{Z'_{\rm iter}} \prod_{i=1}^k \phi^e_{i} \\
\phi^s_{0} & = \frac{e^{\beta\mu(1-r_s)}}{Z'_{\rm iter}}  \prod_{i=1}^k \phi^e_{i} ~\sum_{j=1}^{k} e^{(1-j)\beta V} 
\sum\limits_{\substack{1\leq i_1<\dots<i_j\leq k}}~
\prod_{p=1}^j  \frac{\phi_{i_p}^u e^{\beta\mu r_u}+\phi_{i_p}^s ~e^{- \beta V+\beta\mu r_s} }{\phi_{i_p}^e }\\
& = \frac{e^{\beta\mu(1-r_s)}}{Z'_{\rm iter}}  ~ e^{\beta V}\(( \prod_{i=1}^k \((\phi_{i}^e + e^{-\beta V} 
(\phi_{i}^u e^{\beta\mu r_u}+\phi_{i}^s ~e^{- \beta V+\b\m r_s}) \)) - \prod_{i=1}^k \phi_{i}^e \))
\ .
\end{split}
\eeq
Of course analogous transformations take place in the expression of $Z_{\rm site}$ and $Z_{\rm link}$ given
in Eq.~(\ref{fsite_flink}). This change of parameters was particularly important in the quantum case
discussed below where the population sizes are much more limited by computational costs.

\subsection{Thermodynamics of the classical model}
\label{results_classic}

\begin{figure*}
\centering%
 \includegraphics[width=12cm]{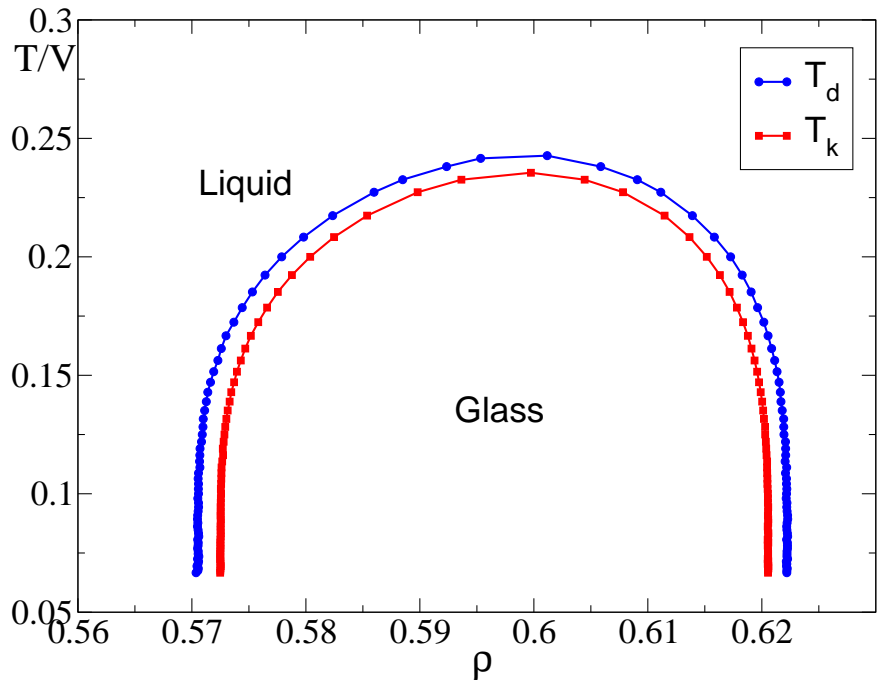}
\caption{Classical phase diagram in the $(\r,T)$ plane for $c=3$ and $\ell=1$. 
The dynamical ($T_d$) and Kauzmann ($T_K$) temperatures are plotted as a function of the density.
See Ref.~\cite{BMS10} for a study of a realistic model that displays a similar phase diagram.
}
\label{class_phase_diagram}
\end{figure*}

In Fig.~\ref{class_phase_diagram} we report the phase diagram for the classical model 
with $c=3$ and $\ell=1$, as a function of density and of temperature (in units of $V$). 
The two lines denote the dynamic and Kauzmann transitions, below which the system is found
in the glass phase. Increasing the density from a low value, for a fixed low enough temperature,
the system encounters first the dynamic transition and then the Kauzmann transition towards
the condensed glass phase.
However if one keeps increasing the density the system returns 
to a liquid phase; this is because if the chemical potential is large enough ($\mu \gg V$),
almost all sites will be occupied leaving only a small fraction of ``holes''. The system
of holes has a short range interaction similar to the original one (i.e. for particles) and is therefore liquid
when the density of holes is small enough, while it becomes glassy when the density of holes
is large enough.
The ``hole glass'' transition which is approached from higher density is therefore due to the fact
that we introduced a finite repulsion $V$ which can be overcome by a large enough $\mu$,
and it is qualitatively equivalent to the transition which takes place at 
lower densities. A very similar phase
diagram has been found in the study of a realistic model for colloidal systems~\cite{BMS10},
which confirms the validity of the Biroli-M\'ezard model to describe the physics of structural
glasses.

\begin{figure*}
  \centering
  \subfigure[][]{ \label{class_mu_rho}
    \includegraphics[width=0.5\textwidth,height=0.45\textwidth]{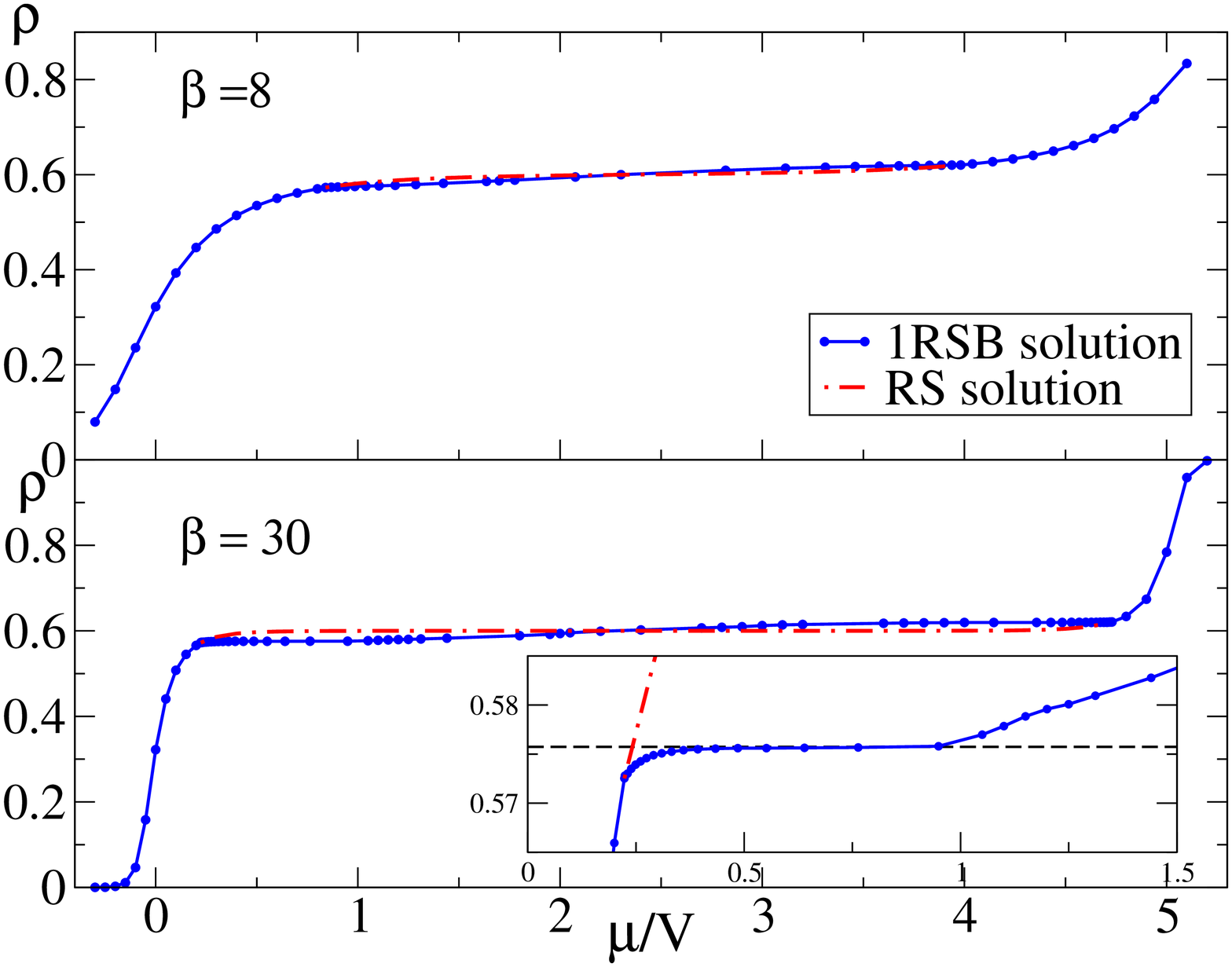}
  }
  \subfigure[][]
  { \label{class_m_rho}
    \includegraphics[width=0.47\textwidth]{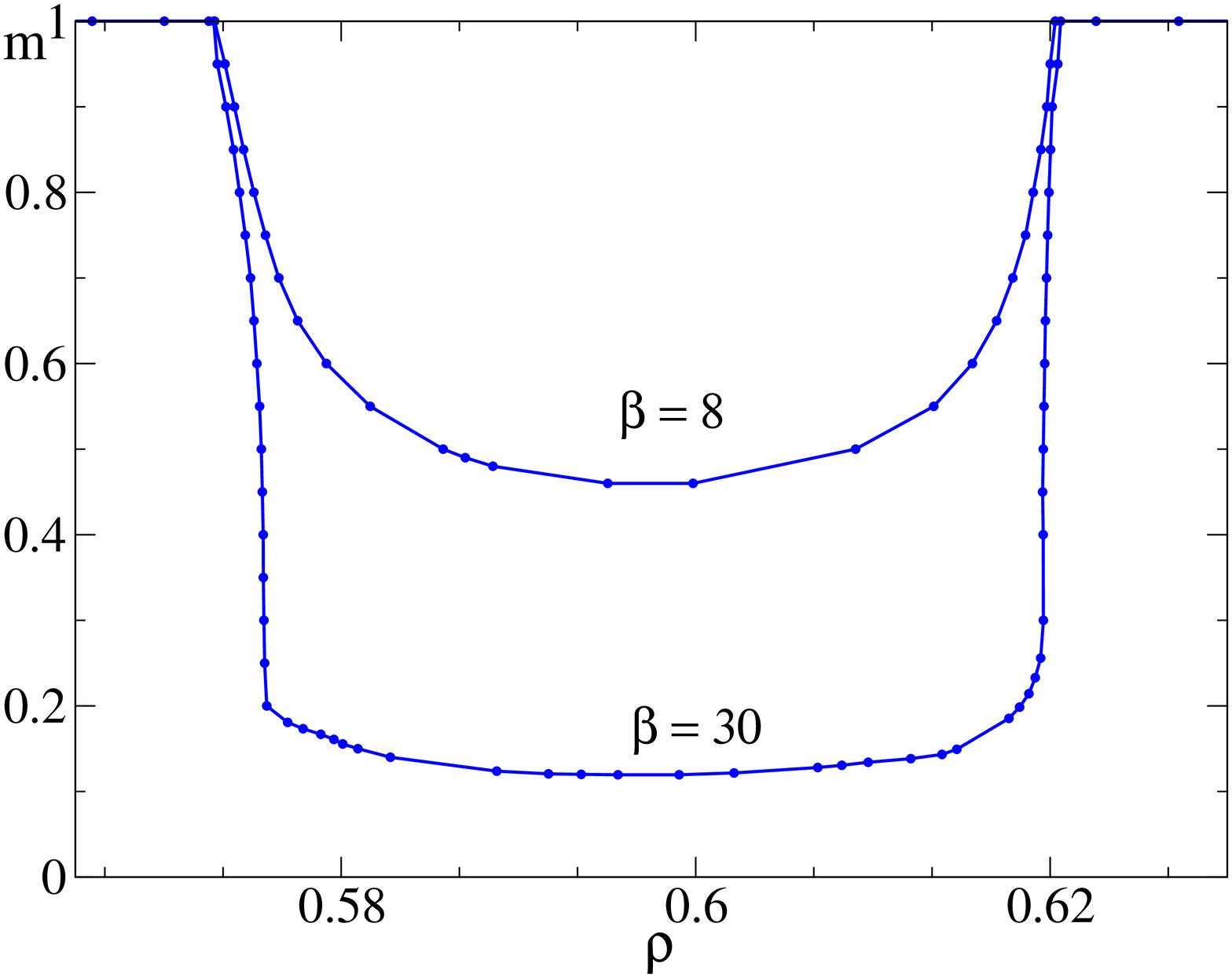}
  }
  \caption{\textit{Left Panel}: Average density as a function of the chemical potential,
  at $\beta=8$ (upper panel) and $\beta=30$ (lower panel). The inset shows a zoom
  of the $\beta=30$ data around the Kauzmann transition. \textit{Right Panel}: Behavior of the parameter $m^\ast$ as a function of 
  the density at $\beta=8$ and $\beta=30$.}
\end{figure*}

The thermodynamics of the system inside the glass phase has to be computed exploiting the 1RSB 
formalism developed in the previous section. Fig.\ref{class_mu_rho} shows in blue the density as a function of the chemical potential at 
$\beta=8$ and $\beta=30$, both in the liquid and in the glass phase. Then, the red dashed line 
represents the RS solution within the glass region, which clearly differs from the true 1RSB solution. 
For the same two temperatures the behavior of the parameter $m^\ast$ as a function of the density is shown in 
Fig.~\ref{class_m_rho}. It differs from one only inside the condensed glass
phase, after the Kauzmann transition. It decreases fast going deeper into the glass phase, down to a value
proportional to the temperature.
Note that the model presents a glass transition also at 
zero temperature, as a function of the density, with
$\r_d =0.5708$ and $\r_K=0.5725$ on the ``particle'' 
side~\cite{RBMM04,KTZ08}.
This is of particular interest in 
view of the extension of the model to the quantum case. 

Although we will not be concerned with this physics in the rest of the paper,
it is worth mentioning at this point that in the limit of zero temperature the
parameter $m^*$ approaches zero at a finite density 
$\r_{\rm RCP} = 0.57574$~\cite{RBMM04,KTZ08}. Below this value of density, the
average energy is zero, while above this value it becomes non-zero (the system
is so dense that it is unable to satisfy the constraint that each particle
has less or equal than $\ell$ neighbors). This point corresponds to what
is called ``random close packing'' or ``J-point'' in the literature on hard
spheres in the continuum (see e.g.~\cite{PZ10} for a review and a list of
relevant references). This is seen in more detail in the inset of 
Fig.~\ref{class_mu_rho} which reports data for $\beta=30$ that are very
close to the zero temperature limit: above the glass transition $\r_K$,
the density in the 1RSB solution increases slower than in the RS solution
and reaches a plateau at the value $\r_{\rm RCP} = 0.57574$ for $\mu<V=1$.
Only when $\mu>V=1$ the constraints are violated at zero temperature and the
density can increase above $\r_{\rm RCP}$. The latter is marked by the horizontal
dashed line in the figure.

One can recognize that the glass transition of the original model proposed in~\cite{BM01,RBMM04} is recovered 
here as the zero temperature limit of the glass transition, and it is here reached in the ``SAT'' (namely zero energy) 
phase. It corresponds indeed, to the limit $\beta V \to \infty$ for a suitable and finite value of $\beta \mu$. 
On the other hand, in that context, the presence of a ``hole'' glass transition is excluded a priori by 
the definition of the model, which does not allow configurations violating the energy constraint.

\section{The quantum model}
\label{quantum_model}

\subsection{The model}
\label{quantum_model_definition}

The quantum version of the model is obtained by adding to its classical potential energy a kinetic
term allowing hoppings of particles between neighboring sites. Since the classical energy forbids configuration
with more than one particle per site, we are in fact dealing with interacting hard-core bosons. The Hamiltonian
is now an operator acting on the Hilbert space spanned by the $2^N$ classical configurations of the occupation
numbers. The potential energy is diagonal in this basis. Denoting $J$ the intensity of the quantum tunneling 
between neighboring sites one obtains the following Hamiltonian:
\beq\label{qH}
 \hat{H} = - J \sum\limits_{\langle i,j\rangle} ( \hat{a}^{\dag}_i \hat{a}_j + \hat{a}^{\dag}_j \hat{a}_i )
 +  V \sum\limits_{i} \, \hat{n}_i \, \hat{q}_i \, \theta(\hat{q}_i) -  \mu \sum\limits_{i}  \hat{n}_i \ ,
 \eeq
where the first sum runs over the pairs of adjacent sites, 
$\hat{a}^{\dag}_i$ and $\hat{a}_i$  are standard hard-core 
bosonic creation and annihilation operators on site $i$, $\hat{n}_i = \hat{a}^{\dag}_i \hat{a}_i$ and 
$\hat{q}_i = \sum_{j\in\partial i} \hat{n}_j - \ell$. 
A thermodynamic study of the model then amounts to the computation of the quantum partition function
$Z = \Tr \[[e^{-\beta \hat{H}}\]]$ and of the average value of the observables $\hat O$ defined as 
$\langle \hat O \rangle = \Tr \[[e^{-\beta \hat{H}} \hat O \]]/Z$.

As in the classical case we consider the model on the Bethe lattice and 
we solve it through the quantum cavity method, a 
generalization of the classical cavity method which has been derived first for spin~\cite{LSS08,KRSZ08}, and then for 
bosonic systems~\cite{STZ09}.
The basic idea of the method is to exploit the Suzuki-Trotter formalism in order to write the 
partition function as a path integral.  The Hamiltonian is naturally split into two non-commuting term 
$\hat{H}=\hat{H}_1+\hat{H}_2$, with $\hat{H}_1$ the kinetic energy and $\hat{H}_2$ the classical potential energy.
The Suzuki-Trotter method amounts to work in the
diagonal basis of one of the two terms of the Hamiltonian, and treat
the non-commutativity of the two terms through the Suzuki-Trotter formula
\beq
Z = \Tr \[[e^{-\beta \hat{H}}\]] = \lim\limits_{N_s\to\infty} \Tr  \[[ \(( e^{-\frac{\beta}{N_s} \hat{H}_1}e^{-\frac{\beta}{N_s}\hat{H}_2} \))^{N_s}\]] \ .
\eeq
The computation which leads to the path integral formulation proceeds with the insertion of $N_s$ copies 
of the identity between the $N_s$ factors of this product. In particular for the purpose of the implementation 
of the cavity method, it turns out that  the most convenient basis to use in this operation is that of the occupation 
numbers. This is due to the discreteness of this basis. Then the quantum system is mapped into a classical one 
with an additional ``imaginary time dimension''. This procedure, which preserves the spatial 
structure of the graph of the interactions, allows to apply the general formalism 
of the classical cavity method. The crucial difference is that one must now consider as basic
variables the whole trajectories in time $[0,\beta]$ which describe the state on each site.
Detailed explanations of the bosonic quantum cavity method can be found in~\cite{STZ09} which studied the 
Bose-Hubbard model on the Bethe lattice. In the following we shall therefore focus on the new features that
arise here due to the specific form of the interaction (\ref{eq_Hclass}) between neighboring sites, which was
absent from the treatment in~\cite{STZ09}
(see however Ref.~\cite{CTZ09} for a study of the extended Hubbard model, that corresponds to the present
model in the special case $\ell = 0$, but has a very different behavior from the $\ell > 0$ in which we are
interested here). 
The next subsection (Sec.~\ref{qcavity_method}) will present
a derivation of the quantum cavity equations which describe this system, while the numerical method to solve
them will be the subject of sections~\ref{qcavity_equations_rrg} and~\ref{resolution_qcavity_equations}. 
This part of the paper is rather
technical and the reader who is not interested in the details of the derivation can safely jump to 
Sec.~\ref{results_quantum} where we shall discuss the results of this procedure and the physical properties of
the model.

\subsection{Derivation of the cavity equations}
\label{qcavity_method}

\subsubsection{Rewriting the partition function as a simple factor graph} 
\label{qcavity_equations}

We use in this section the same notation as in~\cite{STZ09} for the imaginary time copies
of the occupation variables. We consider a discrete number of Suzuki-Trotter slices
labeled by $\a = 1, \cdots, N_s$.
We denote by $n_i^\a$ the occupancy of site $i$ at Suzuki-Trotter time $\a$,
by $\bn_i = (n_i^1,\cdots,n_i^{N_s})$ the imaginary time trajectory of site $i$,
by $\un^\a = (n_1^\a, \cdots, n_N^\a)$ the configuration at time $\a$, 
by $\ubn = (\un^1, \cdots, \un^{N_s})$ the imaginary time trajectory of the full system.

We write the partition function
\beq
Z = \Tr \left[ e^{-\beta \hat{H}} \right] = \lim_{\Ns \to\infty} 
\sum_{\un^1, \cdots, \un^\Ns} \exp\left[-\frac{\beta}{\Ns} 
\sum_{\alpha = 1}^\Ns \sum_{i=1}^N v(n^\alpha_i, \{n^{\alpha}_{j\in\partial i}\})\right]
\prod_{\alpha=1}^\Ns \langle \un^\alpha | 
e^{\frac{\beta}{\Ns} \underset{\ij}{\sum} J
( \hat{a}^\dag_i  \hat{a}_j +  \hat{a}^\dag_j  \hat{a}_i ) } | \un^{\alpha+1} \rangle \
\label{eq_Z1}
\eeq
where 
\beq
\begin{split}
v (n^\alpha_i, \{n^{\alpha}_{j\in\partial i}\})  & =  V~n_i^{\alpha}~q_i^{\alpha}~\theta(q_i^{\alpha}) - \mu n_i^{\alpha} \\ 
q_i^{\alpha} & =  \sum_{j\in\partial i} n_j^{\alpha} - \ell
\end{split}
\eeq
and we write here and in the following $\theta(x)=\I(x\geq0)$ and  $\I(A)=1$ if the condition $A$ is fulfilled,
0 otherwise. It is simple to show that~\cite{STZ09}
\begin{multline}
\langle \un^\alpha | 
e^{\frac{\beta}{\Ns} \underset{\ij}{\sum} J
(  \hat{a}^\dag_i  \hat{a}_j +  \hat{a}^\dag_j  \hat{a}_i ) } | \un^{\alpha+1} \rangle = 
\sum_{\uby} \prod_\ij
\left(\frac{\beta J \sqrt{n_j^{\alpha+1} n_i^\alpha}}
{\Ns}\right)^{y_{i\to j}^\alpha}
\left(\frac{\beta J \sqrt{n_i^{\alpha+1} n_j^\alpha}}
{\Ns}\right)^{y_{j\to i}^\alpha} \\
\prod_{i=1}^N \I\left(n_i^{\alpha+1} = n_i^\alpha + 
\sum_{j \in \di} [y_{j \to i}^\alpha - y_{i \to j}^\alpha ] \right) 
+ O\left(\frac{1}{\Ns^2} \right) \ .
\label{eq_trick}
\end{multline}
where we introduced variables $y$ with the following notations:
we denote by $y_{i\to j}^\a \in \{ 0,1 \}$ the variable that indicates an hopping
event from site $i$ to $j$ at time $\a$; by $y_\ij^\a = \{ y_{i\to j}^\a , y_{j\to i}^\a \}$
the two hopping variables on link $\ij$; and as for occupation variables,
we use bold notation to indicate the imaginary time trajectory of a variable and
an underline to indicate the full configuration of the $y$'s.
Finally, we take by convention $x^{y=0}=1$ for any value of $x$ (including
zero). 

Eq.~(\ref{eq_trick}) can be checked by inspecting the 
behavior of its left and right-hand-side order by order in $1/\Ns$. The
leading term corresponds to $\un^\alpha = \un^{\alpha +1}$, and indeed all
$y$'s must vanish at this order.
We note that the partition function has the following form:
\beq
Z = \lim_{\Ns \to\infty} \sum_{\ubn,\uby} \prod_\ij w_h (\by_\ij)
\prod_{i=1}^N w_s(\bn_i, \{\bn_j \}_{j\in\partial i},\{ \by_\ij \}_{j \in \di}) 
\label{eq_Z2}
\eeq
with
\beq
w_h (\by_\ij) = \prod_{\alpha=1}^\Ns
\left(\frac{\beta J}{\Ns}\right)^{y^\alpha_{i\to j} + y^\alpha_{j \to i}}
\ ,
\label{eq_w_ij}
\eeq
and
\bea
w_s(\bn_i, \{\bn_j\}_{j\in\partial i},\{\by_\ij \}_{j \in \di}) & = &
\exp\left[-\frac{\beta}{\Ns} \sum_{\alpha = 1}^\Ns v (n^\alpha_i, \{n^{\alpha}_{j\in\partial i}\})\right] \notag \\
& &\prod_{\alpha=1}^{\Ns}\left\{ \I\left(n_i^{\alpha+1} = n_i^\alpha + 
\sum_{j \in \di} [y_{j \to i}^\alpha - y_{i \to j}^\alpha ] \right) 
\left(\sqrt{n_i^{\alpha+1}}\right)^{\underset{j \in \di}{\sum} 
y^\alpha_{j \to i}}
\left(\sqrt{n_i^\alpha}\right)^{\underset{j \in \di}{\sum} 
y^\alpha_{i \to j}} \right\} \notag \\
& = & \exp\left[-\frac{\beta}{\Ns} \sum_{\alpha = 1}^\Ns v (n^\alpha_i, \{n^{\alpha}_{j\in\partial i}\})\right] \wt{w}_s(\bn_i,\{ \by_\ij \}_{j \in \partial i}) 
\ .
\label{eq_w_i}
\eea
One can easily recognize that the graph of the interactions in the above 
representation has small loops, so, at first glance, it may
appear not suited for cavity calculations.
In order to avoid these loops and thus to set down an appropriate cavity treatment of the problem
we adopt a trick which consists in copying on each site the variables of the 
neighboring sites and the jumps across the edges connected to the site.
Let us introduce then
\beq
\nu_i^{\alpha} = \{n_i^{\alpha},\{n^{i \alpha}_j\}_{j\in\partial i} \}
\eeq
where $\{n^{i \alpha}_j\}_{j\in\partial i} $ are the copies on site $i$ of the occupation numbers of its neighbors at time $\alpha$.
In addition, we copy each variable $y^\a_{i\to j}$ on the two sites $i, j$, introducing $y_{j\to i}^{i \alpha} = y_{j\to i}^{j \alpha} = y_{j\to i}^{\alpha}$,
and on each site 
we store these variables in a variable $\Y_i^{\alpha}$:
\beq
\Y^{\alpha}_i = \{y_\ij^{i \alpha} \}_{j\in \di}= \{y_{i\to j}^{i \alpha},y_{j\to i}^{i \alpha} \}_{j\in \di} \ .
\eeq
Then we impose the consistency among neighboring sites' jumps and occupation numbers through a constraint
\beq
\wt w_l(\bnu_i,\bY_{i} ; \bnu_j,\bY_{j})= \prod_{\alpha=1}^{N_s}\delta_{n_i^{\alpha},n_i^{j \alpha}}\delta_{n_j^{\alpha},n_j^{i \alpha}} 
 \delta_{y_{i\to j}^{i \alpha}, y_{i\to j}^{j \alpha} } \delta_{y_{j\to i}^{i \alpha}, y_{j\to i}^{j \alpha} } 
\eeq
Putting together this constraint and the hopping weight $w_h(\by_\ij)$ (properly expressed in terms of the new variables $\Y$) 
we obtain a total weight on each link
\beq
w_l(\bnu_i,\bY_{i} ; \bnu_j,\bY_{j}) = w_h(\bY_i,\bY_j) \wt w_l(\bnu_i,\bY_{i} ; \bnu_j,\bY_{j})
\eeq
The partition function now becomes
\beq
Z = \lim_{\Ns \to\infty} \sum_{\ubnu,\ubY} \prod_\ij w_l(\bnu_i,\bY_{i} ; \bnu_j,\bY_{j})
\prod_{i=1}^N w_s(\bnu_i,\bY_i) \ .
\label{eq_Z3}
\eeq
where we have defined $\ubY=\{\bY_i\}$ and the argument of $w_s$ has been expressed in terms of the copied variables $\{\bnu_i,\bY_i\}$, living on 
site $i$. 

The last form of $Z$ is a standard form with local weights and link weights only, which is therefore defined on
the original graph and hence without short loops; it is therefore suitable for a cavity treatment and the cavity equations as well as the Bethe free
energy can be written straightforwardly.
Introducing  $\tilde{\eta}_{i \to j}(\bnu_i, \bY_i) = Z_{i \to j}(\bnu_i, \bY_i)/z_{i \to j}$, 
which represents the marginal probability  of the variables $\bnu_i$ and $\bY_i$ defined on site $i$, in absence of site $j$,
the cavity equations read~\cite{cavity}
\beq
\tilde{\eta}_{i \to j}(\bnu_i,\bY_i) = \frac{1}{z_{i \to j}}
w_s(\bnu_i,\bY_i)
\prod_{k \in \dimj} \sum_{\substack{ \bnu_k,\bY_k }}\tilde{\eta}_{k \to i}(\bnu_k, \bY_k)w_l(\bnu_k,\bY_{k} ; \bnu_i,\bY_{i})\ .
\label{eq_bh_msg1}
\eeq
The Bethe free energy associated to the partition function, expressed in terms of the copied variables as in
Eq.(\ref{eq_Z3}),  is
\bea
F = -\frac{1}{\beta}\ln Z = &-& \frac{1}{\beta}\sum_{i=1}^N \ln\left( 
\sum_{\substack{ \bnu_i, \bY_i} }
w_s(\bnu_i,\bY_i) 
\prod_{j \in \partial i} \sum_{\substack{\bnu_j,\bY_j}}
\tilde{\eta}_{j \to i}(\bnu_j,\bY_j)w_l(\bnu_i,\bY_i ; \bnu_j,\bY_j)
\right) \nonumber \\ &+& \frac{1}{\beta}
\sum_\ij \ln \left( \sum_{\substack{\bnu_i,\bY_i \\ \bnu_j,\bY_j}}w_l(\bnu_i,\bY_i ; \bnu_j,\bY_j)
\tilde{\eta}_{i \to j}(\bnu_i,\bY_i)\tilde{\eta}_{j \to i}(\bnu_j,\bY_j)  \right) \ .
\label{fVAR_sg}
\eea
This expression is exact whenever the underlying graph is a tree while it corresponds to
the Bethe approximation for a general graph.

\subsubsection{Simplification of the cavity equation} 

The cavity equations written above are not very practical to handle, since there is a lot of redundancy
in the copied variables. We can eliminate much of this redundancy by using the delta functions
in the weights $w_l$.

First we make explicit the dependence of the cavity field on the original variables in the following form:
\beq
\tilde{\eta}_{i \to j}(\bnu_i,\bY_i) = \tilde{\eta}_{i \to j}( \bn_i, \bn_j^i, \{\bn^i_k\}_{k \in\dimj}, \by^i_\ij, \{ \by^i_{\langle i,k \rangle}\}_{k\in \dimj} ) \ .
\eeq
Next we introduce a new cavity field:
\beq
\eta_{k \to i}(\bn_k, \bn^k_i, \{\bn^k_l \}_{l \in\partial k\setminus i},\by_{\langle i,k \rangle}^k) =  w_h(\by_{\langle i,k \rangle}^k)
\sum_{\{\by^k_{\langle k,l \rangle}\}_{{l \in \partial k \setminus i}}}\tilde{\eta}_{k \to i}( \bn_k, \bn_i^k, \{\bn^k_l\}_{l \in\partial k\setminus i}, \by^k_{\langle i,k \rangle}, \{ \by^k_{\langle l,k \rangle}\}_{l \in\partial k\setminus i} ) \ ,
\eeq
and we note that using the delta functions
\beq
\sum_{\substack{ \bnu_k,\bY_k }}\tilde{\eta}_{k \to i}(\bnu_k, \bY_k)w_l(\bnu_k,\bY_{k} ; \bnu_i,\bY_{i}) = 
\sum\limits_{ \{\bn^k_l\}_{l\in\partial k\setminus i}} \eta_{k \to i}(\bn_k^i, \bn_i, \{\bn^k_l\}_{l\in\partial k\setminus i} , \by^i_\ik) \ .
\eeq
This allows to write closed equations for the new cavity fields as follows:
\begin{multline}
\eta_{i \to j}(\bn_i, \bn_j^i, \{\bn^i_k\}_{k\in\dimj}, \by^i_\ij) = \frac{1}{z_{i \to j}}
w_h (\by^i_\ij)  \\
\sum_{\substack{\{\by^i_\ik \}_{k\in\dimj} } }
w_s(\bn_i, \bn_j^i,\{\bn^i_k\}_{k\in\dimj}, \by^i_\ij,\{ \by^i_\ik \}_{k \in \dimj})
\prod_{k \in \dimj} \sum\limits_{ \{\bn^k_l\}_{l\in\partial k\setminus i}} \eta_{k \to i}(\bn^i_k, \bn_i, \{\bn^k_l\}_{l\in\partial k\setminus i} , \by^i_\ik) \ .
\end{multline}
We notice at this point that the upper indices of the variables 
in the equation above are redundant: they can all be eliminated by renaming the variables. We obtain
the following equation:
\begin{multline}
\eta_{i \to j}(\bn_i, \bn_j, \{\bn_k\}_{k\in\dimj}, \by_\ij) = \frac{1}{z_{i \to j}}
w_h (\by_\ij)  \\
\sum_{\substack{\{\by_\ik \}_{k\in\dimj} } }
w_s(\bn_i, \bn_j,\{\bn_k\}_{k\in\dimj}, \by_\ij,\{ \by_\ik \}_{k \in \dimj})
\prod_{k \in \dimj} \sum\limits_{ \{\bn_l\}_{l\in\partial k\setminus i}} \eta_{k \to i}(\bn_k, \bn_i, \{\bn_l\}_{l\in\partial k\setminus i} , \by_\ik) \ .
\label{eq_bh_msg2}
\end{multline}
To further simplify the equations, we introduce a variable $s_{i\to j}^{\alpha}$ which indicates if a site $i$ at time $\alpha$ is saturated by all its neighbors but $j$
\beq
q_{i\to j}^{\alpha} = \sum\limits_{ k \in\dimj} n_k^{\alpha} - \ell \ ,
\hspace{1cm}
s_{i\to j}^{\alpha}= \theta(q_{i\to j}^{\alpha}) \ ,
\eeq
where we recall that  we assume $\theta(0)=1$, then we consider
\beq
\eta_{i \to j}(\bn_i, \bn_j, \bs_{i\to j},\by_\ij) = 
\sum\limits_{ \{\bn_k\}_{k\in\dimj}} \eta_{i \to j}(\bn_i, \bn_j, \{\bn_k\}_{k\in\dimj}, \by_\ij) \ \delta_{\bs_{i\to j},\theta(\bq_{i \to j})}
\eeq
with 
$\delta_{\bs_{i\to j},\theta(\bq_{i \to j})} = \prod\limits_{\alpha=1}^{\Ns} \delta_{s_{i\to j}^{\alpha},\theta(q_{i \to j}^{\alpha})} \ $.
Using the relation 
\beq
q_i\theta(q_i) = (q_{i\to j}+n_j) \theta(q_{i\to j}+n_j) = (q_{i\to j}+n_j)\theta(q_{i\to j}) = (q_{i\to j}+n_j) s_{i\to j} \ ,
\eeq
which can be checked by inspection of the cases $q_{i\to j} \lesseqgtr -1$,
Eq.(\ref{eq_bh_msg2}) becomes
\begin{multline}
\eta_{i \to j}(\bn_i, \bn_j, \bs_{i\to j},\by_\ij) = \frac{1}{z_{i \to j}} 
w_h (\by_\ij)  
\sum_{\substack{\{\by_\ik \}_{k\in\dimj} } } 
\tilde{w}_s(\bn_i, \by_\ij,\{ \by_\ik \}_{k \in \dimj}) \\
 \sum\limits_{ \{\bn_k\}_{k\in\dimj}} e^{ - \frac{\beta}{\Ns} 
\sum_{\alpha = 1}^\Ns n_i^{\alpha} \(( V s_{i\to j}^{\alpha} (n_j^{\alpha} + q_{i\to j}^{\alpha}) - \mu \)) }
 \delta_{\bs_{i\to j},\theta(\bq_{i \to j})}
\prod_{k \in \dimj} \sum\limits_{ \{\bs_{k\to i}\}} 
\eta_{k \to i}(\bn_k, \bn_i, \bs_{k\to i}, \by_\ik) \ .
\label{eq_bh_msg3}
\end{multline}
Now we note that the function $\eta_{i \to j}(\bn_i, \bn_j, \bs_{i\to j},\by_\ij)$ has 
a dependence on the variable $n_j$ of the form
$\eta_{i \to j}(\bn_i, \bn_j, \bs_{i\to j},\by_\ij)   = e^{-\frac{\beta}{\Ns} 
\sum_{\alpha = 1}^\Ns V n_i^{\alpha} s_{i\to j}^{\alpha}n_j^{\alpha} }  \hat{\eta}_{i \to j}(\bn_i, \bs_{i\to j},\by_\ij)$,
so, in terms of the marginal $ \hat{\eta}_{i \to j}(\bn_i, \bs_{i\to j},\by_\ij)$ we finally obtain
\begin{multline}
\hat{\eta}_{i \to j}(\bn_i,\bs_{i\to j},\by_\ij) =  \frac{w_h (\by_\ij)  }{z_{i \to j}} 
\hskip-10pt \sum_{\substack{\{\bn_k\}_{k\in\dimj} \\ \{\bs_{k\to i}\}_{k\in\dimj} \\ \{\by_\ik \}_{k\in\dimj} } }  \hskip-10pt
w_{\rm iter} (\bn_i,\bs_{i\to j},\by_\ij,\{ \bn_k,\bs_{k\to i},\by_\ik \}_{k \in \dimj}) 
\prod_{k \in \dimj}
\hat{\eta}_{k \to i}(\bn_k, \bs_{k\to i}, \by_\ik)
\label{eq_bh_msg4}
\end{multline}
where
\beq\label{eq_witer}
\begin{split}
w_{\rm iter} (\bn_i,\bs_{i\to j},\by_\ij,\{ \bn_k,\bs_{k\to i},\by_\ik \}_{k \in \dimj}) &= \tilde{w}_s(\bn_i, \by_\ij,\{ \by_\ik \}_{k \in \dimj}) \  \delta_{\bs_{i\to j},\theta(\bq_{i \to j})} \times \\
&\times e^{ - \frac{\beta}{\Ns} 
\sum_{\alpha = 1}^{N_s} n_i^{\alpha} \(( V q_{i\to j}^{\alpha} s_{i\to j}^{\alpha}  - \mu \))}
\prod_{k \in \dimj} e^{ - \frac{\beta}{N_s} \sum_{\alpha = 1}^{N_s}V  n_i^{\alpha} n_k^{\alpha} s_{k\to i}^{\alpha} }
 \end{split}
\eeq
From Eq.(\ref{eq_bh_msg4}) we see that $s_{i\to j}^{\alpha}$ enters
 in the equation only through its product with $n_i^{\alpha}$. 
This means that in order to characterize the state of each site one has actually to 
consider the variables $\{n_i^{\alpha},  n_i^{\alpha} s_{i\to j}^{\alpha}\}$ and sum over $s_{i \to j}^{\alpha}$
in case the site is empty.
Roughly speaking, defining $e=\{0,0\}$, $u=\{1,0\}$ and $s=\{1,1\}$ we can interpret  
$ \hat{\eta}_{i \to j}(\bn_i, \bs_{i\to j},\by_\ij)$ as the marginal probability
over trajectories of the kind  $[0,\beta] \to \{e,u,s\}$ which describe the state of each ``cavity site''.
Finally one can note that the variable $\bs_{i \to j}$ is completely determined by the neighbors' occupation numbers. 
This ensures that, even if locally in order to write the cavity recursions one needs 
three possible states to describe each site, globally  the Hilbert space has size $2^N$. 

In terms of the cavity marginal probabilities $\hat{\eta}_{i \to j}(\bn_i, \bs_{i\to j},\by_\ij)$ 
the occupation site trajectory on site $i$ is then expressed as
\beq
\eta(\bn_i) = \frac{1}{z_{i}} \sum_{\substack{ \{\bn_j\}_{j\in\partial i} \\ \{\by_\ij \}_{j\in\partial i}}} w_s(\bn_i,\{\bn_j\}_{j\in\partial i} ,\{ \by_\ij \}) 
\prod_{j \in \partial i} \sum\limits_{\bs_{j\to i}} 
\hat{\eta}_{j \to i}(\bn_j, \bs_{j\to i}, \by_\ij)
e^{ - \frac{\beta}{\Ns} \sum_{\alpha = 1}^\Ns V n_i^{\alpha} n_j^{\alpha} s_{j\to i}^{\alpha} } \ ,
\label{eq_bh_msg5}
\eeq
which allows to compute all the local observables.
From (\ref{fVAR_sg}) we can recover the free energy as a function of the last defined
messages:
\bea
F = -\frac{1}{\beta}\ln Z = &-& \frac{1}{\beta}\sum_{i=1}^N \ln\left( 
\sum_{\substack{ \bn_i, \{\bn_j\}_{j\in\partial i} \\ \{\by_{\ij}\}_{j\in\partial i}} }
w_s(\bn_i,\{\bn_j\},\{\by_{\ij}\}) 
\prod_{j \in \partial i}\sum_{\substack{\bs_{j\to i}}}
\hat{\eta}_{j \to i}(\bn_j,\bs_{j\to i},\by_{\ij})e^{ - \frac{\beta V}{\Ns} \sum_{\alpha = 1}^\Ns n_i^{\alpha} n_j^{\alpha} s_{j\to i}^{\alpha}}
\right) \nonumber \\ &+& \frac{1}{\beta}
\sum_\ij \ln \left( \sum_{\substack{\bn_i,\bs_{i\to j} \\ \bn_j,\bs_{j\to i} \\ \by_{\ij}}} 
\frac{e^{ - \frac{\beta V}{\Ns} \sum_{\alpha = 1}^\Ns n_i^{\alpha} n_j^{\alpha}( s_{i\to j}^{\alpha}+s_{j\to i}^{\alpha})}}{w_h (\by_\ij)}
\hat{\eta}_{i \to j}(\bn_i,\bs_{i\to j},\by_{\ij})\hat{\eta}_{j \to i}(\bn_j,\bs_{j\to i},\by_{\ij})  \right)
 \ .
\eea
In the case of a regular Bethe lattice where all sites have degree $c$, denoting $\eta_{\rm cav}$ the
common value of the distributions $\hat{\eta}_{i \to j}$ on all edges, we obtain for the RS free energy 
density:
\bea
f = -\frac{1}{\beta N}\ln Z = &-& \frac{1}{\beta} \ln\left( 
\sum_{\substack{ \bn_0 \\ \{\bn_j,\by_{0j}\}_{j\in\{1,\dots,c\} } } }
w_s(\bn_0,\{\bn_j,\by_{0j}\} ) 
\prod_{j \in\{1,\dots,c\}}\sum_{\substack{\bs_{j\to 0}}}
\h_{\rm cav}(\bn_j,\bs_{j\to 0},\by_{0j})e^{ - \frac{\beta V}{\Ns} \sum_{\alpha = 1}^\Ns n_0^{\alpha} n_j^{\alpha} s_{j\to 0}^{\alpha}}
\right) \nonumber \\ &+& \frac{c}{2\beta} \ln \left( \sum_{\substack{\bn_0,\bs_{0\to 1} \\ \bn_1,\bs_{1\to 0} \\ \by}} 
\frac{e^{ - \frac{\beta V}{\Ns} \sum_{\alpha = 1}^\Ns n_0^{\alpha} n_1^{\alpha}( s_{0\to 1}^{\alpha}+s_{1\to 0}^{\alpha})}}{w_h (\by)}
\h_{\rm cav}(\bn_0,\bs_{0\to 1},\by)\eta_{\rm cav}(\bn_1,\bs_{1\to 0},\by)  \right)
 \ .
\eea

\subsection{Resolution of the cavity equation for the random regular graph} 
\label{qcavity_equations_rrg}

In this section we explain the idea and the method used to solve
the quantum cavity equation for a random regular graph. This task is rather demanding
since already at the RS level it consists in 
a functional self-consistent equation. However, as it was shown in~\cite{KRSZ08,STZ09}, there is
a representation of the equation which allows for its numerical resolution.

Let us define $\bh_i=\{\bn_i,\bs_{i\to j},\by_{\langle i,j\rangle}\}$ the whole set of variables 
argument of the cavity marginals in Eq.(\ref{eq_bh_msg4}), 
which determine the ``field'' on the $i$ site when $j$ is absent. 
In the following we deal with a random regular graph, so for simplicity, we will label with $0$ the root site which is
added in the cavity iteration when we merge $k = c-1$ neighboring sites and their associated
 fields $\{\bh_i\}_{i \in \{1,...,k\}}$. Finally, we label with an index $k+1$ the new site which is missing 
 ``downwards'' with respect to the root.
The calculation is based on the observation~\cite{KRSZ08,STZ09} that for a random 
 regular graph Eq.~(\ref{eq_bh_msg4}) has the form
 \beq\label{qcaveq_rs_rrg}
 \h_{\rm cav}^0(\bh_0) = \frac{1}{z_{\rm cav}} \sum_{\bh_1 \cdots \bh_k} 
\Z(\bh_1,\dots,\bh_{k}) P(\bh_0 | \bh_1 \cdots \bh_k) \h^1_{cav}(\bh_1) \cdots \h^k_{\rm cav}(\bh_k) \ ,
 \eeq
where $P$ is a conditional probability, positive for all arguments and normalized and 
\beq
\Z(\bh_1,\dots,\bh_{k}) = \sum_{\bh_0=\{\bn_0,\bs_{0\to k+1},\by_{\langle 0,k+1\rangle}\} } w_h (\by_{\langle 0,k+1\rangle}) 
w_{\rm iter} (\bn_0,\bs_{0\to k+1},\by_{\langle 0,k+1\rangle},\{\bh_j\}_{j\in\{1,\dots,k\}}) 
\ .
\eeq
We note that at the RS level we look for the homogeneous solution so all $\h_{\rm cav}^i=\h_{\rm cav}$ 
are equivalent, however in equation (\ref{qcaveq_rs_rrg}) we keep the general form  to allow 
the generalization to the 1RSB case.

As detailed in~\cite{KRSZ08,STZ09}, a strategy of resolution is based on the representation
 of $ \h(\bh)$ as a weighted sample of trajectories:
 \beq\label{popdyn}
 \h_{\rm cav}(\bh) = \sum_{i=1}^\Nt g_i \, \delta(\bh - \bh_i ) \ ,
\eeq
where the $\Nt$ weights of the trajectories are normalized according to
\beq 
\sum_{i=1}^\Nt g_i = 1 \ .
\label{eq_norm}
\eeq
One should not confuse this population of quantum trajectories
(that represents the RS solution of the quantum problem) with the population of cavity fields that 
we introduced in the classical case in Eq.~(\ref{class_pop_dyn}) to describe the 1RSB solution.
The form (\ref{popdyn}) provides an approximate representation of $\h_{\rm cav}$
which becomes more and more accurate as $\Nt$ grows.
Starting from an initial weighted sample of $ \h_{\rm cav}$ in the form of (\ref{popdyn}) 
the procedure amounts to iterate the following three steps: 
\begin{enumerate}
\item extract $k$ trajectories $\bh_i=\{\bn_i,\bs_{i\to j},\by_{\langle i,j\rangle}\}$ in $[1,\dots,\Nt]$ from  (\ref{popdyn}) according to their weight $g_i$
\item draw the new trajectory from $P(\bh_0 | \bh_1 \cdots \bh_k)$
\item set $\bh_i'=\bh_0$ and $g'_i=\Z(\bh_1,\dots,\bh_{k})$
\end{enumerate}
where the second point amounts to a single site problem that will be further explained in the next section.
Once a new representation of $\h_{\rm cav}$ in terms of $\bh_i'$ is available
it can be substituted in the r.h.s. of Eq.(\ref{qcaveq_rs_rrg}) and the procedure is iterated until the convergence,
in the sense of the observable expectation values, is reached. 
 
Before entering into the description of the generation of the new trajectory given the $k$ neighbors, we anticipate that, 
as in the classical case, also the quantum model is characterized by a glass phase
at high enough densities, where the RS equations are no more correct (they have the same pathologies
as in the classical case, yielding for instance negative entropies).
The 1RSB treatment is needed also for the quantum problem and within this framework 
we have derived all the results concerning  the glass phase, presented in the following.

At this level however, the generalization of the classical 1RSB equation (\ref{rsb_eq}) is straightforward. 
The only difference is that now the fields $h=\{\psi^e,\psi^u,\psi^s\}$ of Eq.(\ref{rsb_eq}) are probability distributions over paths, namely the ``cavity fields''
$\h_{\rm cav}$ of the RS equations (\ref{qcaveq_rs_rrg}) and $Z_{\rm iter}$ is 
the normalization constant $z_{\rm cav}$ that appears in the same Eq.(\ref{qcaveq_rs_rrg}). 
Then, the 1-RSB cavity equation for the random regular graph reads:
\beq\label{q_rsb}
P[\h_{\rm cav}] \propto \int dP[\h^1_{\rm cav}]\dots dP[\h^k_{\rm cav}]  
\delta\((\h_{\rm cav}-f(\h_{\rm cav}^1,\dots,\h_{\rm cav}^k)\)) z_{\rm cav}^{m}
\ ,
\eeq
where $f(\h_{\rm cav}^1,\dots,\h_{\rm cav}^k)$ is the r.h.s. of Eq.(\ref{qcaveq_rs_rrg}), and a normalization
constant has been hidden in the notation $\propto$.
The numerical resolution of the equation (\ref{q_rsb}) necessarily goes through the representation
of $P[\h_{\rm cav}] $ as a population of  ``fields'' $\h_{\rm cav}$
\beq
P[\h_{\rm cav}] = \sum_{i=1}^{{\cal N}} w_i \, \delta(\h_{\rm cav} - \h_i ) \ ,
\eeq
where the weights derive from the cavity iteration $w\propto z_{\rm cav}^{m}$  (analogously as
in the classical case) and each $\h_{\rm cav}$ is itself a population of trajectories.
This procedure is of course accompanied by a demanding computational
cost which is however still attainable and controllable.
For the interested reader we refer to~\cite{FKSZ10} for more details 
about the 1RSB cavity method in quantum systems and the computation of observables.

\subsection{Generation of a trajectory} 
\label{resolution_qcavity_equations}

In this section we will focus on the problem
of the generation of trajectories for the thermodynamic limit
of a random regular graph of connectivity $c$.
As we said before the strategy of resolution is based on the representation
 of $\h(\bh)$ as a weighted sample of trajectories. Then one will pick $k$ trajectories 
 $\bh_1 \cdots \bh_k$ from the corresponding distributions, and use $P(\bh_0 | \bh_1 \cdots \bh_k)$ to
 draw a new trajectory $\bh_0$, to which a weight $\Z(\bh_1,\dots,\bh_{k})$ is assigned.
 
In order to calculate $\Z$ and to generate a new path we need to 
take into account all events encoded in the neighboring trajectories.
In particular we can distinguish between three of them:
\begin{itemize}
\item a particle jump from or towards the neighboring sites, which is encoded in $\{\bn_i\}_{i \in \{1,...,k\}}$
\item a change in the $\{\bs_{i\to 0}\}_{i \in \{1,...,k\}}$ variables
\item a jump  from or towards the neighboring sites along the edge connecting the site under consideration, 
described by $\{\by_{\langle i,0\rangle}\}_{i \in \{1,...,k\}}$ variables.
\end{itemize}
The last class of events is a subset of the first one, but it is important
to take it into account separately because it induces constraints on the
new occupation number trajectory. All the other events, instead, will 
enter as shifts of the effective chemical potential induced on the added site.
Moreover we stress that in the continuous time limit ($\Ns\to \infty$) the hopping
trajectories $\bh$ typically contain only a finite number (with respect to
$\Ns$) of events. We can thus assume
that they  occur at different values of the discrete time for the different trajectories. 
We call $p$ the total number of hopping events
occuring in $(\bh_{1},\dots,\bh_{k})$. We also denote $\alpha_1<\dots<\alpha_p$
their discrete time of occurence.

We consider the Hilbert space of a single
site, with $a$ and $a^\dag$ the annihilation/creation operators. We introduce
the operator $b_j=a$ (resp. $b_j=a^\dag$) if at  time $\alpha_j$ there is a jump 
towards (resp. outside) the site under consideration, while $b_j=\I$ when there is
an event which does not involve the new vertex, but only the neighboring trajectories.
The former case corresponds to the
occurrence of an event of the third kind, while the latter corresponds to an event of
the first or second kind.
Finally we introduce
$c_\alpha = b_j$ when $\alpha=\alpha_j$, $c_\alpha = \I$ (the identity
operator) otherwise. In terms of these operators we get
the following expression for $\Z$:
\beq
\begin{split}
\Z(\bh_{1},\dots,\bh_{k}) & = 
\sum_{\substack{\bh_{0}}}  w_h (\bh_0) 
w_{\rm iter} (\bh_0,\{ \bh_i \}_{i \in \{1,...,k\}}) \\
& =  \sum_{\bn_{0}} \prod_{\alpha=1}^\Ns \langle n^\alpha_{0}  | e^{\frac{\beta}{\Ns}
 (\mu^{\alpha}(\{n^{\alpha}_{i},s^{\alpha}_{i\to 0}\}_{i\in \{1,...,k\}})a^\dag a+ J(a+a^\dag) )} c_\alpha 
| n^{\alpha+1}_{0} \rangle
\end{split}
\eeq
up to corrections of order $\Ns^{-2}$, where we have introduced an effective chemical potential, which depends on the
state of the neighboring sites and is defined as follows:
\beq
\mu^{\alpha}(\{n^{\alpha}_{i},s^{\alpha}_{i\setminus 0}\}_{i\in \in \{1,...,k\}}) =  
- V \((q_{{0}\to k+1}^{\alpha} \theta(q_{{0}\to k+1}^{\alpha} ) +
\sum\limits_{i=1}^{k} n_{i}^{\alpha} s_{i\to 0}^{\alpha}\)) + \mu
\ .
\eeq
To take the continuous time limit
it is convenient to define $\t_j = \frac{\beta}{\Ns}\alpha_j$, which are
the continuous times of the hopping events in $(\bh_{1},\dots,\bh_{k})$, and
to denote $\mu_i$ the common value of the $\mu_{\alpha}$ for $\alpha \in [\alpha_{i-1},\alpha_i]$.
We also introduce $\hW_i(\lambda) = e^{\lambda(\mu_i a^\dag a+ J(a+a^\dag) )}$, 
the propagator of an imaginary time evolution on an interval of length 
$\lambda$ for a single site Hamiltonian $H_i = -\mu_i a^\dag a - J(a+a^\dag)$. 
This propagator is a two by two matrix and it easily allows to compute $\Z$ 
in the continuous limit according to the relation:
\beq
\Z(\bh_{1},\dots,\bh_{k}) = 
\Tr \left( \hW_1(\t_1) b_1 \hW_2(\t_2-\t_1) b_2 \dots\hW_{p}(\t_p-\t_{p-1}) b_p \hW_{p+1}(\beta-\t_p) \right) 
\ .
\label{eq_Z}
\eeq
We can now look at the process of the generation of a new trajectory $\bh_0$
given the ones of the $c-1$ other neighbors, which respects the following probability law
\beq
P(\bh_0|\bh_1,\dots,\bh_{k} ) = 
\frac{  w_h (\bh_0) 
w_{\rm iter} (\bh_0,\{ \bh_i \}_{i \in \{1,...,k\}})}
{\Z(\bh_1,\dots,\bh_{k})} \\
 \ .
\label{eq_PZ}
\eeq
The general scheme to determine $\bh_0=\{\bn_0,\bs_0,\by_{\langle 0,k+1\rangle}\}$ consists in first drawing the occupation 
number trajectory $\bn_0$ and then deduce $\by_{\langle 0,k+1\rangle}$ from its 
jumps not associated to the events in $(\by_{\langle0,1\rangle},\dots,\by_{\langle 0,k\rangle})$. 
Finally the trajectory of $\bs_0$ is completely determined by the neighboring 
ones $(\bn_{1},\dots,\bn_{k})$ and does not have to be generated.
We keep the notation
$\t_1 < \dots <\t_p$ and $b_1,\dots,b_p$ for the continuous time of the events 
in $(\bh_1,\dots,\bh_{k})$. Let us call $n_0=n(\t=0)$,
$n_i$ (resp. $n'_i$)
the value of $n(\t)$ at a time just after $\t_i$ (resp. just before
$\t_{i+1}$), with the conventions $n'_p=n_0$. The joint probability law
of these occupation numbers which arises from the expressions of $w_{\rm h}$
and $w_{\rm iter}$ given in Eqs.~(\ref{eq_w_ij}) and (\ref{eq_witer}) reads in the
continuous time limit
\beq
P(n_0,n'_0,n_1,n'_1,\dots,n_p|\bh_1,\dots,\bh_{k}) = 
\frac{1}{\Z(\bh_1,\dots,\bh_{k})}
\langle n_0 | \hW_i(\t_1) | n'_0 \rangle  
\prod_{i=1}^p \left\{ \langle n'_{i-1}|b_i|n_i \rangle
\langle n_i | \hW_{i+1}(\t_{i+1}-\t_i) | n'_i \rangle \right\} \ ,
\eeq
with $\t_{p+1}=\beta$. This probability law is well normalized according to
the above expression of $\Z(\bh_1,\dots,\bh_{k})$. 
It follows immediately from the above equation that whenever the operator $b_i$
is non trivial, i.e. when there is a jump along the edge connecting the neighbors
 to the new site, the state of the site is uniquely defined by the direction
 of the jump. This is a trivial consequence of the fact that we are dealing with hard-core bosons.
 Then, we have to generate only those values in the 
 sequence  $(n_0,n'_0,n_1,n'_1,\dots,n_p)$ which are not already fixed.
 This is a task that can be done sequentially, starting from $n_0$, quite easily. 
Once all the intermediate occupation numbers are fixed, the generation
of the rest of the path will proceed independently for every interval $\{\t_{i+1}-\t_i\}$.
Inside each of them, in fact, one has to generate a trajectory with fixed boundary conditions
 according to the path integral representation
of an effective Hamiltonian which depends on time interval 
$H_{i} = -\mu_{i} a^\dag a - J(a+a^\dag)$.
All the details for this procedure are explained in \cite{STZ09} for the Bose-Hubbard 
model. At this level, in fact, the task is completely equivalent to that of the
Bose-Hubbard, despite the interaction between neighboring sites,
 which is here hidden in effective single site Hamiltonians.
In the next section we will give more details on the parameters we used for the computations
presented in this paper.

\section{Results for the quantum model}
\label{results_quantum}

The system investigated here presents a rich phase diagram, which emerges from 
the classical case discussed in section~\ref{classic} when the additional effect of the hopping
is taken into account. In this section we first discuss
the finite temperature phase diagram, which is directly accessible to our method,
and then we argue on how it extrapolates to the zero temperature limit.
We also present exact diagonalization data for the ground state to support the extrapolation
of the cavity data to zero temperature.

Note that the hopping amplitude $J$ plays in this model the role of the quantum parameter
$\G = \L/a$ (the ratio of the De Broglie thermal wavelength and the interparticle distance) 
defined in the introduction.
As we showed in the previous sections, for the classical problem where $J=0$, 
the thermodynamic glass phase is delimited
by the curve $T_K(\r)$ of Fig.~\ref{class_phase_diagram} (while the dynamical glass is delimited by $T_{d}(\r)$)
in the temperature-density plane,
similarly to what happens for more realistic models of structural glasses~\cite{BMS10}.
One can then imagine to approach the glass phase from higher or lower densities.
As we discussed in section~\ref{results_classic}, the two transitions
are qualitatively equivalent and in the following we will focus on the region of 
the phase diagram corresponding to small densities. 
From a physical point of view,
this side of the glass transition is the most interesting one, since it corresponds to a 
packing problem, which we have relaxed through the soft constraint $V$. 
From now on we will focus on the model
defined by $c=3$ and $\ell=1$. As before we will measure energies in units of $V$ throughout this section,
leaving it implicit in the text (but not in the figures).

\subsection{Order parameters}
\label{sec:order_parameters}

\subsubsection{Edwards-Anderson parameter}

As we already discussed for the classical case, a fingerprint of the glass transition is the appearance of a 
local inhomogeneous density profile, signaling the breaking of translation invariance.
If one considers the Gibbs measure over imaginary-time paths which is constructed by the Suzuki-Trotter decomposition,
the transition that separates the liquid phase from the glass is of the same 
kind of the one described in section~\ref{classic_rsb} for the classical system. It happens in two steps: 
at a first transition ({\it dynamical transition}), the Gibbs measure becomes
dominated by an exponential number of states, 
which are counted by a well defined thermodynamic potential, the \textit{complexity}.  
As we said the thermodynamics of the system is not affected by this transition, while the dynamics is.
This has been well established in the classical case~\cite{CC05}. In the quantum case, 
it has been studied in~\cite{CGS01,BC01} for the imaginary time dynamics
and in \cite{CGLL04,CGLLS02} for the real time dynamics. 
An heuristic argument that relates the transition in the real and imaginary time dynamics 
has been proposed in~\cite{BCZ08}.
At a second transition ({\it Kauzmann transition}), the complexity vanishes, the system undergoes 
the true thermodynamic glass transition to the glass phase 
which is characterized by a sub-exponential (in the size of the system) number of dominant states.

As in section~\ref{classic_rsb}, we can label the different glass states by an index $\a$;
each state has free energy $f_\a$ and we can perform the same construction as in 
section~\ref{classic_rsb} to compute the complexity for the Gibbs measure on imaginary-time paths. 
We denote by $\langle \hat O \rangle_\a$ the average
of an observable $\hat O$ in the restriction of the Gibbs measure to a given glass state.
Inside each glass state one has $\langle \hat n_i\rangle_\a \neq \frac{1}{N} \sum_i \langle\hat n_i\rangle =\rho$,
while the average density is the same for each state, $\r = \frac1N \sum_i \la\hat n_i \ra_\a $. 
These density fluctuations are conveniently characterized by the 
Edwards-Anderson order parameter~\cite{MPV87}:
\beq
\label{qea}
q_{EA} = \overline{\frac{1}{N} \sum\limits_{i} \sum_\a W_\a [ \la\hat n_i \ra_\a -\rho]^2} \ , 
\eeq
where we take the average over the states according to their weights $W_\a \propto \exp(-\b N f_\a)$,
and the over bar indicates the average over random graphs.
The order parameter $q_{EA}$ jumps to a finite value at the dynamical transition, where the glassy states
appear for the first time~\cite{CC05}.
This non-zero value of $q_{EA}$ corresponds physically to the long time limit
of the density-density correlation function~\cite{CC05}, as we will discuss later on.
As we already said,
the jump of the order parameter is a peculiarity of the glass transition which still remains second order 
from the point of view of the
singularity of the free energy.
Note that one should not confuse $q_{EA}$ with the following quantity:
\beq
\label{qav}
q_{\rm av} = \overline{\frac{1}{N} \sum\limits_{i} (\la\hat n_i \ra - \rho)^2} = 
\overline{\frac{1}{N} \sum\limits_{i} \sum_{\a,\g} W_\a W_\g \, (\la\hat n_i \ra_\a -\rho)( \la\hat n_i \ra_\g -\rho )} \ ,
\eeq
The latter quantity is non-zero at finite $N$, but it should vanish in the thermodynamic limit,
since in that limit the full Gibbs measure (\ie the average over all glass states with their weights)
must become uniform and $\la\hat n_i \ra \to \rho$ for all $i$.

\subsubsection{Condensate density}

In presence of a finite hopping strength, the system might display Bose-Einstein condensation (BEC). 
This is manifested 
by a finite expectation value of the bosonic operator $\overline{\langle\hat{a}\rangle} \neq 0$
in the grand-canonical ensemble. 
However, since the number of particles is conserved by the Hamiltonian and by the cavity equation, 
in absence of symmetry breaking terms the expectation value $\overline{\langle\hat{a}\rangle}$
is always zero.
More precisely, one possibility to define the order parameter of the BEC transition amounts to add a small 
perturbation to the Hamiltonian which breaks that symmetry $\hat{H}_h = \hat{H} - h \sum_i (\hat{a}_i+\hat{a}_i^\dag)$ 
and then send it to zero after the thermodynamic limit is taken 
\beq
\label{ahdef}
 \overline{\la \hat a \ra} = \lim_{h \to 0}  \lim_{N\to \infty} \overline{\la  \hat a \ra}_h \ ,
\eeq
where $\la \bullet \ra_h$ denotes the thermodynamic average in presence of the perturbation.
Note that this procedure 
does not present any particular difficulty for the cavity method, that gives direct access to 
the thermodynamic limit.
However, an equivalent and more practical way to allow the symmetry breaking, that we used here, 
is to initialize the population dynamics algorithm
with a fraction of ``asymmetric trajectories'', i.e. trajectories such that the number of hoppings from
a given site $i$ to a given neighbor $j$ is not equal to the hoppings from $j$ to $i$.
In absence of BEC, these trajectories disappear 
under the cavity iterations, while in the BEC phase the symmetry breaking is preserved,
thus allowing for a non zero value of $\overline{\la a \ra}$. The mathematical reasons for this have been detailed
in~\cite{STZ09}. The physical intuition is the following: the cavity field represents the effect on a given site, of the part
of the graph that has already been integrated out. In the insulating phase, the hoppings are localized,
therefore loops are irrelevant and if a boson hops off a given site towards one subtree connected to it, it must at some
point come back from the same subtree. On the other hand, in the BEC phase, there are long range exchanges, therefore
a boson can hop off a site to a subtree, travel around the (infinite) loops, and come back from another subtree. The 
``incoming'' and ``outgoing'' currents from a site to a given neighbor do not balance (of course, they balance on average,
but not on each individual trajectory of the population).
Indeed, defining a notion of ``superfluidity'' on the Bethe lattice is tricky, and the idea
of bosons hopping through infinite loops is probably the closest one to the notion of ``winding numbers'' on a regular
finite dimensional lattice~\cite{Ce95}. In the following we will refer to the BEC phase as ``superfluid'', but one should keep
in mind this difficulty.

The order parameter $\overline{\la \hat a \ra}$ defined above also signals the presence of off diagonal long range order 
in the system~\cite{PO56}. 
In a uniform phase such as the liquid phase, one can define the condensate density as~\cite{PO56}
\beq
\label{rcdef}
\rho_c = \lim_{|i-j|\to\infty} \overline{\langle\hat{a}^{\dag}_i\hat{a}_j\rangle} = \overline{| \langle\hat{a}\rangle |^2} \ .
\eeq
If the latter is finite, there is off diagonal long range order; $\rho_c$ represents the number of bosons that are
condensed in the ground state divided by the volume.
In a non-uniform phase, one has
\beq
\rho^\a_c =  \frac1N \sum_i | \langle\hat{a_i}\rangle_\a |^2
\eeq
for each glass state, and one can define the average over states and graphs:
\beq\label{rcnonunif}
\r_c = \overline{\sum_\a W_\a \r_c^\a} \ .
\eeq
In a non-uniform phase,
the relation with the one-body density matrix is more complicated and we will discuss it later on. 
For the moment, we do not need the relation with the one-body density matrix,
since the cavity method gives direct
access to the expectation of $\langle\hat{a_i}\rangle_\a$, and then
to the condensate density defined in (\ref{rcnonunif}), through the procedure described above.

\subsection{Details on the numerical resolution of the cavity equations: \\ 
results for the order parameters, and how to detect the phase transitions}
\label{sec:details_cavity}

\begin{figure*}
\centering
\includegraphics[width=12cm]{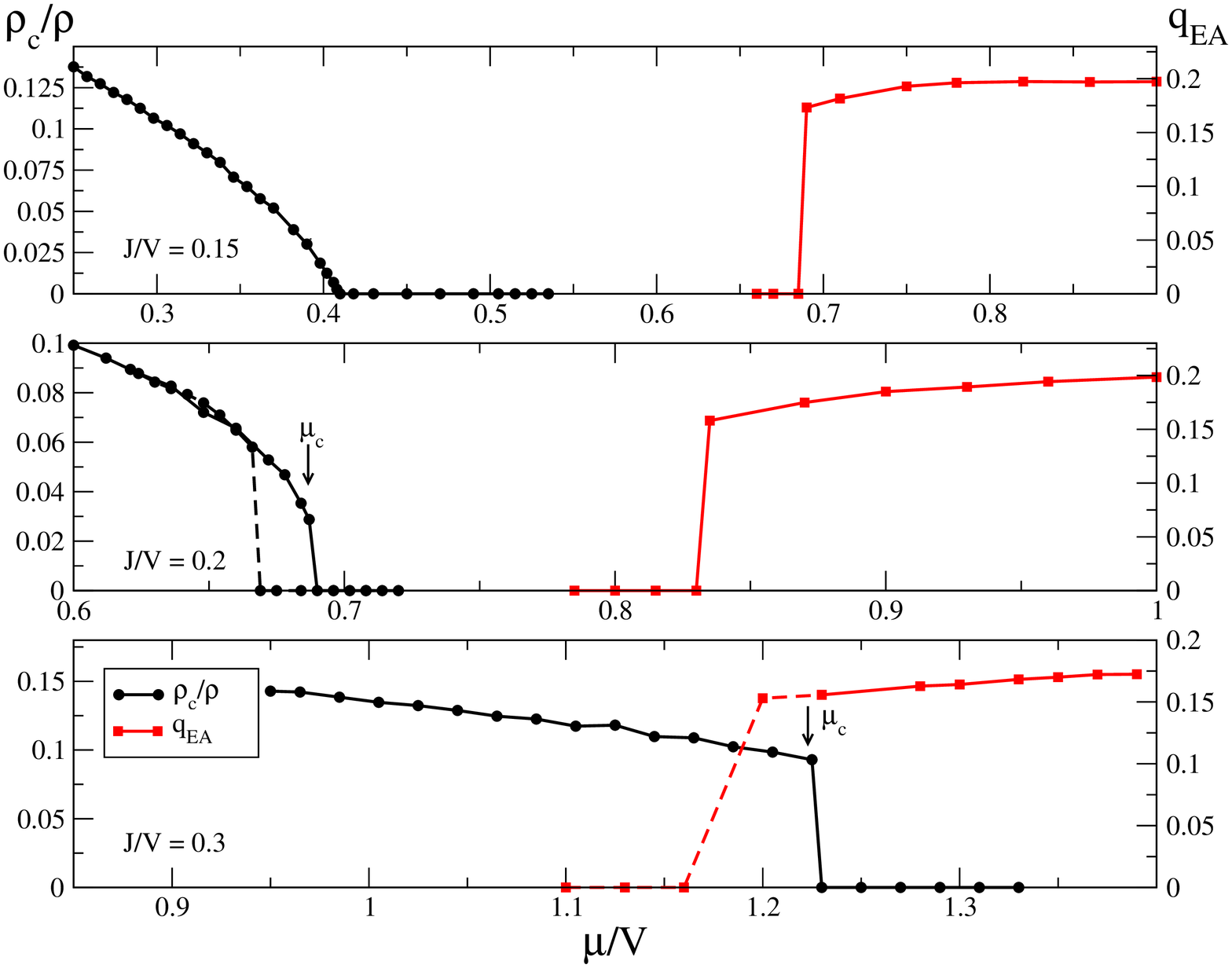}
\includegraphics[width=4.5cm]{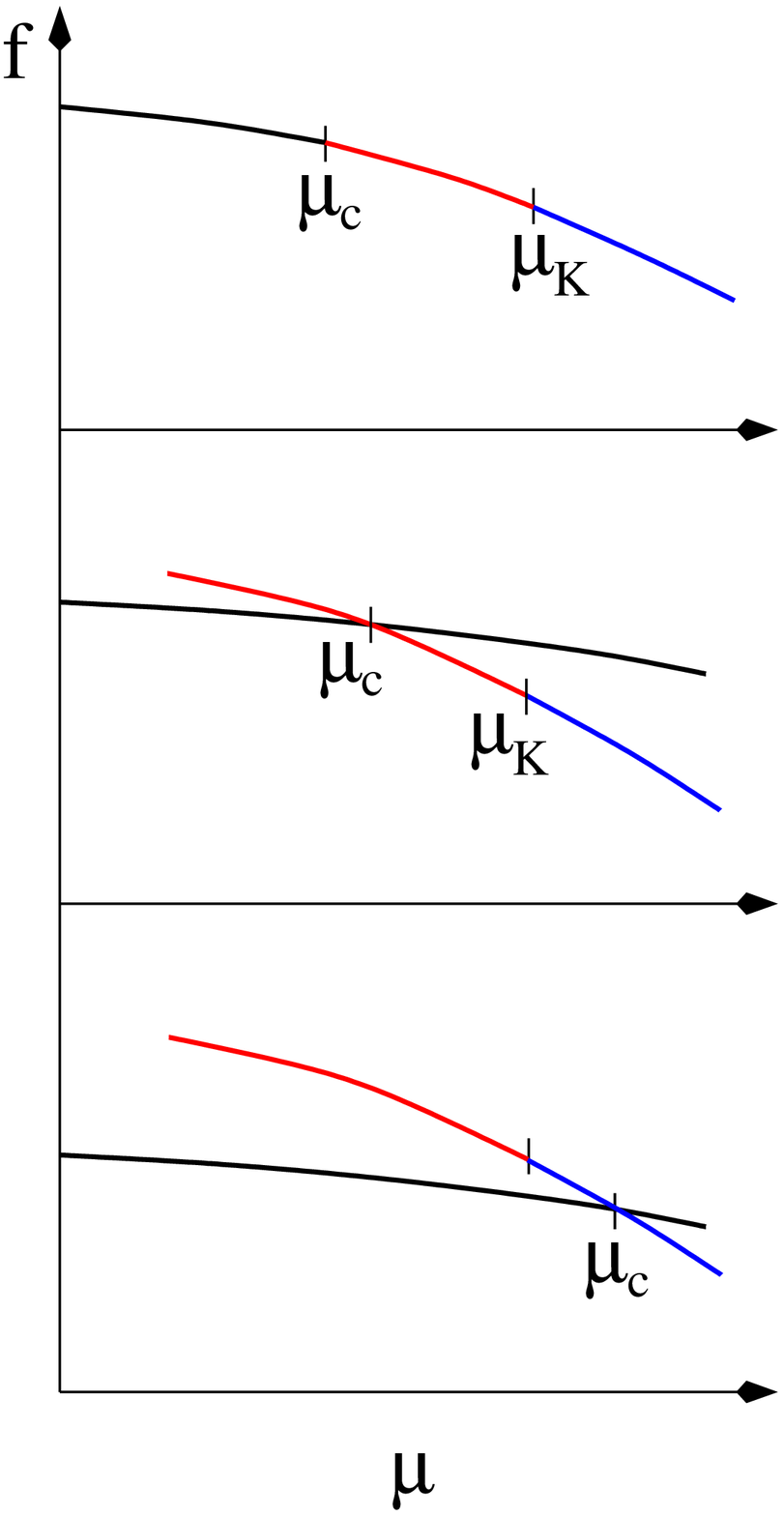}
\caption{
{\it Left Panel:} Order parameters of the superfluid ($\rho_c/\rho$, black circles, left scale, RS cavity method) 
and glass ($q_{EA}$, red squares, right scale, 1RSB cavity method) phases for $c=3$ and $\ell=1$,
for three representative values of the hopping $J$ at the same temperature $\beta V=15$,
as functions of the chemical potential $\mu$. The first order transitions are accompanied by
hysteresis; full lines corresponds to following the solution upon increasing $\mu$,
dashed lines correspond to decreasing $\mu$. Arrows mark the first order transition
points $\m_c$ as determined by the Maxwell construction. See text for a more detailed description.
{\it Right Panel:} Sketches of the free energy as a function of $\mu$ for the corresponding 
left panels: from left to right, black lines represent the superfluid, red lines the normal liquid,
blue lines the glass; $\mu_c$ and $\mu_K$ indicate the phase transitions (we do not report data because
the jump in the derivative of $f$ at the first order transition is very small and would be invisible
in the figure).
}
\label{parametri_ordine}
\end{figure*}

In this section we give the details of our implementation of the procedure described in 
section~\ref{quantum_model} to solve the cavity equations and investigate the properties of
the different phases. Our code is available upon request.
In the following, we will give indicative values for the different parameters
such as population sizes, number of cavity iterations needed to equilibrate the populations, etc.
These parameters are not fixed a priori but changed slightly depending on the region of the phase
diagram under investigation. As far as possible given our computational resources,
the results have been checked to be robust with respect to these parameters by increasing 
the population sizes in some representative points.

\subsubsection{Replica symmetric calculations: the superfluid-normal liquid transition}

We started by performing replica symmetric calculations using a population of $\sim 32000$ 
trajectories. 
The population was initialized at some low density, with most of the trajectories having no jumps,
and a small fraction of trajectories having two jumps towards different sites in order to break
the symmetry and allow a superfluid phase, as discussed above.
We then performed $\sim 100$ cavity iterations to find the fixed point
of the population, and then we performed $\sim 500$ additional iteration to collect statistics
on the averages of the observables.
We fixed $J$ and $T$, and we increased $\mu$ in small steps, in order to reach higher densities, 
repeating the procedure above but starting
from the converged population of the previous step (we inject a small fraction of trajectories
that break the symmetry, again in order to allow a superfluid phase).
In Fig.~\ref{parametri_ordine} we show the behavior of the order parameter of the superfluid
phase, $\rho_c/\rho$ (black circles), as a function of increasing $\mu$ (corresponding to
increasing density). 
We observe two different behaviors depending on the hopping strength:
\begin{itemize}
\item
At high temperature or small hopping (small $\b J$), we observe that $\rho_c/\rho$ vanishes
continuously at some $\mu_c$ corresponding to the superfluid transition 
(upper panel in Fig.~\ref{parametri_ordine}), which is then a second order transition.
The density is also continuous at the transition.
Close to $\mu_c$, we observe a power-law behavior $\overline{\langle\hat{a}\rangle} \sim | \mu - \mu_c |^{1/2}$,
with an exponent $1/2$ typical of the mean field nature of the underlying lattice; correspondingly, the condensate fraction
vanishes linearly. 
\item
At low temperature or large hopping (large $\b J$) we observe that $\rho_c/\rho$ jumps 
abruptly to zero at some value $\mu_c$, indicating that the transition is first order
(middle panel in Fig.~\ref{parametri_ordine}). In this case, we observe
hysteresis when we perform increasing or decreasing $\mu$ scans (see the middle panel of Fig.~\ref{parametri_ordine}).
As a consequence of the first order phase transition in the $\mu$ variables, the density, which is defined 
as $\rho=-\partial f(\m)/\partial \m$, has a jump at 
$\mu_c$, which implies the existence of 
a region of phase coexistence for $\r \in [\r_-,\r_+]$ in the canonical ensemble.
The determination of $\mu_c$ and of $\r_-, \r_+$ can be done in two equivalent ways:
either by looking at the grand-canonical free energy as a function of $\mu$ 
(which is directly accessible within the cavity method) and determining the point
at which the free energies of the two phases cross, or Legendre transforming the free energy as a canonical function of $\rho$
and then performing the Maxwell construction over it.
Since the jump in density is extremely small at the transition, $\r_+-\r_- \ll (\r_++\r_-)/2$,
we found that the Maxwell construction is numerically more precise than looking directly 
at the slope of the free energy $-\partial f(\m)/\partial \m$, which is dominated by the average density.
\end{itemize}

\subsubsection{1RSB calculations: the glass transition}

Next, we turn to the 1RSB equations to investigate the stability of these phases with respect to glassy order.
The 1RSB equations are formulated in terms of a population of populations of trajectory. We typically used
1000 populations, each made of 6000 trajectories, which is already extremely demanding in terms of computational
resources.
Fig.~\ref{parametri_ordine} shows the behavior of the order parameter of the glass phase, $q_{EA}$ (red squares), 
obtained through the 1RSB cavity method, as a function of the chemical potential.
In this case we initialize the population at high density (where we expect a glass phase) in the following
way: almost all trajectories are classical (no jumps), with a small fraction breaking the symmetry as usual,
and we initialize each population of trajectories at a different average
density in order to start with a finite $q_{EA}$. 
We then perform $\sim 150$ cavity iterations to find the fixed point
and $\sim 700$ iterations to collect data.
Since in this case the code is very slow (it can take up to 3 days to compute
one state point), in order to speed up the computations we run the code independently for each $J,T,\mu$ with the 
initialization described above. 
Moreover for each choice of $J,T,\mu$ we run $\sim 5$ computations at different values of $m$ in order to find
the maximum $m^*$ of the free energy.
In some cases we performed scans by decreasing $\mu$ starting at each time by the
previously converged population: we did this in order to observe hysteresis (dashed red line and squares in 
the lower panel of Fig.~\ref{parametri_ordine}).
Generically, we observe as expected that $m^*$ increases on decreasing $\mu$, until it reaches $m^*=1$ at the
Kauzmann transition $\mu_K$. For lower $\mu$, one has $m^*=1$ and by further decreasing $\mu$ the non-trivial 1RSB
solution is lost at the dynamical transition $\mu_d$, where $q_{EA}$ jumps to zero and one gets back the RS solution.
In general, we found that $\mu_K$ and $\mu_d$ are always extremely close to each other, as in the classical case. 
A precise determination of $\mu_K$ can be done by looking at the point where the complexity at $m=1$ vanishes.
On the contrary, the determination of $\mu_d$ is much more difficult since it corresponds to the point 
where the non-trivial 1RSB solution disappears and $q_{EA}=0$; this is a kind of ``spinodal point'' and its precise
location is very sensible to details of the computation such as population size, initialization, number of iterations,
etc. Although we managed to get reliable data for $\mu_d$, in the following we focus only on $\mu_K$, keeping in mind
that the two are very close and follow similar trends.
Finally, it is worth to note at this point that we did not find any superfluid non-trivial 1RSB solution: 
we comment on this in more details in the following.

\subsubsection{Summary}

To summarize, we observed generically three different behaviors upon varying $\mu$, that are summarized in
the three panels of Fig.~\ref{parametri_ordine} (we set $\b=15$ in the figure):
\begin{enumerate}
\item
The upper panel, at small hopping ($J=0.15$), represents a second order superfluid-normal liquid 
phase transition, followed by a liquid-glass transition at higher $\mu$.
The system is superfluid until $\r_c$ vanishes at $\mu_c$, then it is a normal liquid in an interval of $\mu$, until
the parameter $q_{EA}$ jumps from zero to a finite value signaling the dynamic glass transition $\mu_d$. 
Upon further increasing $\mu$, the complexity vanishes at $\mu_K$ and the system enters the glass phase, where
$m^*<1$ (note that $\mu_d$ and $\mu_K$ are indistinguishable on the scale of the figure). 
Both the transitions are second order in the sense of Ehrenfest classification: the first derivative of
the free energy is always continuous and the second derivative is discontinuous at both transition points. 
\item
The middle panel ($J=0.2$) shows a first order transition between 
the superfluid and a normal fluid, followed again by a liquid-glass transition.
Here the condensate fraction $\rho_c/\rho$ jumps suddenly to zero at $\mu_c$.
In this regime the free energy of the superfluid intersects that of
the normal liquid at $\m_c$ with a discontinuity in the first derivative.
As in the previous case, the glass transition happens here at a higher
value of $\mu$ where $q_{EA}$ jumps to a finite value and the free energy of
the glass grows smoothly from that of the liquid with continuous first derivative.
The figure also shows the hysteresis which is a consequence of the
first order transition, when we follow the evolution of the order parameter $\r_c$ 
coming from the superfluid phase or from the normal phase. 
\item
The lower panel, corresponding to a larger hopping $J=0.3$, shows a direct first order phase transition between the
superfluid and the glass phases.
Indeed, in this case the free energy of the superfluid crosses directly the free energy of the glass at $\mu_c$. 
At the transition the condensate fraction jumps to zero while $q_{EA}$ jumps to a non-zero value.
Also in this case we observe a phenomenon of hysteresis, and the exact first order transition point has to be determined
by the crossing of the RS free energy of the superfluid phase and the 1RSB free energy of the glass phase. 
\end{enumerate}
In the next section, we investigate 
the evolution of these phase boundaries by varying hopping, temperature and chemical potential.

\subsection{Phase diagram of the quantum model}
\label{quantum_Tfinite}

The quantum cavity method (at least in its present formulation) 
allows only to access the finite temperature properties of the system.
We therefore start the discussion of the phase diagram from finite temperature,
and then we discuss how to extrapolate the results to the $T=0$ limit.

\subsubsection{Finite temperature phase diagram}

\begin{figure*}
\centering%
 \includegraphics[width=12cm]{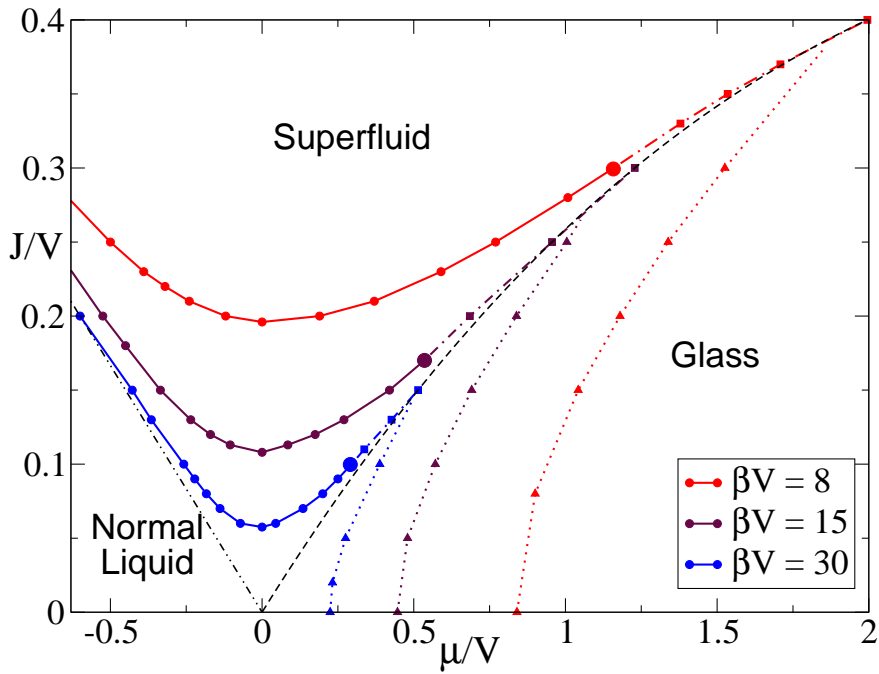}
\caption{Phase diagram for the quantum model in the $(\mu/V,J/V)$ plane, for $c=3$ and $\ell=1$, for three
different temperatures $\b V = 8, 15$ and $30$. The lines divide the phase diagram
into three main regions where the system is found in a glass, superfluid or
(normal) liquid phase. Solid lines and circles indicate the second order 
superfluid transition. The large dots mark the tricritical point 
where the superfluid line changes from second to first order.
Dash-dotted lines and squares represent the first order transition between the superfluid phase
and the normal (glass or liquid) phase. Dotted lines and triangles indicate the Kauzmann glass transition $J_{\rm K}(\mu)$.
The dot-dot-dashed black line indicates the normal liquid-superfluid transition for hard core bosons at zero temperature.
Since below this line $\r=0$ at $T=0$, the interaction is not relevant around it, so this line is the $T=0$ limit of the 
normal liquid-superlfuid transition also for the model
investigated here.
The dashed black line serves as a guide to the eye: it has been obtained by interpolating the large $\mu$ 
superfluid lines at different temperatures (see text for details).
}
\label{quantum_phase_diagram_mu}
\end{figure*}

The phase diagram, at finite temperature 
in the $(\mu,J)$ plane is shown in Fig.~\ref{quantum_phase_diagram_mu}.
For any fixed (and low enough) temperature, 
there is a curve $\mu_c(J)$, or $J_c(\mu)$, which separates a 
superfluid phase from a normal phase. A tricritical point divides this transition
line into two parts. On one side, at small chemical potential, the transition is second order.
Beyond the tricritical point instead, the transition becomes 
first order. This happens at bigger values of $\mu$, close to the glass transition.
We chose different colors to represent the phase diagram at different temperatures. 
Solid lines and circles indicate second order superfluid 
transitions while their continuations, dash-dotted lines and squares, represent first 
order transition between a superfluid phase
and a normal liquid or glass phase. The liquid-glass Kauzmann transition, 
$\mu_K(J)$ or $J_K(\mu)$, is reported using a dotted line and triangles.
It hits the horizontal axis, at $J=0$, 
in correspondence of the classical glass transition; on the other side, it crosses the 
first order superfluid-normal phase transition.
The dynamic transition line
is very close to the Kauzmann transition so we don't report it for clarity.

\begin{figure*}
\centering%
 \includegraphics[width=.45\textwidth]{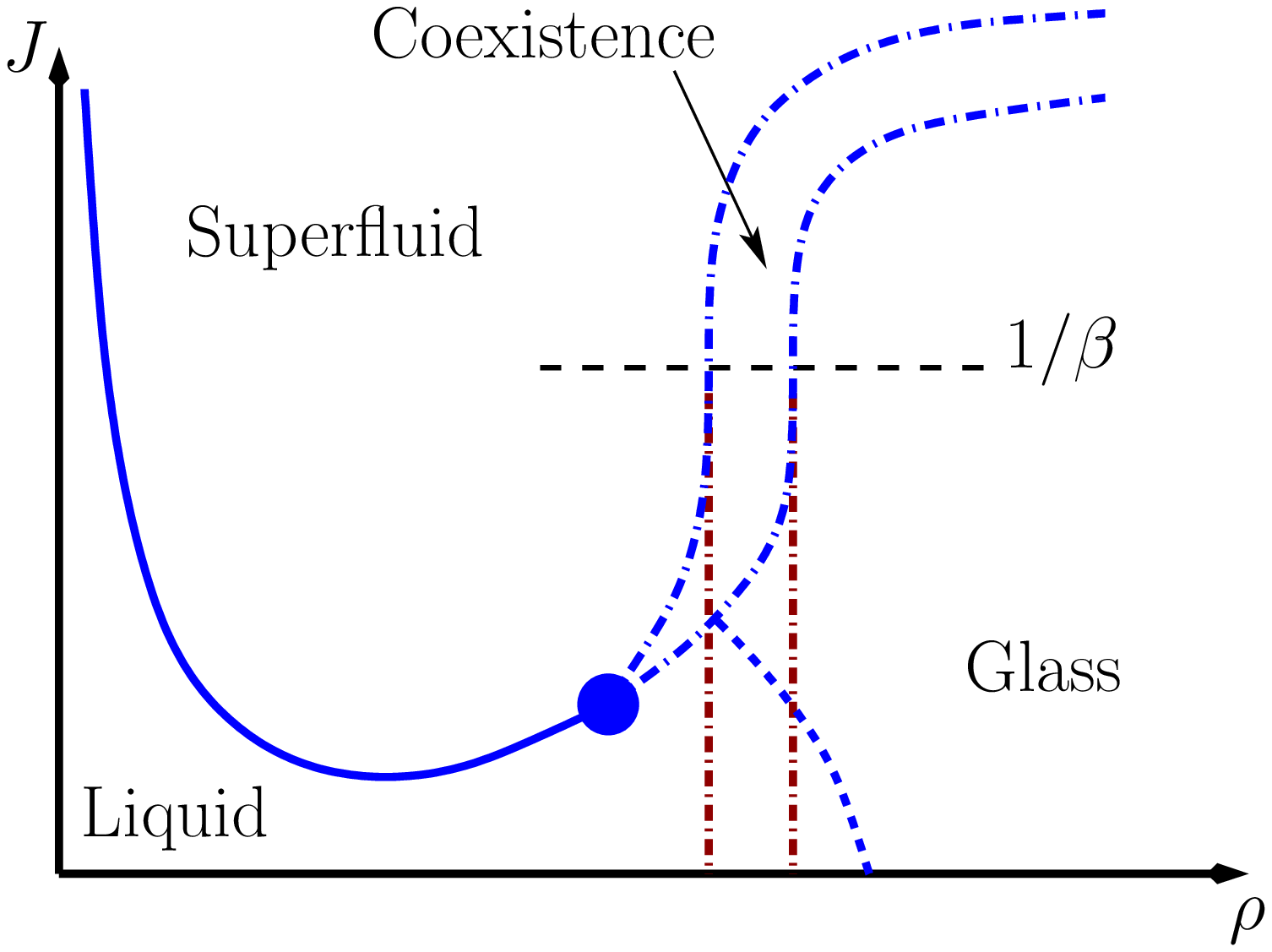}
 \includegraphics[width=.45\textwidth]{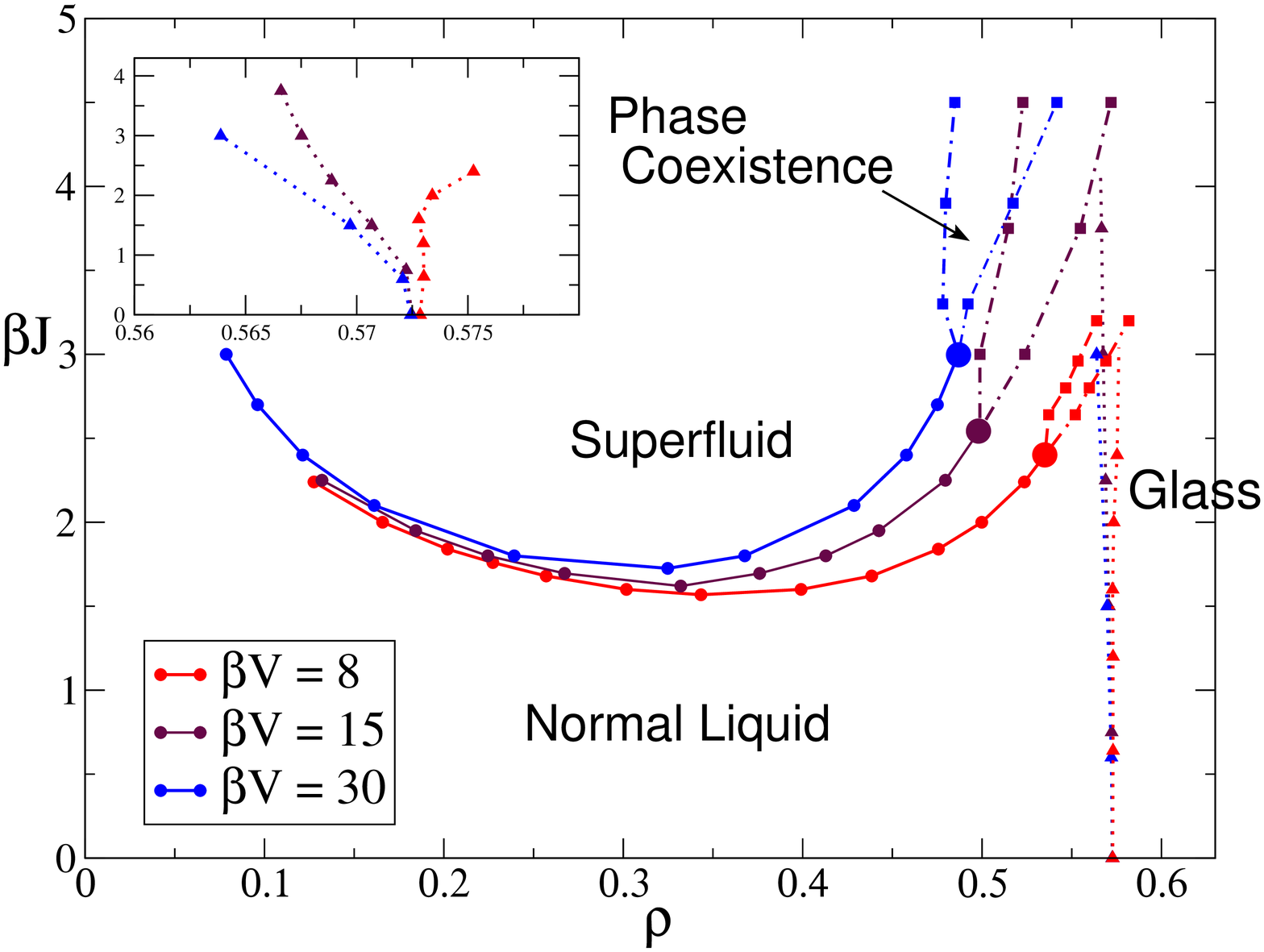}
\caption{
{\it Left Panel:} Schematic $(J/V,\r)$ phase diagram of the model. 
The full line represents the second order superfluid transition $J_c(\r)$, 
separated by a tricritical point
(large dot) from a first order superfluid transition accompanied by phase coexistence
(dot-dashed lines).
The dotted line represents the Kauzmann transition $J_K(\r)$.
In the limit $T\to 0$, the transition lines have distinct behaviors:
the first order transition line has $J \gg T$, and it has a finite limit for $T\to 0$.
On the contrary, the other lines have $J \propto T$ and they shrink to
the $J=0$ axis for $T\to 0$. Therefore, at $T=0$ and for $J>0$, the low-$\r$ part of the phase
diagram contains the superfluid phase while the large-$\r$ part contains the glass. The red lines
indicate the behavior of the superfluid-glass transition at $T= 0$.
{\it Right Panel:} Data for the $(\b J,\r)$ phase diagram, for $c=3$ and $\ell=1$, and for three
different temperatures $\b V = 8, 15$ and $30$. The three regions, glass, superfluid and
normal liquid phase, are here reported. The transition lines are plotted using the same
styles as in Fig.~\ref{quantum_phase_diagram_mu}.
As a consequence of the first order phase transition a region of
phase coexistence -- delimited by dash-dotted lines -- is present. The inset shows the re-entrant behavior of 
the glass transition line, at low enough temperature. We plot $\b J$ on the vertical axis in order
to show that the transition lines are proportional to $T$ in the limit $T\to 0$.
}
\label{quantum_phase_diagram_rho}
\end{figure*}

We now describe the phase diagram as a function of the density.
Since the cavity method works necessarily in the grand-canonical ensemble, we are forced to measure 
the density at fixed $\mu$. Therefore, the density has some fluctuations due to numerical noise.
Unfortunately, for this model the interesting part of the phase diagram is contained in a very small
interval of density, and for this reason the noise is important and prevents us to obtain
transition lines as clean as in the $(J,\m)$ phase diagram of Fig.~\ref{quantum_phase_diagram_mu}.
For this reason, in Fig.~\ref{quantum_phase_diagram_rho}
 we report a schematic $(J,\r)$ phase diagram together with the actual data.
We keep the same code of colors and lines as in Fig.~\ref{quantum_phase_diagram_mu}.

The main difference between the $(J,\m)$ and $(J,\r)$ phase diagrams is that in the latter case,
as already discussed, we observe a phase coexistence between the superfluid phase and a normal
(liquid or glass) phase.
The region of superfluid-glass phase coexistence is particularly interesting since it represents a region which 
manifests both amorphous order and off-diagonal long range order, even if they are phase separated.
Besides the superfluid transition in Fig.\ref{quantum_phase_diagram_rho}  we reported the glass (Kauzmann) transition 
(dotted line) which is also expanded in the inset. From the inset it is clear that the glass transition as a function of 
the ``quantum parameter'' $\beta J$, has re-entrant behavior. 
In fact, looking at the $\beta=30$ or $\beta=15$ curves, one can note that the system reaches the glass 
phase at lower densities when we switch on quantum fluctuations.
This phenomenon, which at first sight can appear very surprising, has been recently found in a related work, focusing 
on the description of the quantum glass transition from the point of view of a microscopic theory of the system 
dynamics~\cite{MMBMRR10}.
Moreover it has also been seen in another recent work, in the context of a simple optimization 
problem \cite{FSZ10}. In general, one can imagine that, beyond the details which pertain to this model, this re-entrant 
behavior may have a more general interpretation, in terms of an order-by-disorder mechanism 
which is induced by quantum fluctuations and in which a particular 
order is selected for entropic reasons, as we will argue in the following.

\subsubsection{Zero temperature limit}
\label{quantum_Tzero}

We now give some hand-waving arguments on the behavior of the transition lines
for $T\to 0$. Although we don't expect our arguments to be completely convincing,
the following is the only consistent scenario we were able to elaborate that
is in agreement with the observed behavior of the lines at low temperatures.

The evolution of the phase transition lines in the $(\m,J)$ plane 
of Fig.~\ref{quantum_phase_diagram_mu}, suggests that in 
the limit $T \to 0$ both the second order superfluid transition line
and the Kauzmann transition line shrink to the origin of the axis.
Note that we are not interested in the region $\mu<0$ since in this region, 
at $T=0$, the system has a small density and it behaves very similarly to weakly interacting
hard core bosons. Note also that in the classical case
$J=0$, we know that $\mu_K$ is proportional to $T$ at low temperature, which 
is consistent with the shrinking of the line $\mu_K(J)$ to the origin. 
Therefore, for $T\to 0$ the glass
phase ``invades'' the lower part of the phase diagram in Fig.~\ref{quantum_phase_diagram_mu}, 
while the superfluid
phase ``invades'' the upper part, and the
first order superfluid-glass
transition line at $\m>0$ extends down to the origin ($J=0$ and $\mu=0$). 

Unfortunately, our 1RSB code becomes very slow and unstable for too large $\beta J$, preventing us from
drawing the superfluid-glass line for $J \gg T$. Still, we observe that the lines at higher temperature
seem to be close to the continuation of the lines at lower temperature. This indicates that, as expected,
for large enough $J$ and $\mu$, the system has reached its zero temperature limit. At the same time, this
allows us to ``extrapolate'' the first order superfluid-glass transition line at $T=0$ by taking, at each
$T$, the largest values of $J$ that we can access, and interpolating these values. The result is shown
as a black dashed line in Fig.~\ref{quantum_phase_diagram_mu}, and we believe that the extrapolation is 
a very reliable representation of the $T=0$ line. 
Therefore, at strictly zero temperature ($T=0$) and positive hopping ($J>0$), the phase diagram contains only a superfluid
and a glass phase separated by a first order transition. On the other hand, at strictly zero hopping $J=0$ and zero temperature,
one recovers the classical model of \cite{BM01}, which displays a RFOT between a liquid and a glass
(however, one has to rescale $\mu$ by the temperature in the classical limit, therefore the transition happens at $\mu=0$ in 
Fig.~\ref{quantum_phase_diagram_mu}). 
Hence, at $T=0$,
the limit of vanishing hopping is extremely singular, the behavior of the quantum model at any $J>0$ being completely different
from that of the classical model at $J=0$.

It is interesting to understand in more details how this singularity develops in the limit $T \to 0$.
Indeed, at large $\beta$, the normal liquid phase exists only in a region
$0\leq\mu\lesssim 1/\b$ and $0\leq J\lesssim 1/\beta$, see Fig.~\ref{quantum_phase_diagram_mu}.
Since both $J$ and $\mu$ go to zero with temperature, in this part of the phase diagram $V$ 
is much larger than any other energy scale and one can consider it as infinite. Then, only
the three energy scales $J, \mu, T$ remain and the phase diagram must be a function of 
$\b J$ and $\b \mu$ only. In the limit $\beta \to \io$, also this energy scale disappears,
and we conclude that at $T=0$, $J_c(\mu) \propto \mu$, \ie the superfluid-glass transition line must be linear
at small $J$ and $\mu$.

When we eliminate the chemical potential and look to the
plane $(\rho,J)$, the scenario described above leads to a phase diagram characterized by two 
 distinct regimes, as shown schematically in the left panel of Fig.~\ref{quantum_phase_diagram_rho}:
\begin{itemize}
\item If $J$ remains finite while $T \to 0$ (\ie if $\beta J \gg 1$), the normal liquid phase disappears.
Only the superfluid and glass phase survive, and they are separated by a phase coexistence region, determined by the 
Maxwell construction, and delimited
by lines that reach a finite limit when $T\to0$.
We expect that when $J/V\gtrsim 1$ the system is superfluid at all densities while 
when $J/V \lesssim 1$ the glass phase appears at large enough densities: therefore the coexistence line
must have the shape reported in the schematic plot (left panel) of Fig.~\ref{quantum_phase_diagram_rho}.
Although our 1RSB code is unable to access the region $J\sim V \gg T$, we could check that these lines have
the expected behavior at least within the RS approximation, which should be a good approximation at least
for not too small temperatures (we don't report these data to avoid confusion in the figures). 
\item
On the contrary, in the interval $J \lesssim 1/\b$ the system is still sensible to finite temperature effects,
but as we argued above, we expect these effects to depend only on the quantity $\b J$ at fixed $\rho$, since 
no other energy scale is relevant here ($J \sim T \ll V$, therefore $V$ is infinite and disappears from the problem).
In this regime a normal liquid phase, a second order superfluid transition $J_c(\rho)$ and a glass transition $J_K(\rho)$ 
still survive, but both critical values of $J$ vanish proportionally to the temperature in the limit $T\to 0$, therefore
confining the liquid phase to a smaller and smaller region of the phase diagram which at $T=0$ reduces to the
classical region $J=0$ alone. 
\item
The phase coexistence boundaries start from the tricritical
point, located at $J \propto T$ for $T \to 0$, and they extend into the large $J$ region of the phase diagram where they
have a finite limit for $T\to 0$. 
The only possibility to match the two regimes
is that the lines become vertical in the $(\rho,J)$ plane in the
region $\beta J \sim 1$; this should be evident from the schematic plot in the left panel of 
Fig.~\ref{quantum_phase_diagram_rho}.
In this way we define the two values of density that delimit the coexistence region at $T=0$ and small $J>0$.
\end{itemize}
 
Looking at the data at finite temperature in right panel of Fig.\ref{quantum_phase_diagram_rho} one sees 
that the superfluid transition far from the glass
satisfies well the scaling with $\beta J$ for $\rho \lesssim 0.3$, where the transition
remains second order for all values of $\b$. 
However Fig.~\ref{quantum_phase_diagram_rho} shows that around densities of the order 
$\rho\simeq0.3$ 
the transition lines do not scale very well. To observe the scaling
with $\b J$ one has to go to lower temperatures which are not easily accessed with our method.
Still, the phase coexistence lines become almost vertical at large $\b J$ around $\r \sim 0.5$
which is an estimate of the coexistence density for $T=0$ and $J \gtrsim 0$, according to the argument
above. The glass transition line is strongly re-entrant at the lowest temperature (we recall that for the 
classical model $\r_K\simeq0.5725$), 
which is consistent with this estimate. This means that in presence of quantum fluctuations a glass can be formed
at zero temperature for densities as low as $\r =0.5$, where the corresponding classical system
is in the liquid phase and quite far from the glass.

\subsubsection{Argument for the re-entrance of the glass transition line}
\label{sec:re-entrance}

The re-entrance of the glass transition shown in the inset of Fig.~\ref{quantum_phase_diagram_rho} is rather 
unexpected and it deserves a more detailed discussion.
It shows that if we consider the classical model at a density slightly below the Kauzmann transition, then it is possible
to make it condense into a glass phase just by switching on quantum fluctuations, for low enough temperatures. 
This means that the quantum dynamics induced by the hopping between neighboring sites
has a profound  effect on the thermodynamical properties of the systems. In fact, the kinetic exchange
selects as equilibrium states some states which are no more exponentially numerous, 
inducing the vanishing of the \textit{complexity} function. 

\begin{figure*}
\centering%
 \includegraphics[width=10cm]{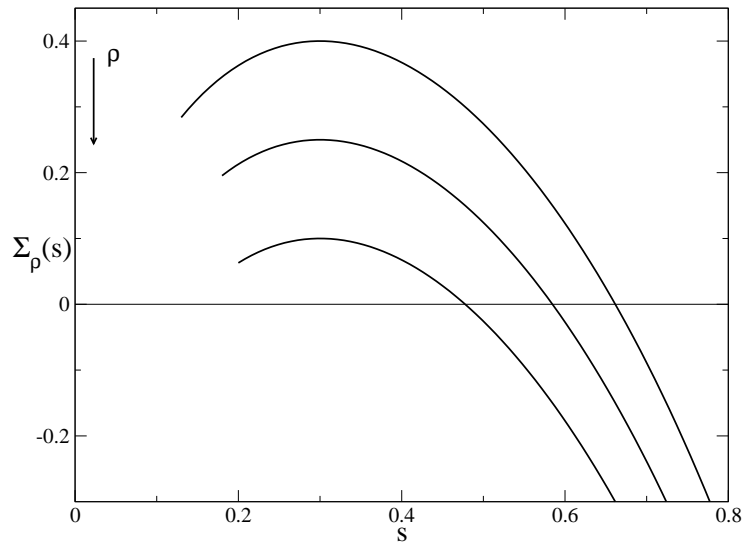}
\caption{A schematic plot of the typical behavior of the complexity as a function of the internal entropy of the states for various densities $\rho$ increasing from top to bottom.
}
\label{fig:sigma}
\end{figure*}

One can devise an explanation of this phenomenon in terms of the entropy of the states~\cite{FSZ10}.
A similar explanation which does not explicitly make use of entropy has been proposed in~\cite{MMBMRR10}.
Since the re-entrance happens close to the classical glass transition, we can for this
discussion consider $V \gg \mu, J, T$, \ie $V = \io$ going back to the original
hard sphere model of Biroli and M\'ezard~\cite{BM01}.
Let us focus first on the classical model at low enough temperatures.
Since $V=\io$, only zero energy configurations contribute to the measure.
Our argument starts from the consideration that beyond the dynamical transition $\r_d$,
the states (or ``clusters'') in which the Gibbs measure splits, are generically 
characterized by a distribution of internal entropies. 
The zero-temperature complexity $\Sigma(s)$ is the logarithm of the number of clusters
that have entropy $s$.
Since the entropy counts the number
of configurations that belong to a given cluster, the clusters with larger entropy have
more configurations, which means that particles are more ``mobile'' inside those clusters.
A second remark is that the complexity $\Sigma(s)$  
is a decreasing (in the branch of interest) convex function. 
Fig.~\ref{fig:sigma} shows its typical behavior 
for systems belonging to the same universality class as the Biroli-M\'ezard model~\cite{RBMM04}. 
While many small states appear in the measure, 
states with higher entropies are less numerous.
The external parameters, such as temperature and density, fix which are the dominant (equilibrium) states
and then which is the equilibrium entropy.

Next we assume that when quantum fluctuations are introduced, if $J$ is sufficiently
small, the distribution $\Si(s)$ is not strongly modified. 
This is due to the fact that configurations belonging to different clusters are at a distance corresponding to
$O(N)$ particle movements.
It is not possible to go ``continuously'' with single particle hoppings from one cluster to the other,
because this requires global rearrangement of the system. If one thinks in terms of perturbation theory 
in the hopping term, this happens only at extremely high orders $J^{O(N)}$.
Along this line of thought, one can safely diagonalize the Hamiltonian restricted
to each state when $V = \io \gg J$.
The major effect of quantum 
fluctuations in this regime is rather to induce, for each cluster, a kinetic energy gain proportional
to its entropy. This gain comes from the fact that higher entropy clusters are larger, therefore
in those clusters particles can delocalize more, allowing to lower the kinetic energy. Conversely, in small
entropy clusters, particles are tightly packed and cannot delocalize to lower their kinetic energy.
In other words, 
the entropy of each cluster measures the number of  ``neighboring'' configurations belonging to the state,
namely configurations which can be reached one from the other through single particle
movements. 

The shift of the quantum energy, proportional to $J$ times the classical entropy, 
has an effect on the selection of the equilibrium states:
it will favor states with bigger entropy because their energy is lowered more.
Since these states are less numerous,
increasing the hopping $J$, one expects that the complexity will be lowered, and 
a condensation transition will be eventually induced at large enough $J$.
All these arguments are not modified when besides the entropy each state is also characterized by a classical energy
(\ie for finite $V$).
A distribution of energies of the states must be included, but this does not change the result, just implying
a more complicated temperature dependence.
Given the complexity of the model under consideration these arguments are simply a qualitative explanation of 
what happens. However we have recovered analytically the same results ``quantizing'' a simpler model whose
classical phase diagram presents the same properties of the model considered here~\cite{FSZ10}.

\subsection{Quantum dynamics}
\label{sec:quantum_dynamics}

From a classical point of view the dynamics of glassy systems has received a lot of attention and 
their slow dynamic behavior has been the subject of a vast part of the literature (see \eg~\cite{KA95a,KA95b}).
Since dynamics and thermodynamics are inevitably intertwined in quantum mechanics, the study of time
dependent equilibrium correlations is particularly interesting to understand to what extent the phenomenology
of glassy quantum systems resembles that of their classical counterpart.
In this section we analyze the local Green function and the local density-density correlation in imaginary time.

\subsubsection{Green function and time scales}

The Green function is defined, for $-\beta/2 \leq \t \leq \beta/2$, by
\beq\label{eq_green_def}
G^i(\t) =  \theta(\t) G^i_>(\t)+\theta(-\t) G^i_<(\t) = \la T \hat{a}_i(\t) \hat{a}^\dag_i(0) \ra ,
\eeq
with the advanced and retarded Green functions
\beq\label{eq_green_ar}
\begin{split}
G^i_>(\t) & = \la\hat{a}_i(\t) \hat{a}_i^\dag(0) \ra = \frac{1}{Z} \Tr \big[ e^{-(\b-\t) \hat{H}} \hat{a}_i e^{-\t \hat{H}} \hat{a}^\dag_i \big] = 
  \frac{1}{Z} \sum_{a,b} e^{-(\b-\t) E_{a} -\t E_{b}} |\langle\psi_{b} | \hat{a}_i^\dag  | \psi_{a} \rangle|^2 \\
G^i_<(\t) & = \la \hat{a}^\dag_i(0) \hat{a}_i(\t) \ra = \frac{1}{Z} \Tr \big[ e^{-(\b+\t) H} \hat{a}^\dag_i e^{\t \hat{H}} a_i \big] 
  = \frac{1}{Z} \sum_{a,b} e^{-(\b+\t) E_{a}+\t E_{b}} |\langle\psi_{b} | \hat{a}_i  | \psi_{a} \rangle|^2  \ ,
\end{split}\eeq
where the many-body eigenvalues and eigenstates satisfy $\hat{H} | \psi_a \rangle = E_a | \psi_a \rangle$.
The way to compute time-dependent correlation functions such as the above within the cavity formalism 
has been detailed in~\cite{STZ09}.

\begin{figure*}
\centering%
 \includegraphics[width=12cm]{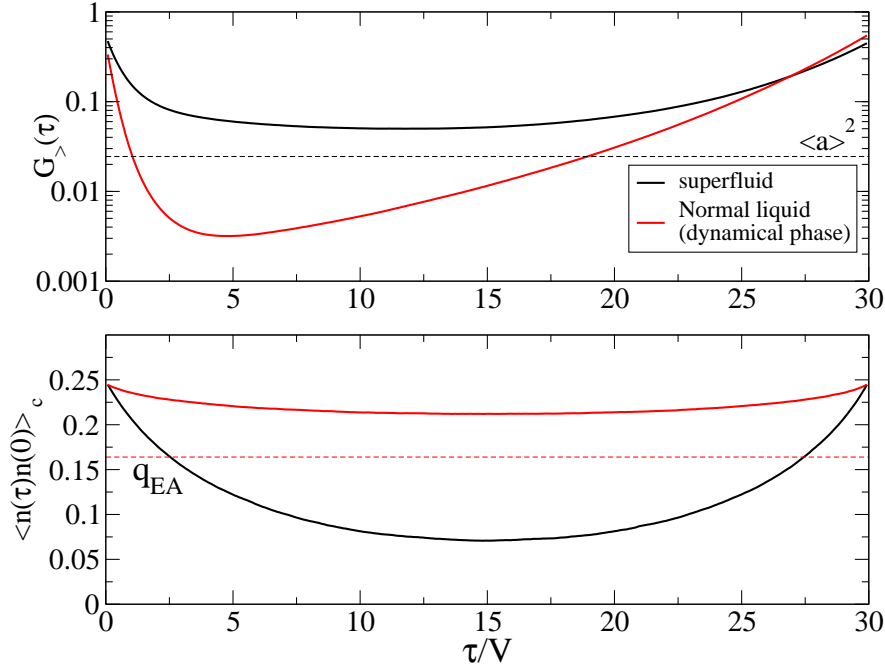}
\caption{
Time-dependent correlation functions obtained with the RS cavity method.
\textit{Upper panel}: Imaginary time advanced Green function $G_>(\t)$, for $c=3$ and $\ell=1$,
at $\b V=30$ and $J/V=0.1$,
in the superfluid phase at $\mu/V=0.275$ and in the liquid phase just
before the Kauzmann transition, at $\mu/V=0.388$. 
The dotted line indicates the square of the expectation value of the bosonic operator $\la \hat a\ra^2$
corresponding to the superfluid phase.
Note that the Green function $G(\t)$ is obtained by the periodic image of period $\b$ 
of this function for $\t \in [-\b/2,\b/2]$.
\textit{Bottom panel}: Time dependent density-density correlation
for the two same regimes. The dotted line indicates the value of the $q_{EA}$ for the normal liquid
(obtained with the 1RSB cavity method).
}
\label{fig:green}
\end{figure*}

The plot in Fig.~\ref{fig:green} shows the advanced Green function $G_>(\t)$ 
in the superfluid phase and in the normal liquid phase, just before
the Kauzmann glass transition and slightly above the dynamical transition.
The Green function follows three main regimes.
First, for $\t\to0^+$ one finds a fast decay of the correlations. The energy scale which fixes the 
particle decay is the interaction potential. This is the dominant contribution
to the average energy of the state $a_i^\dag|\psi_{a} \rangle$ just after the insertion of one particle
on a low energy state $|\psi_{a} \rangle$. On the other hand the contributions from the chemical potential and the kinetic energy
are negligible with respect to $V$. Then for $\t\to0^+$ the process brings the system to highly excited states first.
On the contrary, on the side of holes, when $\t\to\b^-$, the interaction $V$ doesn't play 
any role. Removing a particle cannot lead to a cost in terms of $V$, and for short time the dominant
energy excitation is measured by a loss of chemical potential and the proper time scale is fixed by $\mu$.
Finally, in between the two exponential relaxations, the Green function present a third regime in which the decay is slower. 
Here in the region where the system is not superfluid we expect an exponential decay, 
with a small exponent, of the order of the hopping. 

\subsubsection{Density correlation and ergodicity breaking}

The second time dependent correlation which is interesting from the point of view of glassy 
dynamics is the density-density correlation, defined in imaginary time as follows:
\beq\label{nn_def}
\la \hat{n}_i(\t)\hat{n}_i(0) \ra =  \frac{1}{Z} \Tr \big[ e^{-(\b-\t) \hat{H}} \hat{n}_i e^{-\t \hat{H}} \hat{n}_i \big] \ . 
\eeq
In the limit of large time and small temperatures this quantity (in particular its connected component) decays to
 the overlap parameter $q_{EA}$ defined in (\ref{qea}). 
In fact, rewriting Eq.(\ref{nn_def}) we obtain
\beq\label{nn_def2}
\begin{split}
\la \hat{n}_i(\t)\hat{n}_i(0) \ra & = \frac{1}{Z} \sum_{a,b} e^{-\b E_{a} -\t (E_{b} - E_{a})} |\langle\psi_{a} | \hat{n}_i  | \psi_{b} \rangle|^2 \\
& =  \frac{1}{Z} \sum_{a} e^{-\b E_{a}} |\langle\psi_{a} | \hat{n}_i  | \psi_{a} \rangle|^2 +  
\frac{1}{Z} \sum_{a\neq b} e^{-\b E_{a} -\t (E_{b} - E_{a})} |\langle\psi_{a} | \hat{n}_i  | \psi_{b} \rangle|^2 
\ .
\end{split}
\eeq
The second term in (\ref{nn_def2}) goes to zero when $\beta J\gg\t J\gg1$ (assuming that $J \lesssim V,\mu$ since this
is the interesting regime for the purpose of this discussion), therefore in this limit
\beq
\la \hat{n}_i(\t)\hat{n}_i(0) \ra \to \frac{1}{Z} \sum_{a} e^{-\b E_{a}} |\langle\psi_{a} | \hat{n}_i  | \psi_{a} \rangle|^2  \ .
\eeq
The same result is obtained for real time correlations in the large time limit, for any $\beta$, if one assumes that terms in the sum
with $a \neq b$ vanish due to fast oscillations. In the following we focus on imaginary time correlations in the regime 
$\beta J\gg\t J\gg1$, but the same arguments can be repeated for real time correlations at any $\b$ and large times.
Moreover, in the following we implicitly assume that the thermodynamic limit $N\to \io$ is taken before any other limit.

Let us discuss first what happens in the liquid phase. Here, we expect that the connected correlation functions vanish in the large time
limit (since the liquid phase is assumed to be {\it ergodic}). 
Therefore we expect that the large time limit of $\la \hat n_i(\t) \hat n_i(0) \ra$ equals
$\la \hat n_i \ra^2$, which leads to
\beq\label{eth1}
 \frac{1}{Z} \sum_{a} e^{-\b E_{a}} |\langle\psi_{a} | \hat{n}_i  | \psi_{a} \rangle|^2 = \left[ \frac{1}{Z} \sum_{a} e^{-\b E_{a}} \langle\psi_{a} | \hat{n}_i  | \psi_{a} \rangle \right]^2 \ .
\eeq
 Defining 
\beq
\PP_\b[  n_i  ] =  \frac{1}{Z} \sum_{a} e^{-\b E_{a}}  \d[ \langle\psi_{a} | \hat{n}_i  | \psi_{a} \rangle - n_i ]
\eeq
as the probability distribution of $n_i^a$ over eigenstates sampled at temperature $1/\b$, Eq.~(\ref{eth1}) can hold
if and only if
\beq\label{eth}
{\cal P}_{\b} [ n_i ]  \underset{N\to\infty}{\to}  \d\[[ n_i  -  \la \hat{n}_i \ra \]]
\ .
\eeq
Since the sum is dominated by eigenvectors that have energy $E_a$ of the order of the average energy at temperature $\b$,
the equation above 
states that the quantity $n_i^a = \langle\psi_{a} | \hat{n}_i  | \psi_{a} \rangle$ {\it does not fluctuate} from eigenvalue to eigenvalue in this range of energy.
Indeed, even if up to now we chose the canonical ensemble, this requirement is unessential
and one could have equivalently stated this concentration property of local observables 
also at the level of microcanonical averages. This assumption is known 
as a weak form of the \textit{eigenstate thermalization hypothesis}~\cite{De91,Sr94,PH10,BR10,CRFSS10},
see \cite{ETH_Giulio} for a more detailed discussion.

We now repeat the same discussion in the glass phase.
To this aim, we need to make an important assumption on the spectrum of the Hamiltonian in the 1RSB phase.
Indeed, as we previously discussed, the cavity method allows us to establish that the imaginary-time path Gibbs measure
constructed via the Suzuki-Trotter formalism undergoes a clustering transition as in the classical case. Yet, imaginary
time paths are abstract objects, and the consequences of this clustering transition on the spectrum of the Hamiltonian are
not clear, since a priori clustering might depend on the particular basis that is chosen and on other details of the Suzuki-Trotter
decomposition. In the following, we {\it assume} that in the clustered phase,
the relevant eigenstates of the Hamiltonian (those which dominate the canonical sum) are also arranged in
disconnected {\it clusters} corresponding to the thermodynamic states~\cite{GS98,BCZ08}. 
Other states of the Hilbert space do not belong to any cluster, but their energies
are extensively different from the thermodynamic energy and therefore they are exponentially suppressed in the canonical sum.
We can then decompose the canonical sum as a sum over the clusters
and a sum over the eigenstates belonging to the same cluster.
A reasonable hypothesis that holds in the classical case and that we believe to hold also here,
is that the dynamical behavior of the states inside each cluster is \textit{thermal} (or \textit{ergodic}).
In other words, we assume that the reasoning we made for the liquid can be applied within any cluster $\a$,
leading to
\beq\label{hyp}
\frac{1}{{\cal Z}_{\a}} \sum\limits_{a \in \a} e^{-\b E_{a}} | \la\psi_{a}|\hat{n}_i|\psi_{a}\ra |^2  \underset{N\to\infty}{\to}
 \la \hat{n}_i \ra_{\a}^2 = \left[ \frac{1}{{\cal Z}_{\a}} \sum\limits_{a\in c} e^{-\b E_{a}} \la\psi_{a}|\hat{n}_i|\psi_{a}\ra \right]^2 
\eeq
and
\beq
{\cal P}^\a_{\b} [ n_i ] 
=\frac{1}{{\cal Z}_{\a}} \sum\limits_{a \in \a} e^{-\b E_{a}} \d( n_i - \la\psi_{a}|\hat{n}_i|\psi_{a}\ra)
\underset{N\to\infty}{\to}  \d\[[n_i  -  \la \hat{n}_i \ra_\a \]]
\ .
\eeq
On the other hand, we expect $\la \hat n_i \ra_\a$ to fluctuate from cluster to cluster,
therefore eigenstate thermalization can be assumed to hold at most inside each cluster,
signaling, globally, a breakdown of ergodicity. 
Substituting (\ref{hyp}) in (\ref{nn_def2}) one obtains:
\beq\label{nn_def3}
\la \hat{n}_i(\t)\hat{n}_i(0) \ra \to  \frac{1}{Z} \sum_{a} e^{-\b E_{a}} |\langle\psi_{a} | \hat{n}_i  | \psi_{a} \rangle|^2  =  
\frac{1}{Z} \sum_\a {\cal Z}_{\a} \left[ \frac{1}{{\cal Z}_{\a}} \sum\limits_{a \in \a} e^{-\b E_{a}} | \la\psi_{a}|\hat{n}_i|\psi_{a}\ra |^2 \right]
 \underset{N\to\infty}{\to}
\frac{1}{Z} \sum_\a {\cal Z}_{\a}  \la \hat{n}_i \ra_\a^2  \ .
\eeq
This reduces to the average of the square local density over the amorphous states.
Averaging over different sites 
and subtracting the connected term one recovers the definition of Eq.~(\ref{qea}),
since $W_\a = \ZZ_\a/Z$:
\beq\label{nn_qea}
\la \hat{n}_i(\t)\hat{n}_i(0) \ra - \la \hat{n}_i \ra^2 \underset{\beta J \gg\t J \gg1}{\longrightarrow} 
\frac{1}{Z} \sum_\a {\cal Z}_{\a}  \la \hat{n}_i \ra_\a^2   - \left[ \frac{1}{Z} \sum_\a {\cal Z}_{\a}  \la \hat{n}_i \ra_\a \right]^2
= q_{EA}
\ . 
\eeq
When the quantity in (\ref{nn_qea}) goes to zero, it means that (\ref{eth}) holds in the whole 
Hilbert space and thus the system is ergodic.
We note that a consequence of this discussion is that in the quantum case, the RS solution can
give informations about clustering. Indeed, in the region between the dynamical and thermodynamical transitions,
the Parisi parameter is $m=1$ and in that case even in presence of clustering, local quantities can be computed
using the RS solution. Therefore one can compute $\la \hat{n}_i(\t)\hat{n}_i(0) \ra$ using the RS cavity method,
and detect the presence of clustering using Eq.~(\ref{nn_qea}).
This is not surprising, since already in the classical case the investigation of dynamics at the RS level can give
information about clustering and 1RSB~\cite{CK93}. Similar results have been discussed in \cite{BC01,CGS01}.

In Fig.~\ref{fig:green} we have plotted the time dependent density-density correlation for
the same parameters used for the Green functions, namely $\b=30,J=0.1$ and two
different densities such that the system is in the superfluid and in the liquid phase close to the glass transition.
The plot shows that the function within the glass does not relax at large times. It remains at a high plateau,
higher than $q_{EA}$. The discrepancy is due to the fact that $\b J$ is not sufficiently high to allow the complete relaxation
of the correlations. This is seen also in the superfluid phase where the correlation remains always above zero.

\subsection{Exact diagonalization results}
\label{sec:ED}

To conclude this section, we report some results of exact diagonalization that we performed on this model.
The aim of this investigation was mainly to check that the glass phase is not superfluid: indeed, with 
the cavity method we do not find any non-trivial 1RSB solution with $\la \hat a \ra \neq 0$, but we cannot
reach very low temperatures so we cannot exclude that a tiny condensate fraction might appear at a very low
temperature. Moreover, with the cavity method we cannot access the region of the glass phase at $\beta J \gtrsim 1$ 
where superfluidity might be expected, because the 1RSB code becomes extremely slow and unstable in that region.
Exact diagonalization can give in principle access to the exact ground state of the system, but
the sizes one can study are very small.

There is however one very important problem: in the glass phase, we expect that the ground state is highly degenerate
in the thermodynamic limit, but at finite $N$, this degeneracy should be lifted leaving a unique ground state
and a band of very low energy almost degenerate eigenstates.
Defining, on a given small graph, the glass ground states is a very difficult task since they are
well defined only in the thermodynamic limit, while at finite $N$ they are linear combinations 
of the ground and low-energy eigenstates.
Let us make our expectations more precise. Based on the arguments of~\cite{GS98,BCZ08,FSZ10}, and on the discussion
of section~\ref{sec:re-entrance}, we expect
that in the glass phase, at very large (but finite) $N$, one can construct almost exact low-energy eigenstates 
$|GS_\a\rangle$ of the quantum Hamiltonian, each of them being a linear combination of classical configurations belonging only to one classical glass state (or ``cluster''). 
Indeed, since different glassy states have Hamming distance of order $N$ in the Hilbert space, we expect 
tunnelling between them to be of order $\sim e^{-N}$. 
However, exactly because of this tiny but finite tunnelling rate, the {\it exact} 
eigenstates of the Hamiltonian at finite $N$ are linear combination of the states $|GS_\a \rangle$,
and are spread in an interval of energy $\sim e^{-N}$ above the exact ground state~\cite{GS98}. 
The lowest of them might well be an almost uniform superposition of all the glassy states $|GS_\a\rangle$
(think to the classical example of a double well potential with a finite barrier), and in that case 
the local density would be uniform in the ground state making the detection of glassy order impossible without looking
to the whole band of almost degenerate states.

When $N$ is very small, however, the situation might be very different. Indeed, in this case we expect that the number of
glassy states will be quite small, 
since at high density on a very small graph there will be not so many ways
of packing particles. If a given graph has too many zero-energy configurations, then at such small $N$ particles will be
extremely mobile and delocalized, and the system will be physically closer to the superfluid than to the glass phase.
Conversely, if there are only a few good packings, they will be arranged in clusters that are not completely equivalent:
one of them will have the largest classical entropy (compare the discussion of section~\ref{sec:re-entrance})
and the corresponding state $|GS_\a \rangle$ will have lower energy than all others $|GS_{\a'}\rangle$.
The probability of having two equivalent ``clusters'' that could mix to form an almost 
uniform ground state should be quite small for small values of $N$.
To check this hypothesis we measured the parameter $q_{\rm av}$ (recall that we can't access $q_{EA}$) 
defined in Eq.~(\ref{qav}), in the ground state
for some values of the parameters deep in the glass phase (we give more details below).
We found that $q_{\rm av}$ has a strongly bimodal distribution over the random graphs: for some graphs
$q_{\rm av}$ is extremely small, while for some others it is of the order of the $q_{EA}$ computed by 
the cavity method. We found that the fraction of the latter graphs increases with $N$ in the regime
of small $N$ we can access.
We interpret this as a confirmation of the discussion above: although for $N\to\io$ one must find
$q_{\rm av}=0$ with probability one, the physical mechanism responsible for this is the huge degeneracy
of equivalent clusters, which is still not present for such small $N$. Based on this, we {\it assume}
that the ground state of those graphs that have a large $q_{\rm av}$ is concentrated on a single glass state,
and we focus on them for our analysis, while we disregard those graphs that have $q_{\rm av} \sim 0$.

We stress that the above procedure is quite arbitrary and strongly biased by our expectations
and by the comparison with the cavity method.
Therefore, the results of exact diagonalization might be misleading
and shall be taken with extreme care. 
On the other hand, the cavity method gives the {\it exact} solution
of the problem in the thermodynamic limit (at finite temperature), so 
it is reasonable to use it as a guide to interpret the exact diagonalization results.
Indeed, a systematic study of this model only based of exact diagonalization is impossible
given the technical difficulties discussed above and the complexity of the spectrum of the
Hamiltonian (see also~\cite{FSZ10}). In the following, we give the details of our procedure
and the results, so that the reader will be able to build her/his own interpretation of the data.

\begin{figure*}
\centering%
   \subfigure[][] {
 \includegraphics[width=0.48\textwidth]{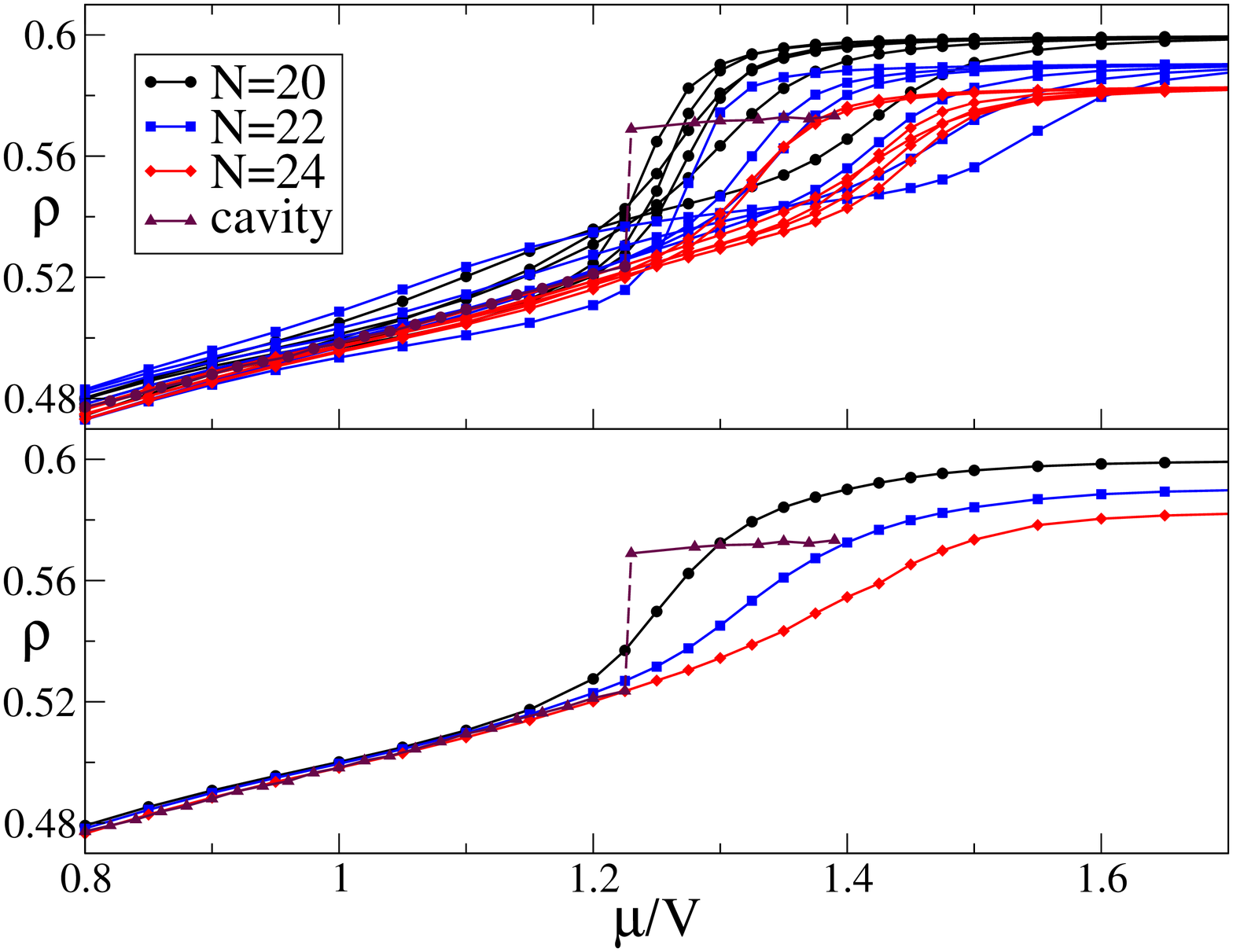}
 }
   \subfigure[][] {
    \includegraphics[width=0.48\textwidth]{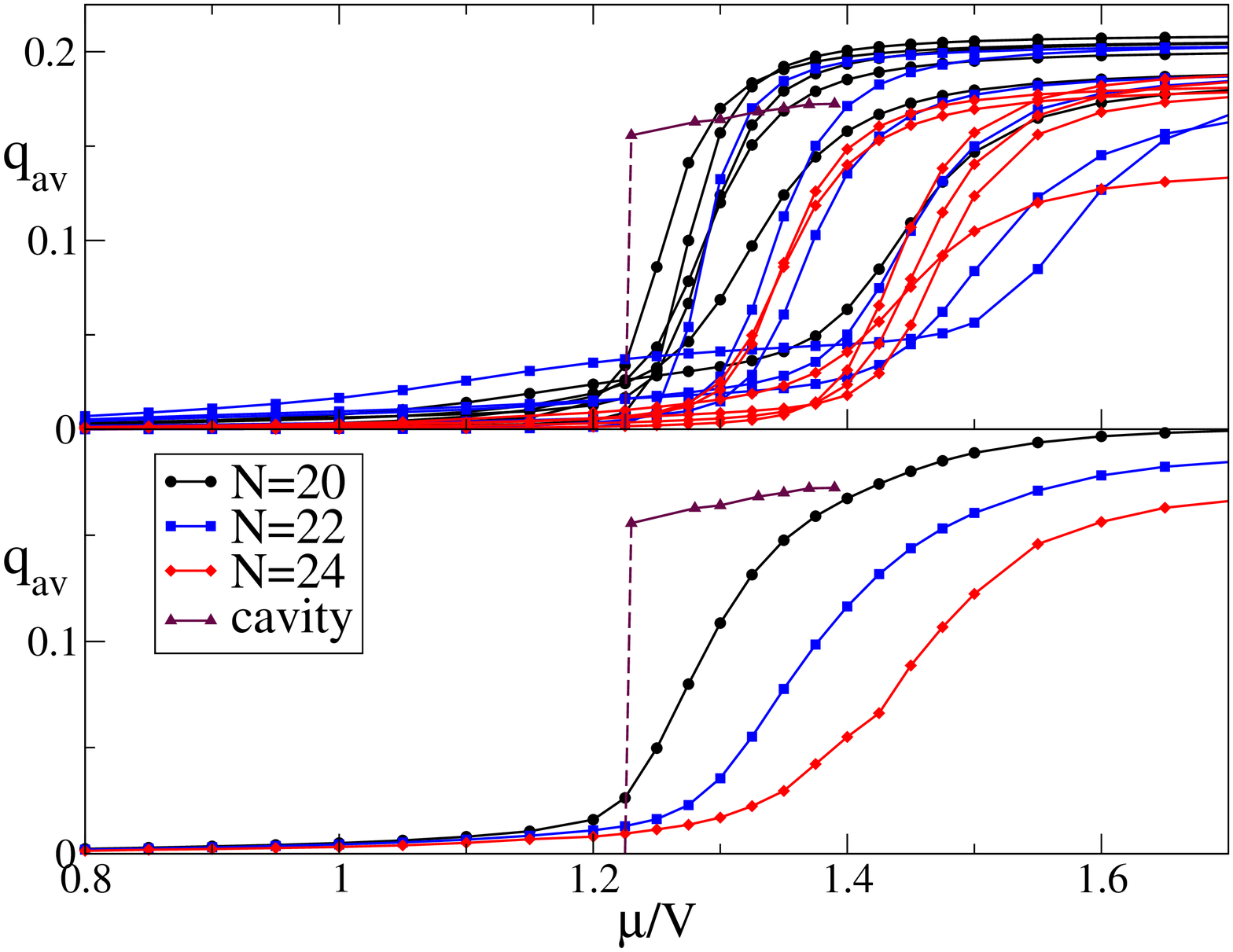}
      }
\caption{
Ground state density (a) and $q_{\rm av}$ (b) as functions of $\m/V$, with $c=3$ and $\ell =1$, 
at hopping $J/V=0.3$, obtained by exact diagonalization in the grand-canonical ensemble.
The maroon triangles show the result of the cavity method for the same parameters
at $\b V=15$. 
In correspondence of the jump there is the first order superfluid-glass phase transition.
{\it Upper panels}: The black circles, blue squares and red diamonds represent the ground state density and $q_{\rm av}$ for some
selected realizations of random graphs of size, respectively, $N=20, 22, 24$.
{\it Lower panels}: Average of the same quantities over 50 random graphs.
}\label{fig:ED_grancan}
\end{figure*}

\subsubsection{Results in the grand-canonical ensemble}

Another technical problem in doing exact diagonalization on this system is that 
the Hamiltonian conserves the number of particles. Therefore, 
its eigenstates are such that
$\la \hat{a} \ra = 0$, and moreover, the density,
which is the control parameter, has jumps of order $1/N$ at some values of the chemical potential.
The latter is a particularly severe problem, because all the relevant transitions happen in a very
small range of densities, of the order of $1/N$ for the values of $N$ we could access.
A way out of both problems is to perform exact diagonalization in the grand-canonical ensemble,
introducing a small field $h$ coupled to the condensate, by adding to the Hamiltonian
a term $-h \sum_i ( \hat{a}_i+\hat{a}^{\dag}_i )$. In the superfluid phase the symmetry is spontaneously
broken, therefore if one takes the limit $h\to 0$ after $N\to \io$, the average of $\la \hat a \ra$ is finite,
as we discussed in Eq.~(\ref{ahdef}). 
In practice, one can scale the field $h$ as $h = h_0/\sqrt{N}$ upon increasing $N$, which would give the
correct result in the thermodynamic limit.

We performed exact diagonalizations in the grand-canonical ensemble at fixed $J=0.3$, with $h_0=0.09$.
We first fixed $\mu=1.6$ deep in the glass phase and investigated the distribution of $q_{\rm av}$ over
the choice of the random graph. We always found a bimodal distribution with a dip around $q_{\rm av}=0.1$.
Therefore we discarded the graphs that had a value of $q_{\rm av} < 0.1$ at this $\mu$.
Table~\ref{tab:1} shows the statistics of the discarded graphs that we found analyzing different 
sizes $N=20, 22, 24$.

\begin{table*}
\begin{tabular}{|c|c|c|c|}
\hline
  $N$ & Total graphs & Discarded & Accepted fraction  \\
\hline
 20 & 1470 & 1420 & 0.034 \\
 22 & 506 & 456 & 0.099  \\
 24 & 206 & 156 & 0.243  \\
 26 & 89 & 39 & 0.562   \\
 28 & 77 & 27 & 0.649   \\
\hline
\end{tabular}
\caption{
Statistics on the number of discarded graphs 
(having $q_{\rm av} < 0.1$ in the ground state)
at different $N$ and $J/V=0.3$. For $N=20, 22, 24$, we did grand-canonical computations
with $\mu/V=1.6$. For $N=26, 28$, we did canonical computations with $15, 17$ occupied
sites (corresponding to $\r = 0.577, 0.607$ respectively).
}
\label{tab:1}
\end{table*}

We then performed, for the accepted graphs, a scan in $\mu$ to study the glass transition.
In Fig.~\ref{fig:ED_grancan} we show a comparison between the ground state density and $q_{\rm av}$
obtained with exact diagonalization and the same quantities obtained with the cavity method at $\b=15$.
Due to the first order phase transition in $\m$, both have a jump at a value $\m_c$.
The upper panels of Fig.~\ref{fig:ED_grancan} show the density and $q_{\rm av}$ for some representative graphs,
while the lower panels show the average over 50 accepted graphs.
The agreement between the exact diagonalization and the cavity method
is very good in the RS region, while it is poor within the glass phase.
The reason is that here, on the one hand exact diagonalization results suffer 
of strong finite size effects, and on the other hand 
the predictions of the cavity method are computed at finite temperature.
The transition predicted by the cavity method at $\b=15$ 
appears at $\mu_c \simeq 1.23$, but a slight residual dependence of $\mu_c$ on $\beta$ is present, and if we extrapolate 
the results of the cavity method computed at $\b=8$ and $\b=15$, for $J=0.3$, we obtain $\mu_c \simeq 1.3$
at $T=0$.
The results of the exact diagonalization instead support a transition at a slightly larger value of $\mu$.  
However, the extrapolation of the results obtained by exact diagonalization within the grand-canonical ensemble 
can be non-monotonic in the size of the system especially in the glass phase where the average density 
at fixed $\m$ does not coincide for graphs of different sizes and is basically fixed by the allowed filling fractions.
The calculation of the condensate fraction
through Eq.~(\ref{ahdef}) turned out to be unreliable even at small densities for such small sizes, so we do not
report it.

\subsubsection{Results in the canonical ensemble}

\begin{figure*}
\centering%
   \subfigure[][] {\label{fig:ED_rhoc}
 \includegraphics[width=0.48\textwidth]{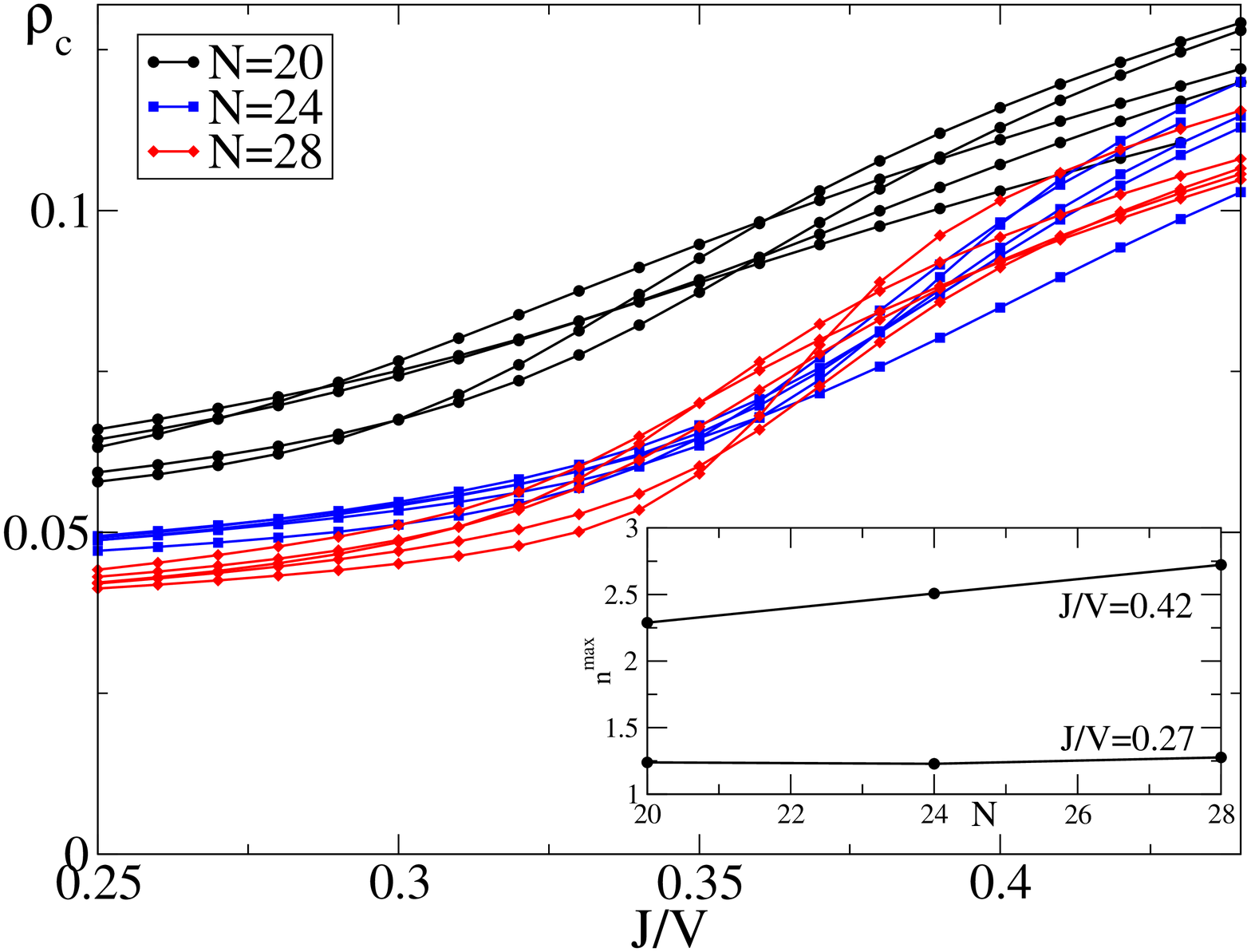}
 }
   \subfigure[][] {\label{fig:ED_qea_canonico}
    \includegraphics[width=0.48\textwidth]{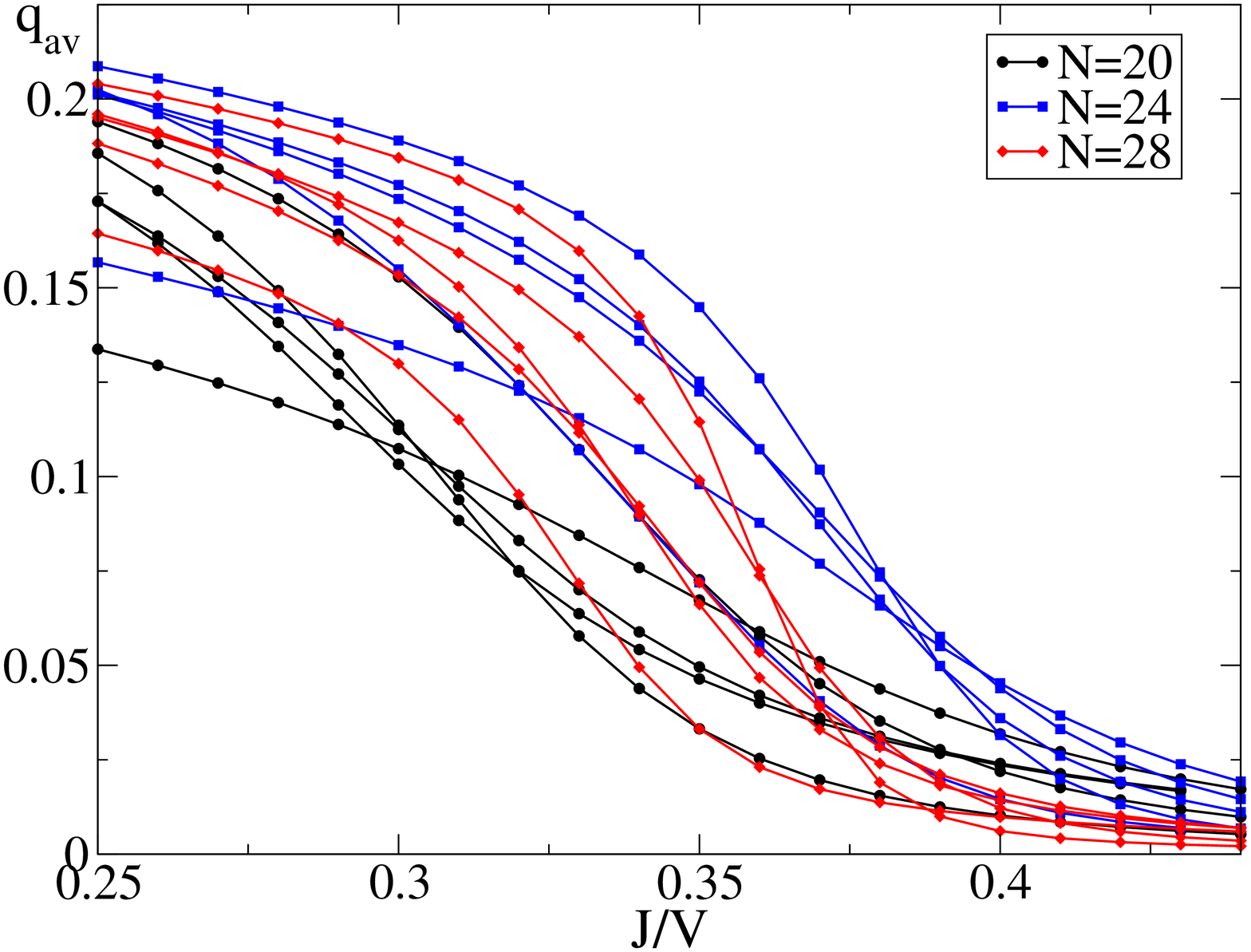}
      }
  \caption{
Exact diagonalization results in the canonical ensemble
for $c=3$ and $\ell=1$, for three different sizes
$N=20, 24, 28$, with $N_p=12, 14, 17$ occupied sites respectively
($\r \sim 0.6$), as functions of $J/V$. Results are shown for some
representative graphs chosen according to the criterion discussed in the text.
(a) The condensate density is seen to decrease with $N$ at small $J$, while it
is almost independent of $N$ at large $J$.
In the inset we show the average (over 50 accepted graphs) 
of the largest eigenvalue $n^{max}$ of the one-body density matrix
for two different hopping strenghts in the superfluid and glass phase,
as a function of $N$.
The former increases linearly with $N$, the latter is independent of $N$.
 (b) Behavior of $q_{\rm av}$ for the same graphs.
}
  \label{fig:ED_scan_J}
\end{figure*}

An alternative way of avoiding the problem of varying the density is to work at fixed density
and look at the superfluid-glass transition as a function of the hopping $J$, in the canonical 
ensemble. The advantage is that we can access slightly larger sizes in this case.
We repeated the same procedure to select the graphs. 
For $N=20, 22, 24$, we kept the same graphs of the grand-canonical study.
For larger $N$, we fixed the hopping at $J=0.3$, and the density
as close as possible to $\r=0.6$, and we again discarded the graphs that had a value of $q_{\rm av} < 0.1$.
Table~\ref{tab:1} shows the statistics of the discarded graphs for different $N$.

The condensate density $\r_c$ can be conveniently estimated in this case by the
maximum eigenvalue $n^{max}$ (divided by $N$) of the one-body density matrix $\hat{\sigma}_{ij}=\langle \hat{a}^{\dag}_{i}\hat{a}_{j} \rangle$,
where the average is taken over the ground state~\cite{PO56}.
Indeed, in presence of Bose-Einstein condensation one has $\rho_c = \lim_{N\to\io} n^{max}/N$, while all the other eigenvalues
are finite for $N\to\io$. This is a well-defined criterion for Bose-Einstein condensation
in interacting systems, proposed in \cite{PO56}. The reduced density matrix  $\hat{\sigma}_{ij}$
is an Hermitian single particle operator that can be diagonalized:
\beq
 \hat{\sigma} = \sum\limits_{\l} n_{\l} |\l\rangle \langle\l |
 \ ,
\eeq
 and its spectrum $\{n_{\l}\}$ can be interpreted as the distribution of the average occupation number
 of the single particle states $| \l \rangle$. 
 From the properties of semi-positivity and of the trace $\text{Tr}[ \hat{\sigma}]=N_p$, where $N_p$ is the
 number of particles in the system,
 each eigenvalue is subject to the constraint $0\leq n_{\l}\leq N_p$. The manifestation of Bose-Einstein condensation appears
 as an extensive occupation of a single particle state, namely $n^{max}={\cal O}(N)$ and it gives an
 estimate of the condensate density of particles in that state as $\r_c=n^{max}/N$.
 The presence of an extensive eigenvalue is related to the appearance of off-diagonal long range order.
 In fact, considering the limit
 \beq
 \lim\limits_{|i-j|\to\infty}  \langle \hat{a}^{\dag}_{i}\hat{a}_{j} \rangle =  \langle \hat{a}_i \rangle \langle \hat{a}_j \rangle
 \to n^{max} \l^{max}_i \l^{max}_j 
  \ ,
  \eeq 
where $ \l^{max}_i=\langle i | \l^{max}\rangle $, one recovers the relation
between the expectation value of $\langle \hat a_i\rangle$ and the condensate density:
\beq
\frac{1}{N} \sum\limits_{i} |\langle \hat{a}_i\rangle|^2 = \frac{n^{max}}{N} \sum_i |\langle \l_i^{max}\rangle|^2 = \frac{n^{max}}{N}  = \r_c
\ .
\eeq
In Fig.~\ref{fig:ED_scan_J} we show the 
behavior of the $q_{\rm av}$ and of the condensate density for $N=20, 24, 28$, as a function of $J$ at fixed density.
The comparison of the results for the three sizes that we show in Fig.~\ref{fig:ED_scan_J} is not straightforward
because they correspond to slightly different densities. Given the small sizes that are tractable by exact diagonalization
it is not possible to change smoothly the density.  However one can note that on average $n^{max}$ grows linearly with $N$
for large $J$, while it is independent of $N$ inside the glass phase, see the inset of Fig.~\ref{fig:ED_rhoc}.
We therefore conclude that the exact diagonalization data support the result that the glass phase is non-superfluid
even at zero temperature and for large hopping.

\section{Conclusions}
\label{sec:conclusions}

In this paper we studied the quantum version of a prototypical mean field lattice model of glass forming
liquid, the Biroli-M\'ezard model~\cite{BM01}. With respect to previous studies of quantum versions
of mean field glass models~\cite{Go90,NR98,BC01,CGS01,JKKM08,WSW03,JKSZ10}, this model has several important
characteristics in the classical case: 
{\it i)} it presents a glass transition at zero temperature as a function of density; and
{\it ii)} it presents a large degeneracy of glassy states with different entropies.
We previously reported on the study of a model with these properties~\cite{FSZ10}, which, although very instructive,
is quite abstract. 
The Biroli-M\'ezard model is instead a realistic model of interacting bosons, which can be quantized adding an hopping term
and therefore allowing for a superfluid phase.
We also stress once again that the classical physics of the Biroli-M\'ezard model is that of a Random First
Order Transition (RFOT), therefore it should be in the same universality class of structural glasses, and quite
different from a standard second order spin glass transition such as the one of the Sherrington-Kirkpatrick
model~\cite{MPV87}. In the latter case, the addition of an hopping term gives rise generically to a superglass
phase~\cite{CTZ09,TGIGM10}; we investigated in this paper whether this remains true for a model in the RFOT class.

We investigated the phase diagram of the model by means of the 1-step replica symmetry breaking quantum 
cavity method, that we conjecture to give the exact solution of the thermodynamics of the model. We showed that the
phase diagram is quite rich, displaying superfluid, normal liquid, and glass phases, separated by different
phase transitions. In particular, we obtained two unexpected and quite interesting results: first of all,
we showed that at low enough temperature the glass transition line is re-entrant as a function of quantum fluctuations, 
implying that one can form a glass by increasing quantum fluctuations at fixed density. 
Similar results have been obtained in~\cite{MMBMRR10} by an extension of Mode-Coupling Theory to quantum hard spheres. 
Since Mode-Coupling Theory is an approximate description of the glass transition based on some uncontrolled approximations, it is very important
to confirm the result of~\cite{MMBMRR10} by means of an exact solution of a mean field model.
Additionally, we showed that the standard RFOT glass transition is replaced by a first order superfluid-glass transition at zero 
temperature, accompanied by phase coexistence between the two phases, while at the same time the glass transition completely
disappears. This shows that for models with such a complex phase space in the classical limit, introducing quantum fluctuations
has a dramatically singular effect, changing completely the nature of the transition between the liquid and the glass phases.
Moreover, the first order superfluid-glass transition is accompanied by a jump in density, implying that there exists an
interval of densities where the two phases would coexist in a finite dimensional version of the model. 
One would therefore
obtain a simultaneous presence of superfluid and glassy ordering, which however would not give rise to a true superglass phase 
since they would be phase separated. 
At variance to what happens in models displaying a second order spin glass transition~\cite{CTZ09,TGIGM10}, we did not find any pure ``superglass'' phase
in the quantum Biroli-M\'ezard model:
in more technical terms this means that, for the range of parameters considered here, we have not found a solution of the quantum 1-RSB equations
within the broken phase $\la \hat{a}\ra\neq0$.

These predictions will hopefully bring new insights on the current theoretical debate on the nature of 
superfluidity in solid Helium~\cite{BC08}. 
Moreover, the present study is quite important, in our opinion, since it opens the way to extend to the quantum regime 
the replica techniques that have been used in the description of classical glasses~\cite{MP99}. These techniques
are based on the assumption that the glass transition is a Random First Order Transition. Since the present study shows
that this is not true in the fully quantum regime, the method of~\cite{MP99} will have to be adapted to describe the crossover
between the RFOT and the first order regime. We expect that this should be possible using the phase diagram of the
model investigated here as a guide. If successful, this program should lead to a quantitative computation of the glassy
phase diagram for realistic potentials such as the one of Helium, which would in turn allow for a systematic numerical
and experimental investigation, in order to link consistently the results of~\cite{BPS06} with the established picture of classical glasses.

\acknowledgments

We warmly thank F.~Krzakala and M.~Tarzia for their useful collaboration in an earlier stage of this work,
and G.~Carleo for performing Monte Carlo simulations on the same model.
We also acknowledge very useful discussions with G.~Biroli, D.~Huse and D.~Reichman.
Part of the numerical calculations have been performed on the cluster ``Titane'' of CEA-Saclay
under the grant GENCI 6418 (2010).


\begin{thebibliography}{58}
\expandafter\ifx\csname natexlab\endcsname\relax\def\natexlab#1{#1}\fi
\expandafter\ifx\csname bibnamefont\endcsname\relax
  \def\bibnamefont#1{#1}\fi
\expandafter\ifx\csname bibfnamefont\endcsname\relax
  \def\bibfnamefont#1{#1}\fi
\expandafter\ifx\csname citenamefont\endcsname\relax
  \def\citenamefont#1{#1}\fi
\expandafter\ifx\csname url\endcsname\relax
  \def\url#1{\texttt{#1}}\fi
\expandafter\ifx\csname urlprefix\endcsname\relax\def\urlprefix{URL }\fi
\providecommand{\bibinfo}[2]{#2}
\providecommand{\eprint}[2][]{\url{#2}}

\bibitem[{\citenamefont{Kob and Andersen}(1995{\natexlab{a}})}]{KA95a}
\bibinfo{author}{\bibfnamefont{W.}~\bibnamefont{Kob}} \bibnamefont{and}
  \bibinfo{author}{\bibfnamefont{H.~C.} \bibnamefont{Andersen}},
  \bibinfo{journal}{Phys. Rev. E} \textbf{\bibinfo{volume}{51}},
  \bibinfo{pages}{4626} (\bibinfo{year}{1995}{\natexlab{a}}).

\bibitem[{\citenamefont{Kob and Andersen}(1995{\natexlab{b}})}]{KA95b}
\bibinfo{author}{\bibfnamefont{W.}~\bibnamefont{Kob}} \bibnamefont{and}
  \bibinfo{author}{\bibfnamefont{H.~C.} \bibnamefont{Andersen}},
  \bibinfo{journal}{Phys. Rev. E} \textbf{\bibinfo{volume}{52}},
  \bibinfo{pages}{4134} (\bibinfo{year}{1995}{\natexlab{b}}).

\bibitem[{\citenamefont{Pusey and Van~Megen}(1986)}]{PV86}
\bibinfo{author}{\bibfnamefont{P.}~\bibnamefont{Pusey}} \bibnamefont{and}
  \bibinfo{author}{\bibfnamefont{W.}~\bibnamefont{Van~Megen}},
  \bibinfo{journal}{Nature} \textbf{\bibinfo{volume}{320}},
  \bibinfo{pages}{340} (\bibinfo{year}{1986}).

\bibitem[{\citenamefont{Debenedetti}(1996)}]{De96}
\bibinfo{author}{\bibfnamefont{P.}~\bibnamefont{Debenedetti}},
  \emph{\bibinfo{title}{{Metastable liquids: concepts and principles}}}
  (\bibinfo{publisher}{Princeton Univ Press}, \bibinfo{year}{1996}).

\bibitem[{\citenamefont{Binder and Kob}(2005)}]{BK05}
\bibinfo{author}{\bibfnamefont{K.}~\bibnamefont{Binder}} \bibnamefont{and}
  \bibinfo{author}{\bibfnamefont{W.}~\bibnamefont{Kob}},
  \emph{\bibinfo{title}{{Glassy materials and disordered solids: An
  introduction to their statistical mechanics}}} (\bibinfo{publisher}{World
  Scientific}, \bibinfo{year}{2005}).

\bibitem[{\citenamefont{Wilks}(1967)}]{Wi67}
\bibinfo{author}{\bibfnamefont{J.}~\bibnamefont{Wilks}},
  \emph{\bibinfo{title}{{The Properties of Liquid and Solid Helium}}}
  (\bibinfo{publisher}{Clarendon Press, Oxford}, \bibinfo{year}{1967}).

\bibitem[{\citenamefont{Bonev et~al.}(2004)\citenamefont{Bonev, Schwegler,
  Ogitsu, and Galli}}]{BSOG04}
\bibinfo{author}{\bibfnamefont{S.}~\bibnamefont{Bonev}},
  \bibinfo{author}{\bibfnamefont{E.}~\bibnamefont{Schwegler}},
  \bibinfo{author}{\bibfnamefont{T.}~\bibnamefont{Ogitsu}}, \bibnamefont{and}
  \bibinfo{author}{\bibfnamefont{G.}~\bibnamefont{Galli}},
  \bibinfo{journal}{Nature} \textbf{\bibinfo{volume}{431}},
  \bibinfo{pages}{669} (\bibinfo{year}{2004}).

\bibitem[{\citenamefont{Boninsegni et~al.}(2006)\citenamefont{Boninsegni,
  Prokof'ev, and Svistunov}}]{BPS06}
\bibinfo{author}{\bibfnamefont{M.}~\bibnamefont{Boninsegni}},
  \bibinfo{author}{\bibfnamefont{N.}~\bibnamefont{Prokof'ev}},
  \bibnamefont{and}
  \bibinfo{author}{\bibfnamefont{B.}~\bibnamefont{Svistunov}},
  \bibinfo{journal}{Physical Review Letters} \textbf{\bibinfo{volume}{96}},
  \bibinfo{eid}{105301} (\bibinfo{year}{2006}).

\bibitem[{\citenamefont{Biroli et~al.}(2008)\citenamefont{Biroli, Chamon, and
  Zamponi}}]{BCZ08}
\bibinfo{author}{\bibfnamefont{G.}~\bibnamefont{Biroli}},
  \bibinfo{author}{\bibfnamefont{C.}~\bibnamefont{Chamon}}, \bibnamefont{and}
  \bibinfo{author}{\bibfnamefont{F.}~\bibnamefont{Zamponi}},
  \bibinfo{journal}{Physical Review B} \textbf{\bibinfo{volume}{78}},
  \bibinfo{eid}{224306} (\bibinfo{year}{2008}).

\bibitem[{\citenamefont{Hunt et~al.}(2009)\citenamefont{Hunt, Pratt, Gadagkar,
  Yamashita, Balatsky, and Davis}}]{HPGYBD09}
\bibinfo{author}{\bibfnamefont{B.}~\bibnamefont{Hunt}},
  \bibinfo{author}{\bibfnamefont{E.}~\bibnamefont{Pratt}},
  \bibinfo{author}{\bibfnamefont{V.}~\bibnamefont{Gadagkar}},
  \bibinfo{author}{\bibfnamefont{M.}~\bibnamefont{Yamashita}},
  \bibinfo{author}{\bibfnamefont{A.~V.} \bibnamefont{Balatsky}},
  \bibnamefont{and} \bibinfo{author}{\bibfnamefont{J.~C.} \bibnamefont{Davis}},
  \bibinfo{journal}{Science} \textbf{\bibinfo{volume}{324}},
  \bibinfo{pages}{632} (\bibinfo{year}{2009}).

\bibitem[{\citenamefont{Balibar and Caupin}(2008)}]{BC08}
\bibinfo{author}{\bibfnamefont{S.}~\bibnamefont{Balibar}} \bibnamefont{and}
  \bibinfo{author}{\bibfnamefont{F.}~\bibnamefont{Caupin}},
  \bibinfo{journal}{Journal of Physics: Condensed Matter}
  \textbf{\bibinfo{volume}{20}}, \bibinfo{pages}{173201}
  (\bibinfo{year}{2008}).

\bibitem[{\citenamefont{Lubchenko and Wolynes}(2007)}]{LW07}
\bibinfo{author}{\bibfnamefont{V.}~\bibnamefont{Lubchenko}} \bibnamefont{and}
  \bibinfo{author}{\bibfnamefont{P.~G.} \bibnamefont{Wolynes}},
  \bibinfo{journal}{Annual Review of Physical Chemistry}
  \textbf{\bibinfo{volume}{58}}, \bibinfo{pages}{235} (\bibinfo{year}{2007}).

\bibitem[{\citenamefont{Singh et~al.}(1985)\citenamefont{Singh, Stoessel, and
  Wolynes}}]{SSW85}
\bibinfo{author}{\bibfnamefont{Y.}~\bibnamefont{Singh}},
  \bibinfo{author}{\bibfnamefont{J.~P.} \bibnamefont{Stoessel}},
  \bibnamefont{and} \bibinfo{author}{\bibfnamefont{P.~G.}
  \bibnamefont{Wolynes}}, \bibinfo{journal}{Phys. Rev. Lett.}
  \textbf{\bibinfo{volume}{54}}, \bibinfo{pages}{1059} (\bibinfo{year}{1985}).

\bibitem[{\citenamefont{Kirkpatrick and Wolynes}(1987)}]{KW87}
\bibinfo{author}{\bibfnamefont{T.~R.} \bibnamefont{Kirkpatrick}}
  \bibnamefont{and} \bibinfo{author}{\bibfnamefont{P.~G.}
  \bibnamefont{Wolynes}}, \bibinfo{journal}{Phys. Rev. A}
  \textbf{\bibinfo{volume}{35}}, \bibinfo{pages}{3072} (\bibinfo{year}{1987}).

\bibitem{MCT}
W.~G\"otze, J.~Phys.:~Condens.~Matter {\bf 11}, A1-A45 (1999);
D.~R.~Reichman and P.~Charbonneau, J.~Stat.~Mech. (2005) P05013. 


\bibitem[{\citenamefont{Kirkpatrick and Thirumalai}(1987)}]{KT87}
\bibinfo{author}{\bibfnamefont{T.~R.} \bibnamefont{Kirkpatrick}}
  \bibnamefont{and}
  \bibinfo{author}{\bibfnamefont{D.}~\bibnamefont{Thirumalai}},
  \bibinfo{journal}{Phys. Rev. Lett.} \textbf{\bibinfo{volume}{58}},
  \bibinfo{pages}{2091} (\bibinfo{year}{1987}).

\bibitem[{\citenamefont{Kirkpatrick et~al.}(1989)\citenamefont{Kirkpatrick,
  Thirumalai, and Wolynes}}]{KTW89}
\bibinfo{author}{\bibfnamefont{T.~R.} \bibnamefont{Kirkpatrick}},
  \bibinfo{author}{\bibfnamefont{D.}~\bibnamefont{Thirumalai}},
  \bibnamefont{and} \bibinfo{author}{\bibfnamefont{P.~G.}
  \bibnamefont{Wolynes}}, \bibinfo{journal}{Phys. Rev. A}
  \textbf{\bibinfo{volume}{40}}, \bibinfo{pages}{1045} (\bibinfo{year}{1989}).

\bibitem[{\citenamefont{Cugliandolo and Kurchan}(1993)}]{CK93}
\bibinfo{author}{\bibfnamefont{L.~F.} \bibnamefont{Cugliandolo}}
  \bibnamefont{and} \bibinfo{author}{\bibfnamefont{J.}~\bibnamefont{Kurchan}},
  \bibinfo{journal}{Phys. Rev. Lett.} \textbf{\bibinfo{volume}{71}},
  \bibinfo{pages}{173} (\bibinfo{year}{1993}).

\bibitem[{\citenamefont{Cavagna}(2009)}]{Ca09}
\bibinfo{author}{\bibfnamefont{A.}~\bibnamefont{Cavagna}},
  \bibinfo{journal}{Physics Reports} \textbf{\bibinfo{volume}{476}},
  \bibinfo{pages}{51} (\bibinfo{year}{2009}).

\bibitem[{\citenamefont{Biroli and Bouchaud}(2009)}]{BB09}
\bibinfo{author}{\bibfnamefont{G.}~\bibnamefont{Biroli}} \bibnamefont{and}
  \bibinfo{author}{\bibfnamefont{J.-P.} \bibnamefont{Bouchaud}},
  \bibinfo{journal}{{\tt arXiv.org:0912.2542}}  (\bibinfo{year}{2009}).

\bibitem[{\citenamefont{Biroli and M\'ezard}(2001)}]{BM01}
\bibinfo{author}{\bibfnamefont{G.}~\bibnamefont{Biroli}} \bibnamefont{and}
  \bibinfo{author}{\bibfnamefont{M.}~\bibnamefont{M\'ezard}},
  \bibinfo{journal}{Phys. Rev. Lett.} \textbf{\bibinfo{volume}{88}},
  \bibinfo{pages}{025501} (\bibinfo{year}{2001}).

\bibitem[{\citenamefont{Rivoire et~al.}(2004)\citenamefont{Rivoire, Biroli,
  Martin, and M\'ezard}}]{RBMM04}
\bibinfo{author}{\bibfnamefont{O.}~\bibnamefont{Rivoire}},
  \bibinfo{author}{\bibfnamefont{G.}~\bibnamefont{Biroli}},
  \bibinfo{author}{\bibfnamefont{O.}~\bibnamefont{Martin}}, \bibnamefont{and}
  \bibinfo{author}{\bibfnamefont{M.}~\bibnamefont{M\'ezard}},
  \bibinfo{journal}{Eur.Phys.J B} \textbf{\bibinfo{volume}{37}},
  \bibinfo{pages}{55} (\bibinfo{year}{2004}).

\bibitem[{\citenamefont{Pica~Ciamarra et~al.}(2003)\citenamefont{Pica~Ciamarra,
  Tarzia, de~Candia, and Coniglio}}]{PTCC03}
\bibinfo{author}{\bibfnamefont{M.}~\bibnamefont{Pica~Ciamarra}},
  \bibinfo{author}{\bibfnamefont{M.}~\bibnamefont{Tarzia}},
  \bibinfo{author}{\bibfnamefont{A.}~\bibnamefont{de~Candia}},
  \bibnamefont{and} \bibinfo{author}{\bibfnamefont{A.}~\bibnamefont{Coniglio}},
  \bibinfo{journal}{Phys. Rev. E} \textbf{\bibinfo{volume}{67}},
  \bibinfo{pages}{057105} (\bibinfo{year}{2003}).

\bibitem[{\citenamefont{M\'ezard and Parisi}(1999)}]{MP99}
\bibinfo{author}{\bibfnamefont{M.}~\bibnamefont{M\'ezard}} \bibnamefont{and}
  \bibinfo{author}{\bibfnamefont{G.}~\bibnamefont{Parisi}},
  \bibinfo{journal}{The Journal of Chemical Physics}
  \textbf{\bibinfo{volume}{111}}, \bibinfo{pages}{1076} (\bibinfo{year}{1999}).

\bibitem[{\citenamefont{Mezard and Parisi}(2009)}]{MP09}
\bibinfo{author}{\bibfnamefont{M.}~\bibnamefont{Mezard}} \bibnamefont{and}
  \bibinfo{author}{\bibfnamefont{G.}~\bibnamefont{Parisi}},
  \emph{\bibinfo{title}{{Glasses and replicas}}} (\bibinfo{year}{2009}),
  \eprint{{\tt arXiv:0910.2838}}.

\bibitem[{\citenamefont{Parisi and Zamponi}(2010)}]{PZ10}
\bibinfo{author}{\bibfnamefont{G.}~\bibnamefont{Parisi}} \bibnamefont{and}
  \bibinfo{author}{\bibfnamefont{F.}~\bibnamefont{Zamponi}},
  \bibinfo{journal}{Rev. Mod. Phys.} \textbf{\bibinfo{volume}{82}},
  \bibinfo{pages}{789} (\bibinfo{year}{2010}).

\bibitem[{\citenamefont{Goldschmidt}(1990)}]{Go90}
\bibinfo{author}{\bibfnamefont{Y.~Y.} \bibnamefont{Goldschmidt}},
  \bibinfo{journal}{Phys. Rev. B} \textbf{\bibinfo{volume}{41}},
  \bibinfo{pages}{4858} (\bibinfo{year}{1990}).

\bibitem[{\citenamefont{Nieuwenhuizen and Ritort}(1998)}]{NR98}
\bibinfo{author}{\bibfnamefont{T.}~\bibnamefont{Nieuwenhuizen}}
  \bibnamefont{and} \bibinfo{author}{\bibfnamefont{F.}~\bibnamefont{Ritort}},
  \bibinfo{journal}{Physica A} \textbf{\bibinfo{volume}{250}},
  \bibinfo{pages}{8} (\bibinfo{year}{1998}).

\bibitem[{\citenamefont{Biroli and Cugliandolo}(2001)}]{BC01}
\bibinfo{author}{\bibfnamefont{G.}~\bibnamefont{Biroli}} \bibnamefont{and}
  \bibinfo{author}{\bibfnamefont{L.~F.} \bibnamefont{Cugliandolo}},
  \bibinfo{journal}{Phys. Rev. B} \textbf{\bibinfo{volume}{64}},
  \bibinfo{pages}{014206} (\bibinfo{year}{2001}).

\bibitem[{\citenamefont{Cugliandolo et~al.}(2001)\citenamefont{Cugliandolo,
  Grempel, and da~Silva~Santos}}]{CGS01}
\bibinfo{author}{\bibfnamefont{L.~F.} \bibnamefont{Cugliandolo}},
  \bibinfo{author}{\bibfnamefont{D.~R.} \bibnamefont{Grempel}},
  \bibnamefont{and} \bibinfo{author}{\bibfnamefont{C.~A.}
  \bibnamefont{da~Silva~Santos}}, \bibinfo{journal}{Phys. Rev. B}
  \textbf{\bibinfo{volume}{64}}, \bibinfo{pages}{014403}
  (\bibinfo{year}{2001}).

\bibitem[{\citenamefont{J\"org et~al.}(2008)\citenamefont{J\"org, Krzakala,
  Kurchan, and Maggs}}]{JKKM08}
\bibinfo{author}{\bibfnamefont{T.}~\bibnamefont{J\"org}},
  \bibinfo{author}{\bibfnamefont{F.}~\bibnamefont{Krzakala}},
  \bibinfo{author}{\bibfnamefont{J.}~\bibnamefont{Kurchan}}, \bibnamefont{and}
  \bibinfo{author}{\bibfnamefont{A.~C.} \bibnamefont{Maggs}},
  \bibinfo{journal}{Phys. Rev. Lett.} \textbf{\bibinfo{volume}{101}},
  \bibinfo{pages}{147204} (\bibinfo{year}{2008}).

\bibitem[{\citenamefont{Westfahl et~al.}(2003)\citenamefont{Westfahl,
  Schmalian, and Wolynes}}]{WSW03}
\bibinfo{author}{\bibfnamefont{H.}~\bibnamefont{Westfahl}},
  \bibinfo{author}{\bibfnamefont{J.}~\bibnamefont{Schmalian}},
  \bibnamefont{and} \bibinfo{author}{\bibfnamefont{P.~G.}
  \bibnamefont{Wolynes}}, \bibinfo{journal}{Phys. Rev. B}
  \textbf{\bibinfo{volume}{68}}, \bibinfo{pages}{134203}
  (\bibinfo{year}{2003}).

\bibitem[{\citenamefont{J\"org et~al.}(2010)\citenamefont{J\"org, Krzakala,
  Semerjian, and Zamponi}}]{JKSZ10}
\bibinfo{author}{\bibfnamefont{T.}~\bibnamefont{J\"org}},
  \bibinfo{author}{\bibfnamefont{F.}~\bibnamefont{Krzakala}},
  \bibinfo{author}{\bibfnamefont{G.}~\bibnamefont{Semerjian}},
  \bibnamefont{and} \bibinfo{author}{\bibfnamefont{F.}~\bibnamefont{Zamponi}},
  \bibinfo{journal}{Phys. Rev. Lett.} \textbf{\bibinfo{volume}{104}},
  \bibinfo{pages}{207206} (\bibinfo{year}{2010}).

\bibitem[{\citenamefont{Foini et~al.}(2010{\natexlab{a}})\citenamefont{Foini,
  Semerjian, and Zamponi}}]{FSZ10}
\bibinfo{author}{\bibfnamefont{L.}~\bibnamefont{Foini}},
  \bibinfo{author}{\bibfnamefont{G.}~\bibnamefont{Semerjian}},
  \bibnamefont{and} \bibinfo{author}{\bibfnamefont{F.}~\bibnamefont{Zamponi}},
  \bibinfo{journal}{Phys. Rev. Lett.} \textbf{\bibinfo{volume}{105}},
  \bibinfo{pages}{167204} (\bibinfo{year}{2010}{\natexlab{a}}).

\bibitem[{\citenamefont{Markland et~al.}(2010)\citenamefont{Markland, Morrone,
  Berne, Miyazaki, Rabani, and Reichman}}]{MMBMRR10}
\bibinfo{author}{\bibfnamefont{T.~E.} \bibnamefont{Markland}},
  \bibinfo{author}{\bibfnamefont{J.~A.} \bibnamefont{Morrone}},
  \bibinfo{author}{\bibfnamefont{B.~J.} \bibnamefont{Berne}},
  \bibinfo{author}{\bibfnamefont{K.}~\bibnamefont{Miyazaki}},
  \bibinfo{author}{\bibfnamefont{E.}~\bibnamefont{Rabani}}, \bibnamefont{and}
  \bibinfo{author}{\bibfnamefont{D.~R.} \bibnamefont{Reichman}},
  \bibinfo{journal}{Nature Physics (published online) {\tt arXiv:}1011.0015}
  (\bibinfo{year}{2010}).

\bibitem[{\citenamefont{M\'ezard et~al.}(1987)\citenamefont{M\'ezard, Parisi,
  and Virasoro}}]{MPV87}
\bibinfo{author}{\bibfnamefont{M.}~\bibnamefont{M\'ezard}},
  \bibinfo{author}{\bibfnamefont{G.}~\bibnamefont{Parisi}}, \bibnamefont{and}
  \bibinfo{author}{\bibfnamefont{M.~A.} \bibnamefont{Virasoro}},
  \emph{\bibinfo{title}{Spin glass theory and beyond}}
  (\bibinfo{publisher}{World Scientific}, \bibinfo{address}{Singapore},
  \bibinfo{year}{1987}).

\bibitem[{\citenamefont{Carleo et~al.}(2009)\citenamefont{Carleo, Tarzia, and
  Zamponi}}]{CTZ09}
\bibinfo{author}{\bibfnamefont{G.}~\bibnamefont{Carleo}},
  \bibinfo{author}{\bibfnamefont{M.}~\bibnamefont{Tarzia}}, \bibnamefont{and}
  \bibinfo{author}{\bibfnamefont{F.}~\bibnamefont{Zamponi}},
  \bibinfo{journal}{Phys. Rev. Lett.} \textbf{\bibinfo{volume}{103}},
  \bibinfo{pages}{215302} (\bibinfo{year}{2009}).

\bibitem[{\citenamefont{Tam et~al.}(2010)\citenamefont{Tam, Geraedts, Inglis,
  Gingras, and Melko}}]{TGIGM10}
\bibinfo{author}{\bibfnamefont{K.-M.} \bibnamefont{Tam}},
  \bibinfo{author}{\bibfnamefont{S.}~\bibnamefont{Geraedts}},
  \bibinfo{author}{\bibfnamefont{S.}~\bibnamefont{Inglis}},
  \bibinfo{author}{\bibfnamefont{M.~J.~P.} \bibnamefont{Gingras}},
  \bibnamefont{and} \bibinfo{author}{\bibfnamefont{R.~G.} \bibnamefont{Melko}},
  \bibinfo{journal}{Phys. Rev. Lett.} \textbf{\bibinfo{volume}{104}},
  \bibinfo{pages}{215301} (\bibinfo{year}{2010}).

\bibitem[{\citenamefont{Laumann et~al.}(2008)\citenamefont{Laumann,
  Scardicchio, and Sondhi}}]{LSS08}
\bibinfo{author}{\bibfnamefont{C.}~\bibnamefont{Laumann}},
  \bibinfo{author}{\bibfnamefont{A.}~\bibnamefont{Scardicchio}},
  \bibnamefont{and} \bibinfo{author}{\bibfnamefont{S.~L.}
  \bibnamefont{Sondhi}}, \bibinfo{journal}{Phys. Rev. B}
  \textbf{\bibinfo{volume}{78}}, \bibinfo{eid}{134424} (\bibinfo{year}{2008}).

\bibitem[{\citenamefont{Krzakala
  et~al.}(2008{\natexlab{a}})\citenamefont{Krzakala, Rosso, Semerjian, and
  Zamponi}}]{KRSZ08}
\bibinfo{author}{\bibfnamefont{F.}~\bibnamefont{Krzakala}},
  \bibinfo{author}{\bibfnamefont{A.}~\bibnamefont{Rosso}},
  \bibinfo{author}{\bibfnamefont{G.}~\bibnamefont{Semerjian}},
  \bibnamefont{and} \bibinfo{author}{\bibfnamefont{F.}~\bibnamefont{Zamponi}},
  \bibinfo{journal}{Phys. Rev. B} \textbf{\bibinfo{volume}{78}},
  \bibinfo{pages}{134428} (\bibinfo{year}{2008}{\natexlab{a}}).

\bibitem[{\citenamefont{Semerjian et~al.}(2009)\citenamefont{Semerjian, Tarzia,
  and Zamponi}}]{STZ09}
\bibinfo{author}{\bibfnamefont{G.}~\bibnamefont{Semerjian}},
  \bibinfo{author}{\bibfnamefont{M.}~\bibnamefont{Tarzia}}, \bibnamefont{and}
  \bibinfo{author}{\bibfnamefont{F.}~\bibnamefont{Zamponi}},
  \bibinfo{journal}{Phys. Rev. B} \textbf{\bibinfo{volume}{80}},
  \bibinfo{eid}{014524} (\bibinfo{year}{2009}).

\bibitem[{\citenamefont{Dawson et~al.}(2002)\citenamefont{Dawson, Lawlor,
  de~Gregorio, McCullagh, Zaccarelli, and Tartaglia}}]{DLGMZT02}
\bibinfo{author}{\bibfnamefont{K.~A.} \bibnamefont{Dawson}},
  \bibinfo{author}{\bibfnamefont{A.}~\bibnamefont{Lawlor}},
  \bibinfo{author}{\bibfnamefont{P.}~\bibnamefont{de~Gregorio}},
  \bibinfo{author}{\bibfnamefont{G.~D.} \bibnamefont{McCullagh}},
  \bibinfo{author}{\bibfnamefont{E.}~\bibnamefont{Zaccarelli}},
  \bibnamefont{and}
  \bibinfo{author}{\bibfnamefont{P.}~\bibnamefont{Tartaglia}},
  \bibinfo{journal}{Physica A} \textbf{\bibinfo{volume}{316}},
  \bibinfo{pages}{115 } (\bibinfo{year}{2002}).

\bibitem[{\citenamefont{Darst et~al.}(2010)\citenamefont{Darst, Reichman, and
  Biroli}}]{DRB10}
\bibinfo{author}{\bibfnamefont{R.}~\bibnamefont{Darst}},
  \bibinfo{author}{\bibfnamefont{D.}~\bibnamefont{Reichman}}, \bibnamefont{and}
  \bibinfo{author}{\bibfnamefont{G.}~\bibnamefont{Biroli}},
  \bibinfo{journal}{The Journal of chemical physics}
  \textbf{\bibinfo{volume}{132}}, \bibinfo{pages}{044510}
  (\bibinfo{year}{2010}).

\bibitem[{\citenamefont{M\'ezard and Parisi}(2001)}]{cavity}
\bibinfo{author}{\bibfnamefont{M.}~\bibnamefont{M\'ezard}} \bibnamefont{and}
  \bibinfo{author}{\bibfnamefont{G.}~\bibnamefont{Parisi}},
  \bibinfo{journal}{Eur. Phys. J. B} \textbf{\bibinfo{volume}{20}},
  \bibinfo{pages}{217} (\bibinfo{year}{2001}).

\bibitem[{\citenamefont{Monasson}(1995)}]{Mo95}
\bibinfo{author}{\bibfnamefont{R.}~\bibnamefont{Monasson}},
  \bibinfo{journal}{Phys. Rev. Lett.} \textbf{\bibinfo{volume}{75}},
  \bibinfo{pages}{2847} (\bibinfo{year}{1995}).

\bibitem[{\citenamefont{Castellani and Cavagna}(2005)}]{CC05}
\bibinfo{author}{\bibfnamefont{T.}~\bibnamefont{Castellani}} \bibnamefont{and}
  \bibinfo{author}{\bibfnamefont{A.}~\bibnamefont{Cavagna}},
  \bibinfo{journal}{Journal of Statistical Mechanics: Theory and Experiment}
  \textbf{\bibinfo{volume}{2005}}, \bibinfo{pages}{P05012}
  (\bibinfo{year}{2005}).

\bibitem[{\citenamefont{Berthier et~al.}(2010)\citenamefont{Berthier, Moreno,
  and Szamel}}]{BMS10}
\bibinfo{author}{\bibfnamefont{L.}~\bibnamefont{Berthier}},
  \bibinfo{author}{\bibfnamefont{A.~J.} \bibnamefont{Moreno}},
  \bibnamefont{and} \bibinfo{author}{\bibfnamefont{G.}~\bibnamefont{Szamel}},
  \bibinfo{journal}{Phys.~Rev.~E {\bf 82}, 060501(R)}  (\bibinfo{year}{2010}).

\bibitem[{\citenamefont{Krzakala
  et~al.}(2008{\natexlab{b}})\citenamefont{Krzakala, Tarzia, and
  Zdeborov\'{a}}}]{KTZ08}
\bibinfo{author}{\bibfnamefont{F.}~\bibnamefont{Krzakala}},
  \bibinfo{author}{\bibfnamefont{M.}~\bibnamefont{Tarzia}}, \bibnamefont{and}
  \bibinfo{author}{\bibfnamefont{L.}~\bibnamefont{Zdeborov\'{a}}},
  \bibinfo{journal}{Phys. Rev. Lett.} \textbf{\bibinfo{volume}{101}},
  \bibinfo{eid}{165702} (\bibinfo{year}{2008}{\natexlab{b}}).

\bibitem[{\citenamefont{Foini et~al.}(2010{\natexlab{b}})\citenamefont{Foini,
  Krzakala, Semerjian, and Zamponi}}]{FKSZ10}
\bibinfo{author}{\bibfnamefont{L.}~\bibnamefont{Foini}},
  \bibinfo{author}{\bibfnamefont{F.}~\bibnamefont{Krzakala}},
  \bibinfo{author}{\bibfnamefont{G.}~\bibnamefont{Semerjian}},
  \bibnamefont{and} \bibinfo{author}{\bibfnamefont{F.}~\bibnamefont{Zamponi}},
  \bibinfo{journal}{in preparation}  (\bibinfo{year}{2010}{\natexlab{b}}).

\bibitem[{\citenamefont{Cugliandolo et~al.}(2004)\citenamefont{Cugliandolo,
  Grempel, Lozano, and Lozza}}]{CGLL04}
\bibinfo{author}{\bibfnamefont{L.~F.} \bibnamefont{Cugliandolo}},
  \bibinfo{author}{\bibfnamefont{D.~R.} \bibnamefont{Grempel}},
  \bibinfo{author}{\bibfnamefont{G.}~\bibnamefont{Lozano}}, \bibnamefont{and}
  \bibinfo{author}{\bibfnamefont{H.}~\bibnamefont{Lozza}},
  \bibinfo{journal}{Phys. Rev. B} \textbf{\bibinfo{volume}{70}},
  \bibinfo{pages}{024422} (\bibinfo{year}{2004}).

\bibitem[{\citenamefont{Cugliandolo et~al.}(2002)\citenamefont{Cugliandolo,
  Grempel, Lozano, Lozza, and da~Silva~Santos}}]{CGLLS02}
\bibinfo{author}{\bibfnamefont{L.~F.} \bibnamefont{Cugliandolo}},
  \bibinfo{author}{\bibfnamefont{D.~R.} \bibnamefont{Grempel}},
  \bibinfo{author}{\bibfnamefont{G.}~\bibnamefont{Lozano}},
  \bibinfo{author}{\bibfnamefont{H.}~\bibnamefont{Lozza}}, \bibnamefont{and}
  \bibinfo{author}{\bibfnamefont{C.~A.} \bibnamefont{da~Silva~Santos}},
  \bibinfo{journal}{Phys. Rev. B} \textbf{\bibinfo{volume}{66}},
  \bibinfo{pages}{014444} (\bibinfo{year}{2002}).

\bibitem[{\citenamefont{Ceperley}(1995)}]{Ce95}
\bibinfo{author}{\bibfnamefont{D.~M.} \bibnamefont{Ceperley}},
  \bibinfo{journal}{Reviews of Modern Physics} \textbf{\bibinfo{volume}{67}}
  (\bibinfo{year}{1995}).

\bibitem[{\citenamefont{Penrose and Onsager}(1956)}]{PO56}
\bibinfo{author}{\bibfnamefont{O.}~\bibnamefont{Penrose}} \bibnamefont{and}
  \bibinfo{author}{\bibfnamefont{L.}~\bibnamefont{Onsager}},
  \bibinfo{journal}{Phys. Rev.} \textbf{\bibinfo{volume}{104}},
  \bibinfo{pages}{576} (\bibinfo{year}{1956}).

\bibitem[{\citenamefont{Deutsch}(1991)}]{De91}
\bibinfo{author}{\bibfnamefont{J.~M.} \bibnamefont{Deutsch}},
  \bibinfo{journal}{Phys. Rev. A} \textbf{\bibinfo{volume}{43}},
  \bibinfo{pages}{2046} (\bibinfo{year}{1991}).

\bibitem[{\citenamefont{Srednicki}(1994)}]{Sr94}
\bibinfo{author}{\bibfnamefont{M.}~\bibnamefont{Srednicki}},
  \bibinfo{journal}{Phys. Rev. E} \textbf{\bibinfo{volume}{50}},
  \bibinfo{pages}{888} (\bibinfo{year}{1994}).

\bibitem[{\citenamefont{Pal and Huse}(2010)}]{PH10}
\bibinfo{author}{\bibfnamefont{A.}~\bibnamefont{Pal}} \bibnamefont{and}
  \bibinfo{author}{\bibfnamefont{D.~A.} \bibnamefont{Huse}},
  \bibinfo{journal}{Phys. Rev. B} \textbf{\bibinfo{volume}{82}},
  \bibinfo{pages}{174411} (\bibinfo{year}{2010}).

\bibitem[{\citenamefont{Berkelbach and Reichman}(2010)}]{BR10}
\bibinfo{author}{\bibfnamefont{T.~C.} \bibnamefont{Berkelbach}}
  \bibnamefont{and} \bibinfo{author}{\bibfnamefont{D.~R.}
  \bibnamefont{Reichman}}, \bibinfo{journal}{Phys. Rev. B}
  \textbf{\bibinfo{volume}{81}}, \bibinfo{pages}{224429}
  (\bibinfo{year}{2010}).

\bibitem[{\citenamefont{Canovi et~al.}(2010)\citenamefont{Canovi, Rossini,
  Fazio, Santoro, and Silva}}]{CRFSS10}
\bibinfo{author}{\bibfnamefont{E.}~\bibnamefont{Canovi}},
  \bibinfo{author}{\bibfnamefont{D.}~\bibnamefont{Rossini}},
  \bibinfo{author}{\bibfnamefont{R.}~\bibnamefont{Fazio}},
  \bibinfo{author}{\bibfnamefont{G.~E.} \bibnamefont{Santoro}},
  \bibnamefont{and} \bibinfo{author}{\bibfnamefont{A.}~\bibnamefont{Silva}}
  (\bibinfo{year}{2010}), \eprint{{\tt arXiv:1006.1634}}.

\bibitem{ETH_Giulio}
G.~Biroli, C.~Kollath, and A.~M.~L\"auchli,
Phys.~Rev.~Lett. {\bf 105}, 250401 (2010).

\bibitem[{\citenamefont{{Gaveau} and {Schulman}}(1998)}]{GS98}
\bibinfo{author}{\bibfnamefont{B.}~\bibnamefont{{Gaveau}}} \bibnamefont{and}
  \bibinfo{author}{\bibfnamefont{L.~S.} \bibnamefont{{Schulman}}},
  \bibinfo{journal}{Journal of Mathematical Physics}
  \textbf{\bibinfo{volume}{39}}, \bibinfo{pages}{1517} (\bibinfo{year}{1998}).

\end{thebibliography}
\end{document}